%% file: SHOCK_5a.tex
\documentclass[showpacs,showkeys,preprintnumbers,amsmath,amssymb,times,graphicx]{revtex4}

\usepackage{graphicx} 
\usepackage{dcolumn}
\usepackage{bm}

\usepackage{color}
\usepackage{amsmath}
\usepackage{amsfonts}
\usepackage{amssymb}
\usepackage{amsbsy}
\usepackage[figuresright]{rotating}
\usepackage[font=small,labelfont=bf]{caption}
\usepackage{natbib}

\input{calbook_a_defs}

\begin{document}

\title{Fundamentals of  Non-relativistic Collisionless Shock Physics: \\  IV. Quasi-Parallel Supercritical Shocks}

\author{R. A. Treumann$^\dag$ and C. H. Jaroschek$^{*}$}\email{treumann@issibern.ch}
\affiliation{$^\dag$ Department of Geophysics and Environmental Sciences, Munich University, D-80333 Munich, Germany  \\ 
Department of Physics and Astronomy, Dartmouth College, Hanover, 03755 NH, USA \\ 
$^{*}$Department Earth \& Planetary Science, University of Tokyo, Tokyo, Japan
}%

\begin{abstract} 1. Introduction, 2. The (quasi-parallel) foreshock; Ion foreshock, Ion foreshock boundary region; Diffuse ions;Low-frequency upstream waves; Ion beam waves; The expected wave modes; Observations; Diffuse ion waves; Electron foreshock; Electron beams; Langmuir waves; stability of the electron beam; Electron foreshock boundary waves; Nature of electron foreshock waves; Radiation; Observations; Interpretation; 3. Quasi-parallel shock reformation; Low-Mach number quasi-parallel shocks; Turbulent reformation; Observations; Simulations of quasi-parallel shock reformation; Hybrid simulations in 1D; Hybrid simulations in 2D; Full particle PIC simulations; Conclusions; 4. Hot flow anomalies; Observations; Models and simulations; Solitary shock; 5. Downstream region; 6. Summary and conclusions.
\end{abstract}
\maketitle 

\section{Introduction}\noindent
At a first glance it is surprising that a change in the shock-normal angle $\thetabn$ by just a few degrees from, say, $\thetabn=50^\circ$ to $\thetabn=40^\circ$ should completely change the character of the supercritical shock. We have seen in the last chapter, when having discussed the conditions for reflection of ions from  the shock ramp that such a change in the shock properties is theoretically predicted. The critical shock-normal angle $\thetabn=45^\circ$ does indeed separate two completely different phases of a supercritical shock. At this angle the shock experiences a `phase transition' from the quasi-perpendicular to the quasi-parallel shock state, with the shock-normal angle $\thetabn$ having the property of a `critical control parameter'. \index{shocks!`phase transition'}

We have already learned that the reason for the different behaviour of the two shock phases is that in a quasi-perpendicular shock all reflected particles from the foot region of the shock return to the shock after not more than a few gyrations when they have picked up sufficient energy in the upstream convection electric field ${\bf E}_1 = -{\bf V}_1\times{\bf B}_1$ to ultimately overcome the shock ramp potential, pass the shock ramp and to merge into the downstream flow. We have not discussed what happens to the accelerated ions in the downstream region as this is not of primary importance in the shock formation mechanism which to good approximation depends only on the upstream conditions. This question will be treated in a separate section on shock-particle acceleration. 

In contrast, in a quasi-parallel shock the combined geometries of the upstream magnetic field and generally curved shock surface prevent the shock-reflected particles from immediate return to the shock. The reason is that their gyro-orbits, after having suffered reflection from the shock ramp, lie completely upstream, outside the shock ramp, such that they do not touch the shock ramp again after reflection. Since, in addition, their upstream velocities have a large component parallel to the upstream magnetic field, which increases the more the shock-normal turns parallel to the upstream magnetic field, the reflected particles are enabled to escape upstream from the shock along the magnetic field thereby forming fast upstream particle beams. 

A quasi-parallel supercritical collisionless shock thus populates the upstream space with a reflected particle component. This population moves a long distance away from the shock along the magnetic field. Was the upstream flow, in the case of the quasi-perpendicular shock, completely uninformed about the presence of the shock up to a distance of the mere width of the shock foot so,  at the quasi-parallel shock, it receives a first signal of the presence of a shock already at quite a large upstream distance when the first and fastest reflected particles arrive on the magnetic field lines that connect the flow to the shock. It is, in fact, only these particles that can inform the flow about the presence of a supercritical shock, because any low-frequency plasma wave cannot propagate far upstream for supercritical Mach numbers ${\cal M>M}_c$, while any electromagnetic radiation that is generated at the shock has frequency $\omega_{rad}>\omega_{pe}$. For it the plasma flow presents a vacuum. The upstream flow recognises the reflected particles in its own frame of reference as a high-speed magnetic-field aligned beam. Thereby a beam-beam configuration is created which leads to a number of beam-driven instabilities. These excite various plasma waves that fill the space in front of the shock and modify its properties. 

\begin{figure}[t!]
\hspace{0.0cm}\centerline{\includegraphics[width=0.95\textwidth,clip=]{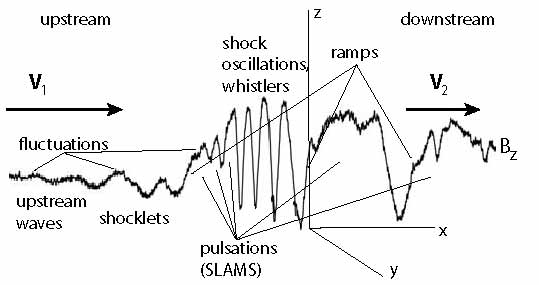} }
\caption[Sketch of quasi-parallel shock profile]
{\footnotesize Schematic one-dimensional profile taken along the nominal instantaneous shock normal of a supercritical quasi-parallel shock as seen in the magnetic field component $B_z$. This is the analogue to the quasi-perpendicular shock profile. It shows the main features in the vicinity of the quasi-parallel shock transition: the large amplitude upstream waves with the turbulent fluctuations on top of the waves, the formation of shocklets, i.e. steep flank formation on the waves exhibiting small-scale fluctuations on top of the wave, which act already like small shocks, very-large amplitude pulsations  (magnetic pulsations or {\SL}) which turn out to be the building blocks of the shock, multiple shock-ramps at the leading edges of the pulsations belonging to diverse ramp-like steep transitions from upstream to downstream lacking a clear localisation of the shock transition (Note that the entire figure is, in fact, the shock transition, as on this scale no clear decision can be made where the shock ramp is located.), and their attached phase-locked whistlers. Not shown here are the out of plane oscillations of the magnetic field that accompany the waves. Also not shown is the particle phase space.}\label{chap5-fig-qpashock}
\end{figure}

Figure\,\ref{chap5-fig-qpashock} shows a sketch of the magnetic profile of a supercritical quasi-parallel shock which contrasts the profile of a quasi-perpendicular shock that has been given in the previous chapter. The quasi-parallel magnetic shock profile is much stronger distorted than that of a quasi-perpendicular shock, such that it becomes difficult to identify the location of the genuine shock ramp on the profile. 

The main difference between quasi-perpendicular and quasi-parallel shocks is that quasi-perpendicular shocks possess a narrow $\sim 1 r_{ci}$ wide foot region that is tangential to the shock surface, while quasi-parallel shocks possess an extended {\it foreshock}\index{shocks!foreshock} region. Interestingly, in curved shocks which arise, for instance, in front of spatially confined obstacles, both phases of a supercritical shock can co-exists at the same time, being spatially adjacent to each other. An example is shown in Figure\,\ref{Tsurutani-f1} in the sketch of the curved Earth's bow shock. Dealing with quasi-parallel shocks means to a large extent dealing with the processes that are going on in the foreshock.  It will thus be quite natural to start with a discussion of the properties of the foreshock. This discussion will occupy a substantial part of this chapter.

However, before continuing we point out that in spite of the strict distinction between the quasi-perpendicular and quasi-parallel shocks there is also a close relation between the two. Both, being supercritical, can exist only because they reflect ions; and both possess an upstream region in front of the shock transition that is populated by the reflected ions. That this region is narrow in the case of quasi-perpendicular shocks is a question of the ions being tied to the magnetic field, which also holds in the case of the quasi-parallel shock. At quasi-perpendicular shocks the reflected ions do readily return to the shock. At the quasi-parallel shock they ultimately do also return to the shock, but only after having been processed far away from the shock in the foreshock, having coupled to the flow, having passed several stages in this processing, and having become the energetic component of the main flow. As such they finally arrive at the shock together with the stream. In between their main duty was to dissipate the excess energy, which they possessed when arriving for the first time at the shock and which could not be dissipated in the narrow shock-ramp transition region. This could been achieved only in the broad extended foreshock which, seen from this point of view, {\it is already the shock}. It belongs inextricably to the quasi-parallel shock transition. Here, a substantial fraction of the energy of the incident flow is dissipated in a way which is completely different from the flow being shocked. 

It is these dissipation processes that cause the main difference between the quasi-perpendicular and quasi-parallel states of a collisionless supercritical shock. To stress the analogy with phase transitions a little further, we may say that quasi-perpendicular shocks are in the solid -- or ordered -- shock state, while quasi-parallel shocks are in the fluid -- or partially disordered -- shock state. \index{shocks!states of} 
\begin{figure}[t!]
\hspace{0.0cm}\centerline{\includegraphics[width=1.0\textwidth,clip=]{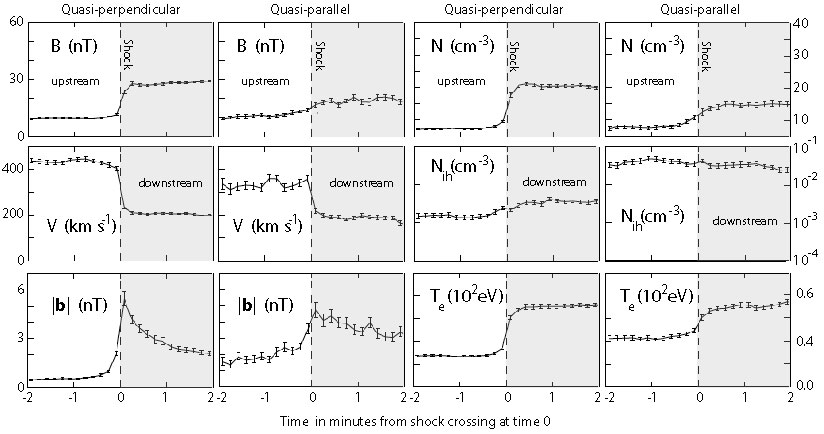} }
\caption[The Shock-foreshock Relation]
{\footnotesize Comparison of average bulk plasma parameters in quasi-perpendicular and quasi-parallel shocks \citep[after][]{Czaykowska2000}. The figure shows (on the left) the mean magnetic field $B$, bulk flow velocity $V$, average magnetic fluctuation amplitude at ultra-low frequencies, and (on the right) plasma density $N$, density of high energy ions $N_{ih}$, and electron temperature $T_e$. The shaded regions are the downstream parts of the AMPTE IRM crossings of the bow shock. The data have been obtained by normalising the time with respect to crossing the nominal shock ramp by using the measured normal component (not shown) of the bulk flow velocity. This for the many observations included in this figure implies a stretching (or squeezing) of each individual shock crossing, causing some uncertainty, in particular for quasi-parallel shock crossings as there the shock ramp is not well defined. However, this figure serves for an immediate overview of the differences in both types of shocks. }\label{chap5-fig-average}
\end{figure}
\section{The (Quasi-parallel Shock) Foreshock}\label{foreshock}\noindent
Quasi-parallel shocks are abundant in space because in most cases when shocks develop the flow direction is independent of the direction of the magnetic field. Moreover, as pointed out earlier, when a bow shock forms around an obstacle (planet, magnetosphere, moon ...) this bow shock is curved around the obstacle, and the shock is quasi-perpendicular only in a certain region on the shock surface that is centred around the point where the upstream magnetic field touches the shock tangentially. Farther away the shock turns gradually to become quasi-parallel. On the other hand, when a shock survives over a long distance in space as for instance in supernova remnants then it sweeps the upstream magnetic field and pushes it to become more tangential to the shock surface. In this case the shock is about quasi-perpendicular.  We will later provide arguments that any quasi-parallel supercritical shock on the small scale close to the shock surface, i.e. on the electron scale, behaves quasi-perpendicularly while on the larger ion scale it remains to be quasi-parallel. This has consequences for the differences in the dynamics of electrons and ions during  their interaction with quasi-parallel shocks.

Figure\,\ref{chap5-fig-average} shows at one glance the main differences in the (average) bulk plasma parameters between quasi-perpendicular and quasi-parallel shocks as measured with the {\IRM} spacecraft at many crossings of the bow shock. The data used in this figure have been stapled, averaged and plotted with respect to the time normalised to the shock ramp crossing. For such a normalisation one uses the shock-normal upstream velocity to recalculate the time. This procedure is not very certain for quasi-parallel shocks since -- as we will see later -- the shock ramp is ill defined in a quasi-parallel shock. However, for a simple comparison of the main differences this uncertainty is less severe. 

The shaded area in the figure corresponds to the downstream region. Shown are -- in pairs of quasi-perpendicular/quasi-parallel values -- the magnetic field $B$, bulk velocity $V$, average fluctuation amplitude in the ultra-low frequency waves $|{\bf b}|$, plasma density $N$, high-energy ion density $N_{ih}$ of energy $>15$\,keV, and electron temperature $T_e$. The general conclusion from this figure is that all quantities in the quasi-perpendicular case exhibit a much sharper transition than in the quasi-parallel case. Moreover, the quasi-perpendicular averages are quieter than those of the quasi-parallel case. Also, in general, the quasi-parallel levels are higher than the quasi-perpendicular. In almost all cases the pre-shock levels are enhanced in the quasi-parallel shock case with over the pre-shock levels of quasi-perpendicular shocks. This is seen most impressively in the energetic ion density, which is nearly constant over this distance/time scale at quasi-parallel shocks and much higher than that in quasi-perpendicular shocks, signifying on the one hand the importance of energetic particles in quasi-parallel shock dynamics, on the other hand the capability of quasi-parallel shocks to accelerate particles to substantial energies. The presence of energetic ions (particles) far in front of the quasi-parallel shocks and the enhanced pre-shock levels indicate the importance of foreshocks in quasi-parallel shock dynamics. In the following we will therefore first concentrate on the foreshock.

The physics of quasi-parallel shocks cannot be understood without reference to the foreshock. The foreshock is that part of the upstream shock region that is occupied with reflected particles. At a curved shock, like the Earth's bow shock, the foreshock starts on the shock surface at the location where the upstream magnetic field shock-normal angle exceeds $\thetabn\gtrsim 45^\circ$. From that point on electrons and ions escape along the magnetic field in upstream direction. Since electrons generally move at a larger parallel velocity than ions they are less vulnerable to the convective motion of the upstream magnetic field line to which they are tied, and so there is generally a region closer to the foreshock-boundary magnetic field line where only upstream electrons are found. This region is confined approximately between the line that marks the electron foreshock boundary and the more inclined line line that marks the ion foreshock boundary. An example of this geometry was depicted in Figure\,\ref{Tsurutani-f1} for the bow shock of the Earth. \index{foreshock!boundary}\index{foreshock!electron, ion}
\begin{figure}[t!]
\hspace{0.0cm}\centerline{\includegraphics[width=0.8\textwidth,height=0.4\textheight,clip=]{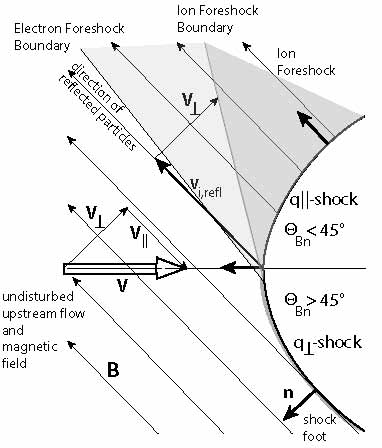} }
\caption[The Shock-foreshock Relation]
{\footnotesize Schematic of the the relation between a curved shock and its foreshock in dependence on the direction of the upstream magnetic field ${\bf B}$, shock-normal ${\bf n}$, and shock-normal angle $\thetabn$ for the special case when the magnetic field is inclined at $45^\circ$ with resepct to the symmetry axis of the shock. In this case the upper half of the shock becomes quasi-parallel ($\thetabn<45^\circ$), the lower half is quasi-perpendicular ($\thetabn>45^\circ$). The velocity of reflected particles is along the magnetic field. However, seeing the flow the field-line to which they are attached displaces with perpendicular velocity. This velocity shifts the foreshock boundary toward the shock as shown for electrons (light shading) and ions (darker shading). The ion foreshock is closer to the shock because of the lower velocity of the ions than the electrons. For the electrons the displacement of the electron foreshock boundary is felt only at large distances from the shock.}\label{chap5-fig-foreschem}
\end{figure}

More schematically this is shown in a simplified version in Figure\,\ref{chap5-fig-foreschem} for the particular case that the upstream magnetic field forms an angle of $45^\circ$ with the symmetry axis of the shock. In this case half of the shock is quasi-perpendicular and the other half is quasi-parallel. The figure also shows the directions of three shock normals, the narrow foot region in front of the quasi-perpendicular shock, and the two (electron and ion) foreshocks. Particles escape from the quasi-parallel shock along the upstream magnetic field. The magnetic field is convected toward the shock by the perpendicular upstream velocity component ${\bf V}_\perp$ as shown in the figure. This component adds to the velocity of the upstream particles leading to an inclined foreshock boundary. Since the ions have much smaller speed than the electrons, the ion foreshock boundary is more inclined than the electron foreshock boundary. 

In discussing the properties of the foreshock one thus has to distinguish of which foreshock is the talk. However, the properties of the electron foreshock are not as decisive for the formation of a quasi-parallel shock as are the properties of the ion foreshock. Because of this reason we will, in the following, refer to the ion foreshock as the foreshock. The electron foreshock properties we will mention only later.

\subsection{Ion foreshock}\noindent
The ion foreshock is not a homogeneous and uniform region. The reflected ion component evolves across the ion foreshock from the ion foreshock boundary to the centre of the ion foreshock and from there towards the shock. Speaking of a reflected ion component that can unambiguously identified as being reflected, i.e. streaming into the upward direction, makes sense only in the immediate vicinity of the ion foreshock boundary. Here the reflected ions appear as a fast ion beam\index{foreshock!boundary} the source of which can be traced back to the shock. Deeper in the foreshock the beam component cannot be identified anymore.  
\subsubsection*{The ion foreshock boundary region}\noindent
First identifications of reflected beam protons in space in the magnetic flux tube connected to the Earth's bow shock wave were reported by \citet{Gosling1978} and \citet{Paschmann1981} who distinguished those beams by their poorly resolved distribution functions from more diffuse protons deeper in the foreshock. Interestingly, observations in the foreshock of interplanetary travelling shocks did not show any indication of such beams but only the diffuse ion component.
\begin{figure}[t!]
\hspace{0.0cm}\centerline{\includegraphics[width=0.9\textwidth,clip=]{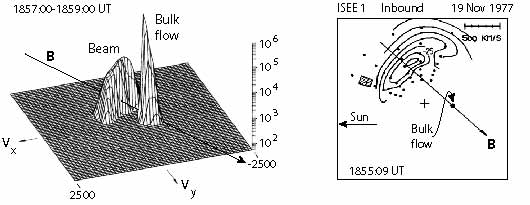} }
\caption[Reflected ion beam ISEE 1]
{\footnotesize ISEE 1 observation of a reflected ion beam on November 19, 1977 propagating along a magnetic field line that  was connected to the quasi-parallel Earth's bow shock. {\it Left}: A two-minute average pseudo-threedimensional ion velocity-space profile in the $(v_x,v_y)$-plane showing the undisturbed and cold (narrow)\index{foreshock!ion beam}\index{shocks!bow shock} plasma inflow in negative $v_x$-direction, and the fast and warm (broad) beam of reflected ions propagating in positive $v_x$-direction and spreading in $v_y$. This beam is quite anisotropic in temperature. Velocities are in km\,s$^{-1}$. The scale on the right is count rates, and background count rates were suppressed by choosing only values above 50 s$^{-1}$. {\it Right}: Contour plot of a similar beam a little earlier showing that the beam is centred on the magnetic field that connects to the shock, is quite narrow along the field and about 2-3 times as broad perpendicular to the field \citep[after][]{Paschmann1981}.  The cross indicates the origin (zero velocity), the dot the bulk flow centre. The $10^{-25}$\ s$^3$\,cm$^{-6}$ level flux contour has been marked. }\label{chap5-fig-pasch81a}
\end{figure}
Figure\,\ref{chap5-fig-pasch81a} gives an observational example of such a reflected ion beam that propagates very close to the foreshock boundary upstream away from the shock. The bulk flow is the narrow cold beam in the left part of the figure which is displaced in negative $v_x$-direction (note that in this figure the positive direction points away from the shock). The reflected beam is less dense (lower count rates) but much more energetic. It is displaced in $+v_x$-direction, i.e. streaming away from the shock, and has also a $-v_y$-component, i.e. it constitutes a gyrating bunch of ions moving away from the shock. In the right part of the figure it is seen that the beam is moving away along the magnetic field line that is connected to the shock, while the bulk of the plasma flows in positive direction.

These beams along the foreshock boundary play some role in the foreshock dynamics as they seem to represent a source population for the entire ion foreshock. Whether and why this is really so is not yet been fully understood as the shock should reflect ions at almost every place in its quasi-parallel state. However, it seems as that only the group of ions that escape from the shock along the foreshock boundary can form such beams. This points on a further interesting relation between quasi-perpendicular and quasi-parallel shocks at a curved shock surface with a smooth transition from quasi-perpendicular to quasi-parallel as sketched in Figure\,\ref{chap5-fig-foreschem} and realised in space, for instance, at planetary bow-shocks. It seems as so these beams escape from the quasi-perpendicular region of the shock along the nearly tangential field lines. This would also be in agreement with the observation \citep{Gosling1984} that the foreshocks of extended interplanetary shocks\index{shocks!interplanetary} do not show any signs of reflected ion beams. They are only very weakly curved being nearly planar, and not possessing a recognisable quasi-perpendicular area on the surface. 

\begin{figure}[t!]
\hspace{0.0cm}\centerline{\includegraphics[width=0.9\textwidth,clip=]{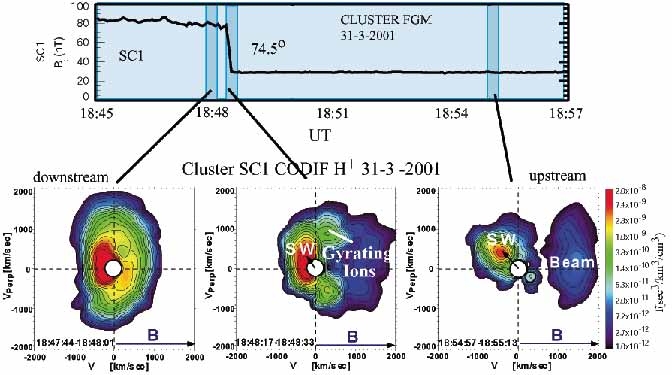} }
\caption[Reflected ion beam ISEE 1]
{\footnotesize Ion phase space at three locations along the shock-connected magnetic field line in a high-Mach number supercritical quasi-perpendicular shock the magnetic field of which is shown in the top panel. The shock-normal angle is $\thetabn=74.5^\circ$.  The three lower plots show  downstream, shock ramp/shock foot, and distant upstream phase-space plots. The ramp/foot plot shows the presence of the inconing flow (SW) and gyrating ions in the foot. The upstream plot shows the usptream field-aligned relatively hot (large velocity spread) beam well separated from the inflow (SW)  \citep[from][]{Kucharek2004}.  }\label{chap5-fig-kucha2004}
\end{figure}
\cite{Kucharek2004} analysed ion distributions along magnetic field lines that were connected to the quasi-perpendicular area of Earth's bow shock. Investigating the origin of those beams these authors found that,  indeed, the observed ion beams at the foreshock boundary result from reflection at the quasi-perpendicular shock and that, without the presence of a quasi-perpendicular region, there would presumably be no distinct foreshock boundary and no ion beams escaping into the foreshock. If this is really the case, then the population in the quasi-parallel ion foreshock is indeed provided by two different  ion sources, the beams from the quasi-perpendicular shock region and the genuine foreshock ion population. The latter has no beam character but is rather a diffuse ion component \citep[see also][]{Meziane2004}. 

\cite{Kucharek2004} estimate that roughly 2\% of the ion inflow leaves the shock ramp upstream in the form of a beam along the magnetic field. They argue that the ions which escape along the magnetic field, are reflected from the very ramp/overshoot region where they have been in resonance with low-frequency plasma waves, which they assume to be large amplitude Alfv\'en-whistler waves. These ions experience pitch-angle scattering and pitch-angle diffusion towards small pitch angles, and subsequently can escape along the magnetic field in the upstream direction. Since the conditions for escape depend in the first place on the pitch-angle scattering process, the beams should be highly variable in time and location. The mechanism might still sound a bit speculative as long as no simulation proves its reality, but any mechanism which is able to pitch-angle scatter ions along the magnetic field in a quasi-perpendicular shock will naturally cause ion beams to escape from the ramp both in the upstream and in the downstream directions. Such simulations require a three-dimensional treatment which is not in reach yet.
\begin{figure}[t!]
\hspace{0.0cm}\centerline{\includegraphics[width=1.0\textwidth,clip=]{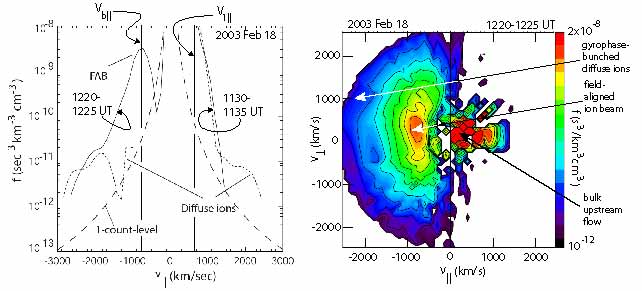} }
\caption[Ion Foreshock Boundary Beam Phase-space Distribution]
{\footnotesize {\it Left}: The reduced field-aligned ion distribution functions in the foreshock region showing the field-aligned reflected ion beam (FAB) at 1220-1225 UT on February 18, 2003 which arrives at the spacecraft location along the magnetic field line that is connected to the quasi-perpendicular region of the Mach number ${\cal M}_A\sim 8$ supercritical shock (smooth solid line). The upstream bulk velocity was $V_1\simeq 660$\,km\,s$^{-1}$. Also shown is the reduced parallel distribution function (dashed) for the diffuse ion distribution observed at 1130-1135 UT the same day (for the geometry see Figure\,\ref{chap5-fig-kis2}). It is clearly seen that the latter is about symmetric to the magnetic field direction indicating the symmetry of the ring of diffuse ions, while the foreshock-boundary field-aligned beam is flowing in the direction away from the shock into the upstream medium (negative velocities). Note also that the diffuse distribution appears as a smooth tail on the full ion distribution. The small gap on the left is uncertain as it dips into the 1-count level. {\it Right}: The two-dimensional phase space plot for the time interval 1220-1225 UT when the field-alighned beam was observed. Indicated are the bulk upstream flow, field-ligned ion beam, and the gyro-phase bunched residue of the diffuse upstream ions \citep[after][]{Kis2007}.  }\label{chap5-fig-kisdist}
\end{figure}

A measured example of the ion foreshock-boundary field-aligned ion beam distribution \citep{Kis2007} is shown in Figure\,\ref{chap5-fig-kisdist} for an upstream flow velocity $V_1\simeq 660$\,km\,s$^{-1}$, Mach number ${\cal M}_A=8$, and average shock-normal angle $\thetabn\simeq15^\circ$ \citep{Archer2005}, at an upstream  distance from the shock (in this case again the Earth's bow shock). This distribution is a so-called reduced distribution; it is the integrated over pitch-angle $\phi$ and perpendicular velocity $v_\perp$ magnetic field-aligned phase space distribution function $f(v_\|)\sim2\pi\int v_\perp^2{\rm d}v_\perp f(v_\perp,v_\|)$ that has been appropriately binned and smoothed. The information to be taken out of this figure is that the reduced ion-beam distribution (solid line)  is narrow in velocity, maximising at a speed (in absolute terms $|v_{b\|}|\simeq 800$\,km/s) that is only slightly larger than the parallel flow velocity ($V_{\|1}\simeq 640$\,km/s, when taking into account the shock-normal angle), i.e. $|v_{b\|}|\simeq 1.25\, V_{1\|}$. It is directed opposite to the flow. For comparison a reduced parallel diffuse ion distribution is shown in the same plot taken deeper in the foreshock. This distribution is about symmetric to the magnetic field direction, indicating the about circular phase-space distribution of the diffuse ion component which appears as an energetic tail on the main ion distribution (note that the small gap on the left of the dashed diffuse-ion distribution curve is questionable as it dips below the 1-count level). \cite{Meziane2004} reported {\CL} observations of foreshock-boundary ion beams simultaneously with diffuse ions. They found that the nominal ion-beam velocity $V_b\simeq 1.7\, V_1$ had no relation to any known shock-reflection mechanism like specular reflection, a conclusion which supports pitch-angle scattering as the beam-injection process as this is independent of the speed of the inflow.   

\begin{figure}[t!]
\hspace{0.0cm}\centerline{\includegraphics[width=0.98\textwidth,clip=]{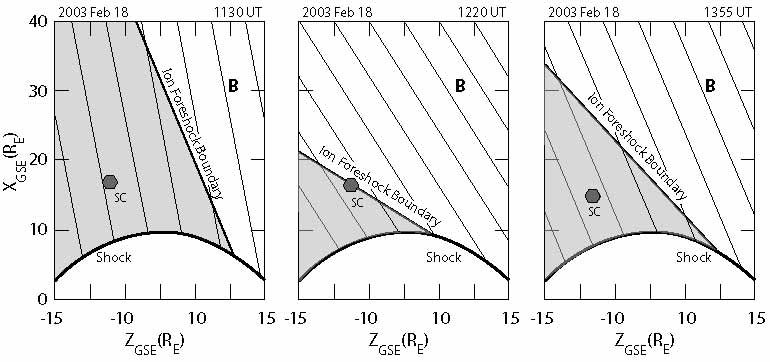} }
\caption[Ion Foreshock Boundary Reconstruction]
{\footnotesize Reconstruction of the shape and location of the ion-foreshock boundary from the measured upstream plasma properties (direction of magnetic field, speed and density) for three successive times in order to explain the ion-phase space observations in Figure\,\ref{chap5-fig-kisdist} \citep[from][]{Kis2007}. The cold magnetic field-aligned ion beam is observed when the spacecraft is located in the vicinity of the ion-foreshock boundary at 1220\,UT, while at 1130\,UT and 1355\,UT no beam was detected. It is assumed that the beam is generated along the foreshock boundary at the position where at the quasi-perpendicular shock surface the shock-normal angle is roughly about $\thetabn\sim 60^\circ$.}\label{chap5-fig-kis2}
\end{figure}
\cite{Kis2007} have carefully analysed the relation between the observed ion distribution at the upstream spacecraft position and the calculated shock-normal angle $\thetabn$, determined from the local upstream magnetic field direction and the predicted shape of the bow shock. The result is shown in Figure\,\ref{chap5-fig-kis2} for three successive times on February 18, 2003 when the {\footnotesize CLUSTER}  spacecraft\index{spacecraft!CLUSTER} was outside the bow shock. At 1130\,UT and 1355\,UT the shock normal angles $\thetabn\sim 15^\circ$ and shock-spacecraft distances $\sim 7.5$ resp.\, $\sim6.7$\,R$_{\rm E}$ were similar. Despite this similarity the observed ion phase-space distributions were completely different. At 1130\,UT no low-energy gyrating ions were observed, while they were present at the later time 1355\.UT. Hence, at 1130\,UT the spacecraft must have been closer to the ion foreshock boundary, i.e. the foreshock boundary was more inclined than at the later time such that beam particles scattered from the foreshock boundary have not arrived at the location of the spacecraft. Due to the velocity filter effect they have been separated out at spacecraft distance. The reconstruction of the position of the ion-foreshock boundary using the measured upstream conditions and shape of the shock for this period is shown in Figure\,\ref{chap5-fig-kis2}. Indeed, at 1220\,UT the spacecraft was close to the foreshock boundary and, as expected, detected the ion beam (as seen in Figure\,\ref{chap5-fig-kisdist}). 

This observation supports the above advocated view that the ion-foreshock beam is generated in the transition region from quasi-perpendicular to quasi-parallel shock. The remaining questions to be answered are: what pitch-angle scattering mechanism is responsible for the generation of such a beam, and what is the fate of the foreshock-boundary beam-ions during their propagation along the foreshock boundary? Do they contribute to the foreshock ion population and if, in what way? Currently we are not able to answer either of these questions definitely. In particular, the pitch-angle scattering mechanism is unknown or at least uncertain. When discussing wave generation, we will touch on the problem of the fate of the beam. Below we present evidence for the scattering of the ion-beam ions and merging into the upstream foreshock diffuse-ion population.

\subsubsection*{Diffuse ions}\noindent\index{foreshock!diffuse ions}
Reflected upstream magnetic field-aligned ion beams are observed at the foreshock boundary only. The second (and main) ion component encountered in the foreshock is the diffuse ion population which is detected there as the energetic extension of the inflow plasma. It is widely assumed that the origin of this component is also at the shock as there is no other energetic particle source available. However, close investigation of the diffuse ion component in the foreshock of the Earth's bow shock wave has demonstrated that these ions are not produced in a specular reflection process at the shock. Rather their origin is of diffusive nature. 

So far it has not been possible to identify the source of these ions though some models have been proposed which we will briefly discuss below in relation to the appearance of foreshock waves. We will also return to these particles in the next chapter on shock particle acceleration. Here, we merely discuss some of their properties and provide evidence that they are indeed coming from the shock, proving that the shock is an energetic ion source. 
\begin{figure}[t!]
\hspace{0.0cm}\centerline{\includegraphics[width=0.8\textwidth,clip=]{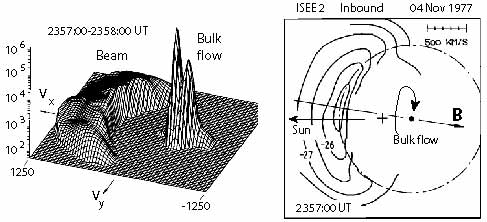} }
\caption[Reflected diffuse ions ISEE 1-first part]
{\footnotesize The evolution of the ion foreshock-boundary ion beam during its convection downstream into the foreshock as has been seen by ISEE 2\index{spacecraft!ISEE} on 04 November 1977. {\it Left}: A one-minute average pseudo-threedimensional ion velocity-space profile in the $(v_x,v_y)$-plane showing the spreading of the beam in angle around the bulk flow without merging into the bulk flow. Velocities are in km\,s$^{-1}$. The scale on the right is count rates, and background count rates were suppressed by choosing only values above 50 s$^{-1}$. {\it Right}: Contour plot of the partial ring distribution. The direction of the magnetic field is also shown  \citep[after][]{Paschmann1981}.  The $10^{-26}$ and $10^{-27}$\ s$^3$\,cm$^{-6}$ level flux contours have been marked. }\label{chap5-fig-pasch81c}
\end{figure}

\begin{figure}[t!]
\hspace{0.0cm}\centerline{\includegraphics[width=0.9\textwidth,clip=]{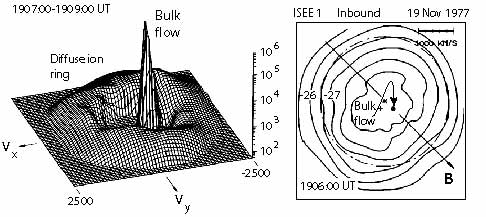} }
\caption[Reflected diffuse ions ISEE 1-second part]
{\footnotesize ISEE 1 observation of diffuse ion beam on November 19, 1977 propagating deep inside the foreshock away from the ion foreshock boundary. {\it Left}: A two-minute average pseudo-threedimensional ion velocity-space profile in the $(v_x,v_y)$-plane showing the undisturbed and cold (narrow) plasma inflow in negative $v_x$-direction about centred  and surrounded by a ring distribution of  fast and warm (broad) diffuse ions that have been reflected from the shock but have been processed in the foreshock region when propagating from the foreshock boundary into the foreshock. These ions are hollow in the sense that they separate from the bulk distribution but have a nearly isotropic distribution function. Velocities are in km\,s$^{-1}$. The scale on the right is count rates, and background count rates were suppressed by choosing only values above 50 s$^{-1}$. {\it Right}: Contour plot of a similar ring a little earlier showing that the ring centre (star) is slightly displaced from the bulk flow (dot) on the magnetic field that connects to the shock and from the centre (cross) of the phase space frame \citep[after][]{Paschmann1981}.  The $10^{-26}$ and $10^{-27}$\ s$^3$\,cm$^{-6}$ level flux contours have been marked. Note the near isotropy of the ring distribution.}\label{chap5-fig-pasch81b}
\end{figure}

The foreshock-boundary ion beams also merge into the foreshock particle distribution by scattering from the foreshock boundary off their self-generated wave spectrum and subsequent being convected downstream by the bulk flow. \cite{Paschmann1981} have mapped this merging process by following the evolution of the foreshock-boundary beam distribution shown in Figure\,\ref{chap5-fig-pasch81a} during the convection. In that figure the beam was detected close to the foreshock boundary. Figure\,\ref{chap5-fig-pasch81c} show its form deeper in the foreshock when it has spread substantially in angle evolving into half of a ring distribution already.  Even deeper inside the foreshock the reflected  ion distribution assumes the shape of a full ring around the bulk distribution as is shown in Figure\,\ref{chap5-fig-pasch81b}. 

The three observations depicted in  Figures \ref{chap5-fig-pasch81a}, \ref{chap5-fig-pasch81c} and \ref{chap5-fig-pasch81b} are from different times; it has, however, been checked that they are at distances corresponding to increasing distance from the foreshock boundary such that the assumption of the convectively processed beam evolution is well founded (or at least reasonable) even though it has not been directly proven. It is interesting to note that a gap remains between the original ion-phase space beam distribution and the bulk-flow distribution, which is another indication that the evolving distribution is part of the evolution of the ion-foreshock beam. Of a distribution that does not evolve out of a beam one expects a less regular behaviour and, generally, no such well expressed gap between bulk flow and  beam in velocity space. In fact, the main energetic ion component in the foreshock is irregular and lacks a well expressed gap. 

The discrepancy between the smooth no-gap foreshock distributions \citep{Sentman1981a} and the gap-observations of \cite{Paschmann1981} has, in fact, been noted much earlier \citep{Sanderson1981,Scholer1985} without giving an explanation but suggesting a continuous ion source at the parallel shock. Below we provide further arguments for the two-source, foreshock-boundary beam and continuous extended shock-surface source hypothesis. Nonetheless, this conclusion must be taken with care because Figure\,\ref{chap5-fig-pasch81b} shows the gap progressively closing. Being sufficiently far, i.e. even farther away from the ion foreshock boundary then in this figure, it will not anymore be possible to distinguish between beam-evolved and genuine diffusive-ion distributions. Ultimately, both distributions will have merged indistinguishably. Still there is no agreement whether and where, i.e. at what distance from the ion foreshock boundary this merging of the two populations occurs.

\cite{Trattner1994}, using {\footnotesize AMPTE IRM}\index{spacecraft!AMPTE IRM} measurements of diffuse ion densities upstream of the quasi-parallel (bow) shock found that the diffuse ion density decreases exponentially with shock distance. This investigation was substantially improved by \cite{Kis2004} in an attempt to infer about the source of the upstream diffuse energetic ions. These authors determined the partial-density gradient of diffuse ions in the energy range from 10 to 32 \,keV as a function of distance from the (bow) shock. This investigation was made possible due to the availability of the {\CL} spacecraft, a four identical-spacecraft mission which during this measuring period had an inter-spacecraft separation distance between $\sim1$ and $\sim1.5$\,R$_{\rm E}$. They used a nominal bow shock model \citep{Peredo1995}, based on the measured upstream flow parameters (basically the dynamic pressure of the flow), providing the shock-spacecraft distances along the magnetic field flux tube for the individual {\CL} spacecraft and the local shock-normal angles $\thetabn$. The average $\thetabn$ over the whole 10 hours of observation time was $20^\circ\pm8^\circ$ proving that {\CL} was in front of the quasi-parallel shock far away from the ion-foreshock boundary, and the Mach number was ${\cal M}_A\sim 8$. 

Diffuse ion partial densities were determined as function of distance from the shock and in four consecutive energy bands every 32\,s at two spacecraft. This allowed to determine the partial diffuse ion-density gradients along the magnetic field as function of energy from the density differences between two spacecraft and the differences of the spacecraft distances along the magnetic field to the shock intersection point. Perpendicular density gradients were neglected. The obtained parallel gradients were attributed to the average {\CL} location.  These gradients were then used to find the e-folding distance of the density variation. 

\begin{figure}[t!]
\hspace{0.0cm}\centerline{\includegraphics[width=1.0\textwidth,height=0.3\textheight,clip=]{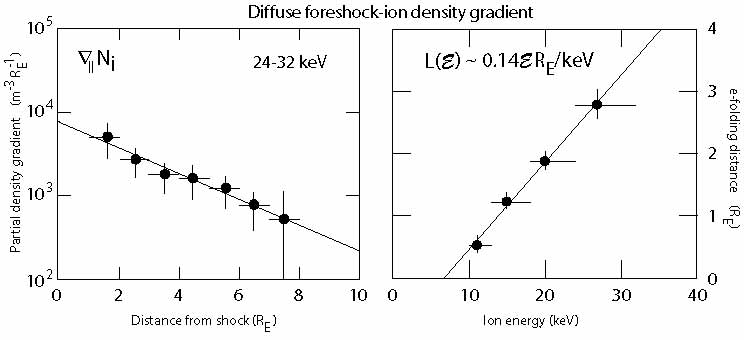}}
\caption[Diffuse ion e-folding distance]
{\footnotesize Partial-density gradient of the diffuse foreshock-ion component along the shock-connected magnetic field, and density e-folding lengths, determined from {\scriptsize CLUSTER} observations in the Earth's bow shock foreshock \citep[after][]{Kis2004}. {\it Left}:  The parallel partial diffuse-ion density as function of distance from the shock along the magnetic field flux tube connecting the spacecraft to the shock.  The diffuse ion density falls off exponentially with distance suggesting an ion-energy dependent  diffusive process being responsible for transport of the ions upstream of the shock. {\it Right}:  The e-folding distance of diffuse ions along the magnetic field as determined as function of energy from the exponential decay of the ion density. The e-folding distance increases linearly with ion energy. }\label{chap5-fig-kis3}
\end{figure}

The results of this investigation are shown in Figure\,\ref{chap5-fig-kis3}.
It is learned from this figure that in the deep foreshock, i.e. at distances far away from the ion-foreshock boundary, the diffuse ion component is densest close to the shock with density decaying exponentially with increasing upstream distance from the shock along the magnetic field. This spatial decay of the diffuse partial ion density $N_i({\cal E},z)\sim \exp[-z/L({\cal E})]$ is different for particles of different energy ${\cal E}$. The e-folding distance $L({\cal E})\sim {\cal E}$ turns out to increase linearly with energy, i.e. low energy particles are confined to the shock. The higher the ion energy  the deeper can the ions penetrate into the upstream plasma. The proportionality constant determined from these data under the special conditions of the Earth's bow shock is $\sim 0.14$\,R$_{\rm E}$/keV. This behaviour of the energetic foreshock ions provides indisputable evidence for the extended parallel shock-surface origin of the diffuse ion component. The source of the diffuse ions lies at the quasi-parallel shock. In order to be found at a distance upstream of the shock the ions undergo a diffusion process along the magnetic field. These ions\index{foreshock!ion dissipation length} \index{dissipation!foreshock scale}\index{collisions!foreshock dissipation length}\index{resistivity!foreshock dissipation}\index{foreshock!dissipation scale} are thus completely different from the beam ions found at the ion-foreshock boundary. 

The e-folding distance\index{foreshock!e-folding distance} for the diffuse ions is given by $L({\cal E})=\kappa_\|({\cal E})/V_1$, with spatial diffusion coefficient $\kappa_\|({\cal E})=\frac{1}{3 }v\ell_\|({\cal E})$, where $\ell_\|$ is the diffusion length (parallel ion mean free path) and $v$ the particle velocity (note that the diffusion coefficient has the correct dimension $[\kappa_\|]={\rm m}^2{\rm s}^{-1}$; justification of the diffusion assumption will be given in the chapter on particle acceleration). From balance between convective inflow and diffusion into upstream direction, one can write $\ell_\|({\cal E})=3L({\cal E})\sqrt{{\cal E}_1/{\cal E}} \sim \sqrt{{\cal E\cdot E}_1}$. The diffusion length increases as the root of the product of particle energy ${\cal E}$ and upstream flow energy ${\cal E}_1$. In the solar wind the flow energy is a few keV, and a 20-keV diffuse ion will have a typical parallel diffusion length (or mean free path) of $\ell_\| \sim (1-2)\,{\rm R_E}$. This is a rather short distance, orders of magnitude shorter than the collisional mean free path of an ion. Hence, the diffusion estimate suggests that strong wave-particle interactions can be held responsible for the scattering and acceleration of the diffuse particle component, which enables it to diffuse and escape upstream from the shock and populate the foreshock. The diffusion process is energy dependent with the most energetic ions diffusing fastest. \index{diffusion!anomalous}\index{collisions!anomalous}

These interactions should take place in the quasi-parallel shock transition because, as we have shown above, the diffuse upstream-ion density maximises closest to the shock. It is interesting to estimate the corresponding upstream-ion collision frequency $\nu_{c,ui}\simeq v/\ell_\|$. For the 20\,keV-upstream ions this yields $\nu_{c,ui}\sim 0.2$\,Hz. This value is comparable to the ion cyclotron frequency $\omega_{ci}/2\pi=(0.1-0.3)$\,Hz in the $B\simeq$\,8\,nT upstream to $B\simeq$\,30\,nT shock ramp magnetic field \citep{Kis2004} during the time of observation. 

It seems that waves, electromagnetic and/or electrostatic, related to the ion-cyclotron frequency are involved into the process of upstream ion diffusion. Since this diffusion is energy dependent, this process is not a simple pitch-angle diffusion as in the case of the generation of the ion beam that propagates along the foreshock boundary. The diffuse ion component must have experienced a substantial acceleration in this process, and this acceleration is located at or around the shock transition and contrasts with the ion-beam acceleration which is a scattering process followed up by pick-up acceleration when the upstream-propagating beam ions are subject to the effect of the main-bulk-stream convection-electric field in which they become accelerated in the direction perpendicular to the magnetic field to roughly four times the energy of the bulk flow, thereby evolving into the ring distribution that characterises their phase space distribution.

\subsection{Low-frequency upstream waves}\noindent
In the frame of the upstream bulk flow the two ion components that populate the ion foreshock carry a substantial amount of free energy which is subject to dissipation. Since this dissipation is collisionless it can proceed only through the excitation of waves and wave turbulence through instability upstream of the quasi-parallel supercritical shock. On the other hand it is obvious that the presence of neither of the components can be understood without complete knowledge of the waves in the foreshock and their interaction with the particles. 

Since their first detection by \cite{Olson1969}, \cite{Russell1971} and \cite{Fairfield1974}, observation of shock-upstream waves has been a long-standing issue. Their existence was predicted by \cite{Tidman1968}, followed by hydromagnetic \citep{Barnes1970} and kinetic \citep{Hasegawa1972} theories of electromagnetic wave excitation and propagation upstream of a collisionless quasi-parallel supercritical shock. \cite{Wu1972} suggested that they might develop into discrete wave-packets as had been inferred from observation by \cite{Russell1971}. The {\footnotesize ISEE\,1-3} spacecraft allowed for a more elaborate investigation of the properties of upstream waves \citep[cf., e.g.,][among others]{Hoppe1981,Hoppe1983,Russell1981,Russell1983,Sentman1981a,Sentman1981b,Sentman1983,Hoppe1982,Thomsen1985,Mellott1986,Russell1988,Le1992}. More recently, {\CL} measurements have been used to investigate the temporal and spatial structure of upstream waves and wave turbulence \citep[cf., e.g.,][among others]{Mazelle2000,Mazelle2003,Eastwood2002,Eastwood2003,Meziane2004,Narita2004,Narita2006,Narita2007}. We will briefly review the properties of the upstream waves in the ion-beam and diffuse ion region in view of the observations and mechanisms of their generation. Wave generation is coupled to particle-energy loss and to particle scattering both playing a substantial role in particle acceleration. It will therefore be quite natural that in the next chapter on particle acceleration at shocks we will return to the upstream-wave problem. 

\cite{Burgess1997} gave a comprehensive review of the various types of waves encountered upstream of quasi-parallel shocks in the ion foreshock. In his words, ``upstream particles cause upstream waves .... Once a wave is created .., it then propagates, and so its continued existence relies on it remaining in a region where it is undamped. Its properties might even change as it propagates. Observationally, the wave propagation is superimposed on the convection of the plasma frame, which introduces Doppler shifts in frequency, and possible reversal of polarisation sense.... And that is not the end of the story, since one must take account of the feedback of the waves on the particle distribution function..., and even the possibility that the ... shock injection of particles into the foreshock os modulated either by intrinsic processes or even by the foreshock waves themselves." 

Burgess' review was organised by the observed wave frequencies. He distinguishes between Low Frequency Waves (5\,mHz- few 100\,Hz) and High Frequency Waves ($>1$\,kHz), the latter covering the electrostatic waves from ion-sound to electron plasma waves, as well as radiation. Radiation generation will be discussed in detail later. Here, we only note the almost continuous presence of waves in the ion-acoustic band which have been known since \cite{Rodriguez1975} to populate the complete foreshock region. These spectra might be composed of several different modes, ion-sound, electron-acoustic, Buneman modes, electron-cyclotron harmonics, and others. Their generation mechanism is not clear yet. They might, via a number of different instabilities,  be the result of the presence of the hot foreshock-electron component, which also invades the ion foreshock, or they are excited by unresolved narrow electron beamlets that emanate from the quasi-parallel shock. They might also be excited by plasma inhomogeneities, spatial inhomogeneities in the electron distribution, or they are the result of nonlinear wave-wave interaction which is expected to take place in the foreshock. Currently these questions are difficult to answer and await further observation, simulation and theory. Little has changed so far since Burgess' remarks concerning high frequency waves.

However, there has been substantial progress in the understanding of the low frequency waves and their role in quasi-parallel shock dynamics. The waves that are most important in shock formation propagate in the ultra-low frequency range $<0.1$\,Hz. Usually they have large (magnetic) wave amplitudes, around $|{\bf b}|/B\sim 0.2-1.0$, which identifies them as highly nonlinear. These large amplitude waves had already been observed by \cite{Russell1971} to have wave forms from monochromatic to solitary waves, frequently with steep edges resembling shocklets and suggesting that the waves have experienced nonlinear steepening during their evolution and propagation. They sometimes show the typical signs of fluctuations that are connected to these edges and obviously propagate in the whistler mode. Thus these forms are indeed little shock-like structures. 

In the same frequency window, large-amplitude pulsations have been identified. These are very typical for quasi-parallel shocks. \cite{Schwartz1991} and \cite{Schwartz1992}, identifying them in the {\footnotesize AMPTE} magnetic field measurements, coined the (somewhat ugly as in German it means mud) name {\SL} for them, which stands for `Short, Large Amplitude Magnetic Structures'. We prefer to call them upstream pulsations, here. Their duration is 10-20\,s; they have very large amplitudes $|{\bf b}|/B \sim 5$, indeed, but appear as a more coherent structure that is embedded in the ultra-low frequency wave turbulence. Like the ultra-low frequency waves, they propagate in upstream direction in the plasma rest frame while being swept toward the shock by the convective flow. Their polarisation is mixed with -- possibly --  left-hand polarisation (in the plasma rest frame) slightly dominating, suggesting their ultra-low-frequency wave origin. Sometimes the polarisation is different on both sides of the upstream pulsation, indicating that they have evolved by some process which produces both kinds of polarisation, which is similar to a solitary wave. 

The propagation velocity of the upstream pulsations exhibits an interesting amplitude dependence. The upstream directed pulsation speed increases with amplitude, which also is a solitary wave-like property. Moreover, they grow when approaching the shock and entering the increasing density gradient of diffuse ions, and they play an important role in shock reformation. \cite{Schwartz1992} suspect that these pulsations are the `building blocks' \citep[an expression used by][]{Schwartz1991} of quasi-parallel shocks. 

Another interesting property is that upstream pulsations contain thermal plasma with properties of the upstream flow, while being surrounded by the hot foreshock plasma. One would therefore believe that their source region is located at the ion-foreshock boundary. Being created there by an ion-ion beam plasma interaction they might grow nonlinearly until reaching quasi-equilibrium like solitary structures, having captured the upstream plasma, and afterwards being convected toward the shock into the heart of the ion foreshock. We will return to these interesting structures when discussing simulations below.

\subsubsection*{Ion-beam waves}\noindent
Each of the two different upstream-ion populations is responsible for the excitation of its own instabilities. In this section we deal only with those waves which are excited by the foreshock-boundary ion beam. From Figure\,\ref{chap5-fig-kisdist} we obtain that the situation is that of an ion-ion beam (when for the moment neglecting the gyrophase-bunched diffuse-ion component). In the frame of the upstream flow the reflected foreshock-boundary ion beam propagates upstream along the magnetic field at parallel speed $v_{b\|}\simeq -(2-3)V_{1\|}$, where $V_{1\|}\simeq 600$\,km\,s$^{-1}$. The upstream ion temperature (in energy units) for this case was $T_i\simeq (1-2)$\,keV, yielding roughly a thermal ion velocity of $v_i\simeq 200$\,km\,s$^{-1}$. Moreover, the beam can be taken as warm with thermal speed $v_{b,th}\lesssim v_{b\|}$. Hence, $v_{b\|}>v_i$ with magnetised background and beam ions. On the other hand, the electrons are hot with $T_e\simeq 100$\,eV. 
\paragraph{The expected wave modes.} 
Since the large parallel speed of the beam corresponds to a large parallel temperature anisotropy, it is clear that the beam can excite long-wavelength negative-helicity Alfv\'en waves via the firehose instability. In addition, because the beam is mildly warm, it can excite the resonant left-hand ion-ion beam instability, which is possible for $v_{b\|}>V_A\sim 100$\,km\,s$^{-1}$. And under conditions, when the beam thermal speed can be considered to be small, it excites the right-hand resonant ion-ion beam mode. Both waves propagate with the beam upstream along the magnetic field on the background of the upstream flow. They are not as fast as the beam, however, and are thus subject to downstream convection with the flow towards the shock. The firehose mode, at the contrary, moves against the beam and thus by itself moves downstream in the direction of the shock position, when excited. 

In all three cases the foreshock-boundary\index{foreshock!ion boundary} beam will lead to the excitation of low-frequency Alfv\'en and ion cyclotron waves which in the shock frame approach the shock while having their source on the foreshock-boundary field line. During this shock-directed convection and/or propagation they populate the foreshock region with low frequency electromagnetic fluctuations, which might further interact with the diffuse foreshock-ion component. On the other hand, since the phase and group velocities $\omega/k_\|\sim(\partial\omega/\partial k_\|)\sim V_A\ll v_{b\|}$ of these waves are of the order of the Alfv\'en velocity and are, thus, much less than the beam and flow velocities, downstream convection will quickly remove them from the foreshock-boundary source region. Hence, their further evolution in the foreshock is determined by the competition between nonlinear wave steepening and interaction with the diffuse foreshock-ion component. 
\begin{figure}[t!]
\hspace{0.0cm}\centerline{\includegraphics[width=0.7\textwidth,clip=]{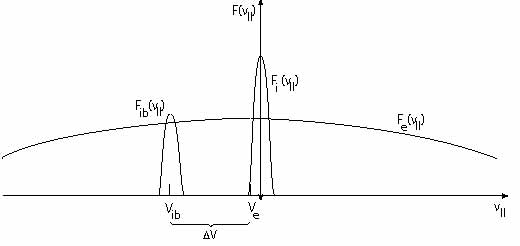}}
\caption[Beam configuration FSB]
{\footnotesize Reduced parallel distribution functions along the ion-foreshock boundary showing the cold main flow ion distribution $F_i(v_\|)$, the hot main flow electron distribution $F_e(v_\|)$ shifted to the left into the direction of the foreshock ion beam for keeping zero-current conditions, and the ion-foreshock ion beam distribution $F_{ib}(v_\|)$. This configuration is unstable with respect to ion-ion beam instabilities and the ion-beam driven ion-acoustic instability.}\label{chap5-fig-ifb}
\end{figure}

Including the electrons (while so far neglecting the electron foreshock component) leaves us with an ion-acoustic unstable phase-space configuration. In the upstream ion-plasma frame a relatively dense foreshock-boundary ion-beam is propagating upstream on a cold ion-hot electron plasma. In order to keep the plasma current-free the electron component is slightly retarded creating conditions under that ion-acoustic waves can be excited. On the other hand, the configuration is not able to excite neither the Buneman-two stream instability nor -- because the beam propagates solely parallel to the magnetic field -- the modified-two stream instability. 

This is shown in Figure\,\ref{chap5-fig-ifb}. The instability is excited by the velocity difference between the ion foreshock-boundary ion beam $F_{ib}(v_\|)$ and the slightly shifted to the left hot ($T_e\sim 100$\,eV) electron distribution because the velocity difference $\Delta V$ between ion beam and electron component $v_e>\Delta V\sim 10^3\,{\rm km\,s}^{-1}>c_{ia}\sim 100$\,km\,s$^{-1}$  exceeds the ion acoustic speed while being less than the electron thermal velocity $v_e$. These ion acoustic waves occupy a relatively broad spectrum with downstream parallel phase velocities $<v_{ib}$ and become swept towards the shock. Since they quickly leave the foreshock boundary and since the beam is not hot, ion-Landau damping in the source region plays no role. However, when entering the diffuse-ion foreshock region these waves encounter the hot diffuse ion component, will interact with it, and will thereby experience Landau damping. 
\begin{figure}[t!]
\hspace{0.0cm}\centerline{\includegraphics[width=1.0\textwidth,clip=]{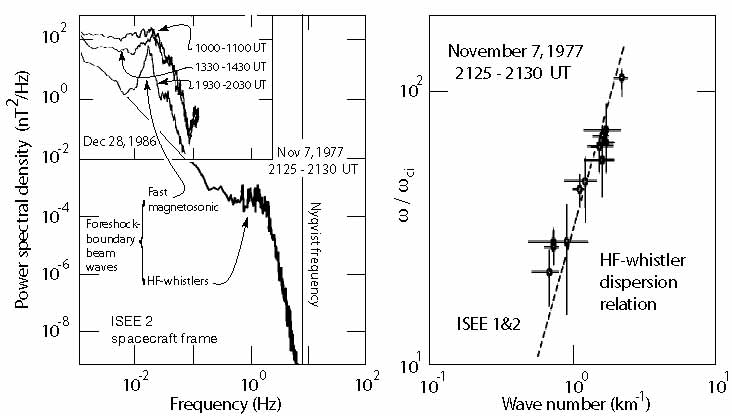}}
\caption[Foreshock ion beam waves]
{\footnotesize {\it Left}: The combined power spectral density of magnetic fluctuations excited by the ion foreshock-boundary ion beam with central frequency around $\sim 1$\,Hz \citep[lower part, after][]{Hoppe1982} and at $<0.1$\,Hz \citep[left upper part, after][]{Le1992} as measured by the ISEE spacecraft. This spectrum is obtained in the ISEE spacecraft frame. The low frequency spectrum is shown at three different times corresponding to (from below upward) increasing distance from inside the foreshock to the foreshock boundary. It is about five orders of magnitude more intense than the high-frequency waves. It is worth noting that the high frequency spectrum of \cite{Hoppe1982} toward lower frequencies in the overlap with the \cite{Le1992} spectrum does not show any indication of the lower frequency peak. As indicated by the thin straight line, it would smoothly continue into the 1930\,UT branch at 0.01\,Hz. This lack of the low frequency peak is probably accidental and due to the particular conditions at the time of measurement.  {\it Right}: The wave dispersion relation for the high frequency $\sim 1$\,Hz whistler-mode waves determined from a tentative estimate of the relavant wave numbers and transformed into the plasma rest frame. It is seen that these waves in the observation range have about linear dispersion and frequencies $\omega\sim (20-100)\,\omega_{\,ci}$, far above the ion-cyclotron frequency $\omega_{\,ci}$ \citep[after][]{Hoppe1982}. }\label{chap5-fig-ifbws}
\end{figure}
\paragraph{Observations.} \index{waves!foreshock-ion beam}
Observationally, it is difficult to distinguish between foreshock-boundary waves and diffuse-ion generated waves. Two types of waves that can be related to the presence of the upstream beams have been reported in the vicinity of the bow shock. \cite{Fairfield1974} found monochromatic large-amplitude discrete wave packets in the frequency range $\omega/2\pi\sim0.4$\,Hz which \cite{Hoppe1980} could show to be right-hand polarised whistlers at $\omega\simeq 10\,\omega_{ci}$ propagating on the plasma rest frame and being unable to escape far upstream. The other class of waves is of smaller amplitude $|b|/B\sim 0.1$ and frequency of the order of $\omega/2\pi\sim 1$\,Hz. These waves form trains which are directly tied to the upstream foreshock-boundary ion beams \citep{Hoppe1982}. In fact, as \cite{Hoppe1982} demonstrated they occur only in the presence of the observed reflected $2-5$\,keV foreshock-boundary ion beams \citep{Eastman1981} which \cite{Kis2007} have shown to evolve from a beam into a gyrating particle component that in the deep foreshock is superimposed on the diffuse ion component before it merges into it. However, as \cite{feldman1983} have shown from measuring the electron distribution function, it is not the ion beam who excites these whistlers. The probability that they are driven by the particular electron distribution in the foreshock is rather higher.

This is a very favourable case as these waves carry information about the generation mechanism. The waves propagated obliquely ($<60^\circ$) with respect to the magnetic field, at an average angle of $\sim45^\circ$. The observed spectrum and the dispersion relation determined from the measurements are shown in the high-frequency part of Figure\,\ref{chap5-fig-ifbws}. This figure has been combined from the high-frequency observations of \cite{Hoppe1982} and the low-frequency observations of \cite{Le1992}, both obtained from the {\footnotesize ISEE} spacecraft. Tentative wave numbers have been determined for the high-frequenc waves from measuring the time delay of the wave front arrivals at the two spacecraft {\footnotesize ISEE 1} and {\footnotesize ISEE 2} spacecraft yielding (surprisingly) short wavelengths $<100$\,km. Knowing the wave number, the frequency has been back-Doppler shifted $\omega=\omega_{\,\rm ISEE}-{\bf k\cdot V}_1$ into the plasma frame yielding unusually high frequencies $\omega\simeq (20-100)\omega_{ci}$. This procedure determines the dispersion relation (naturally with large errors as indicated by the bars), which is found to be about linear in the narrow high frequency range of the waves in the rest frame of the plasma. The waves that have been left-hand polarised in the spacecraft frame turn out to become right-hand polarised in the plasma rest frame. 

From these properties it was initially concluded that the waves propagate in the whistler branch and have been excited by the cool ion-ion beam instability, which would be consistent with our initial discussion. This conclusion is, however, questionable. The high rest-frame frequencies are not in agreement with model calculation \citep{Sentman1981a,Gary1993} using the realistic observed beam properties. These yield wave frequencies of the order of $\omega\sim 0.1\,\omega_{ci}$ for both the ion-ion beam and firehose modes. In fact, \cite{Eastwood2005} report {\CL} observations of similar waves but with much lower frequency $\omega/\omega_{ci}\sim0.1$ and wavelength the order of $\sim 1\,{\rm R_E}\sim6000$\,km, which is in excellent agreement with the theoretical predictions. These waves are cold ion beam-excited fast-kinetic (magnetosonic) whistlers, in the terminology of \cite{Gary1993} and \cite{Krauss1994}.  The high-frequency waves $\sim 1$\,Hz whistlers are instead most probably driven by the particular electron distribution in the foreshock \citep{Feldman1983} rather than by the cold-ion beam instability. This claim is also supported by the fact that deeper in the ion foreshock, where the electron distribution becomes more isotropic, these waves do not occur separate from smaller magnetic field structures (large-amplitude magnetic pulsations of the {\SL} type or shocklets). 

\begin{figure}[t!]
\hspace{0.0cm}\centerline{\includegraphics[width=0.8\textwidth,clip=]{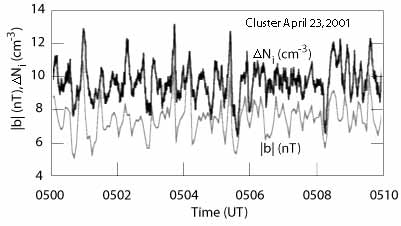}}
\caption[Field-density correlation of MS waves]
{\footnotesize {\it Left}: The excellent correlation between the density and magnetic field variations in the ion-beam excited foreshock-boundary low-frequency waves as measured by CLUSTER \citep[after][]{Eastwood2002}.  The variations are practically in phase thus identifying the fluctuations as fast magnetosonic.}\label{chap5-fig-east1}
\end{figure}

In quite good agreement with the {\CL} measurements are the {\footnotesize ISEE} foreshock waves analysed by \cite{Le1992} who found the spectral peak at frequencies $\omega\sim 10^{-2}$\,Hz with the spectrum broadening with increasing distance from the ion-foreshock boundary. This is shown in the upper left part of the spectrum in Figure\,\ref{chap5-fig-ifbws} where the measurements of \cite{Hoppe1982} and \cite{Le1992} have been combined. Even though the conditions on the two observation times were not identical, one sees that the high-frequency whistlers identified by \cite{Hoppe1982} occur on the approximate high-frequency extension of the \cite{Le1992} 1930-2030\,UT spectrum which is closest to the ion foreshock boundary. Note, however, that the low-frequency part of the \cite{Hoppe1982}-spectrum (not shown here other but indicated by a thin straight line in the figure) was flatter and in the (relatively short) overlap region with the \cite{Le1992}-spectrum did not show any indication of the magnetosonic foreshock-boundary waves, which must be due to the particular conditions prevailing during this observation period. Thus the \cite{Hoppe1982} waves are a (high frequency) wave species that is different from these more common low-frequency/long-wavelength fast-magnetosonic waves. These latter waves are also of much higher ($\sim$\,5 orders of magnitude) spectral density as is seen from Figure\,\ref{chap5-fig-ifbws} and has been confirmed by the later {\CL} measurements. Moreover, the magnetosonic waves are left-hand polarised in the plasma frame, and their compressive fast-magnetosonic character is proved from the in-phase variation of the density and magnetic field fluctuations shown in Figure\,\ref{chap5-fig-east1}. 

The presence of these low-frequency ion-beam generated waves implies that the ion beams interact with the fluctuating electromagnetic field. In this process they become scattered and diffuse in phase-space. This is the reason for the spreading of the ion beam in phase space and the final merging into the diffuse background distribution. \cite{Kis2007} have followed this evolution of the beam as we have described above. It is responsible also for the gradual spreading of the spectrum in the upper left part of Figure\,\ref{chap5-fig-ifbws}. A spectrum similar to that given there can be found in the paper of \cite{Kis2007}. \cite{Archer2005}, from {\CL} observations, inferred the nature of these ultra-low frequency (ULF) waves and showed that their correlation lengths along the wave vector direction ${\bf k}$  is of the order of $1-3\,{\rm R_E}$, while it can be a factor of three larger in the direction perpendicular to ${\bf k}$, rendering these waves oblate though nearly planar.\index{foreshock!ion waves} 

During times of hot beams \cite{Eastwood2003} observed upstream propagating left-hand low-frequency waves which have been excited on the Alfv\'en-ion-cyclotron branch of the kinetic dispersion relation. These waves are kinetic Alfv\'en waves which have been excited by the hot upstream propagating beam similar to the one shown on the right in Figure\,\ref{chap5-fig-kisdist}. In the spacecraft frame these waves because they are swept downstream by the flow, appear as right-handed waves. \cite{Eastwood2003} report that with onset of the waves the beam gets more diffuse. This can either be interpreted as the reaction of the waves on the beam or that the spacecraft enters the region where the initially cold beam enters the foreshock, spreads in velocity and after becoming hot enough generates the observed waves. 

The question arises whether there are simulation studies available of the evolution of the reflected ion beam and the generation of upstream waves. This question cannot be definitely answered at present. It is clear that a simulation of this kind should reconstruct the region of the foreshock boundary which requires that a curved shock surface must be assumed from the beginning. In other word, this problem can be investigated only in a two-dimensional simulation. Two-dimensional simulations are available in hybrid form but have have not been applied to curved shocks. Hence, this problem remains an open simulation problem. On the other hand, in the following we will extensively discuss the role diffuse particles play in shock formation and upstream wave generation for planar shocks in one and two dimensions and hybrid as well as PIC simulations.
\begin{figure}[t!]
\hspace{0.0cm}\centerline{\includegraphics[width=0.9\textwidth,clip=]{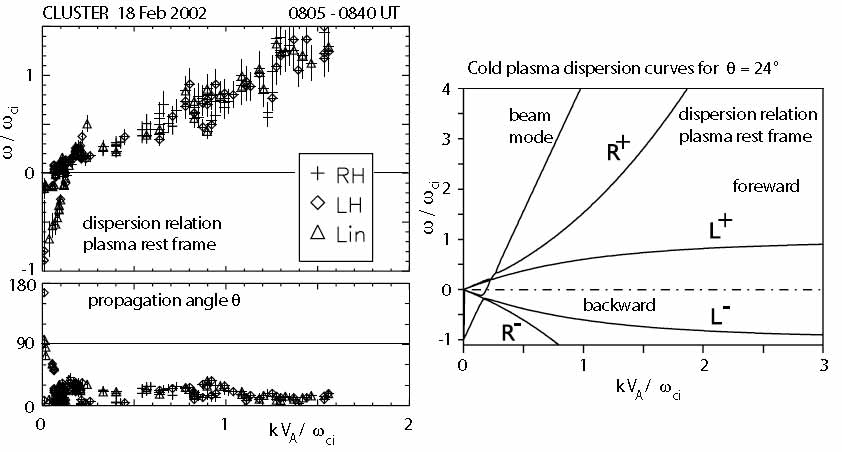}}
\caption[FS waves disprel]
{\footnotesize {\it Left}: The dispersion relation of low-frequency waves as measured by CLUSTER in the foreshock away from the foreshock boundary \citep[after][]{Narita2005}, determined from the ``wave-telescope" analysis method.  The upper panel shows the dispersion relation $\omega(k)$ consisting of several parts and distinguished by the sense of polarisation as right-hand, left-hand and linear. The different branches cluster. There are forward (positive frequency parallel to the magnetic field) and backward (negative frequency antiparallel) branches. The lower panel shows the wave propagation angle for the shorter wavelengths being mostly oblique ear $\theta\approx (30-25)^\circ$, for longer wavelengths being more perpendicular to the magnetic field. {\it Right}: The five low-frequency modes in a cold beam plasma system of similar conditions as the foreshock plasma and for the propagation angle $\theta=24^\circ$ showing similarity to the observed dispersion relation \citep[][]{Narita2003}.}\label{chap5-fig-naritadisprel}
\end{figure}

\subsubsection*{Diffuse ion waves}\noindent
Long-period $\sim 30$\,s waves are the rule in the foreshock. They occur together with the diffuse foreshock-ion component \citep[cf., e.g.,][]{Sanderson1983, Mellott1986} which is located deeper inside the foreshock. Because of this reason, any waves that are excited by the diffuse ion component are restricted to the interior of the foreshock. One even has defined some fuzzy boundary of these waves called ULF-wave boundary \citep{Russell1983,Greenstadt1986}, which is even more inclined against the upstream magnetic field than the ion-foreshock boundary and outside of which  the ULF-wave activity should be weak. The reason is that ULF waves, if propagating upstream, can move at most with fast-magnetosonic velocity which for a supercritical shock is smaller than the stream and also less than the reflected ion beam velocities. The advection by the flow will blow them downstream towards the shock and confine them to a region relatively close to the shock \citep{Meziane1998} bounded to upstream by the ULF-boundary.  

\cite{Narita2005} determined the dispersion\index{foreshock!dispersion relation} relation of the low-frequency waves in the foreshock from the three-dimensional observations of {\CL}. Their result is shown in Figure\,\ref{chap5-fig-naritadisprel}. 
The left part of the figure shows the dispersion relation in the plasma rest frame and the angle of wave propagation with respect to the local average upstream magnetic field. The scatter is quite large. Nevertheless it is surprising to find a well expressed nearly linear part on the dispersion relation even though the dispersion relation is composed from the contributions of several distinct modes with different  polarisation. The representation is linear, which, on the $k$-axis, emphasises the short wavelengths (larger $k$ values). From linear theory one expects that the long wavelengths (small $k$) should be more pronounced. Unfortunately, this region could not be resolve sufficiently at the available {\CL} separation distances at that time \citep{Narita2003}.  These low frequency waves, in the plasma frame, seem to propagate close to perpendicular to the magnetic field and, therefore, are probably in the fast magnetosonic mode. Moreover, it seems that a straight beam dispersion relation contributes to the perpendicular propagating waves. It also seems that in the low-frequency (magnetosonic) waves have negative frequency. Linear dispersion theory for the parameters during the observation time yields the curves on the right in the figure showing the coupling between the presumable beam mode and the four plasma modes  at wave numbers $kV_A/\omega_{ci}<0.3$. The experimentally determined dispersion relation resembles this clustering of couplings at small $k$ and very low frequencies $\omega$. However, application of linear theory is dangerous if not questionable because the foreshock plasma is highly disturbed, the low frequency waves have rather large amplitudes and can barely be related linearly, and the plasma is very inhomogeneous exhibiting steep gradients in density and field. For an Alfv\'en speed $V_A\simeq 30$\,km\,s$^{-1}$ and an ion-cyclotron frequency $\omega/2\pi\simeq 1$\,Hz, wavelengths around the linear wave coupling are of the order of $k/2\pi\simeq 100$\,km and should thus be affected by the plasma inhomogeneities (note that decreasing the reference cyclotron frequency decreases the wavelength even further). Nevertheless, at the very low frequencies and very long waves the waves are probably in the fast magnetosonic wave band. 
\begin{figure}[t!]
\hspace{0.0cm}\centerline{\includegraphics[width=0.95\textwidth,clip=]{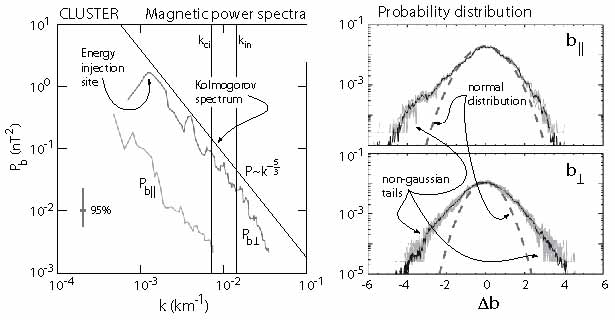}}
\caption[FS waves disprel]
{\footnotesize {\it Left}: Short wavelength spectra for the magnetic fluctuations measured by CLUSTER in the foreshock away from the foreshock boundary. These spectra are obtained by application of the ``wave-telescope" analysis method. The perpendicular fluctuations are more than one of magnitude more intense than the parallel fluctuations while the spectra decay about according to the Kolmogorov $-\frac{5}{3}$-law of stationary inertial turbulence. Indication of a cut-off is seen fat wave numbers larger than the inertial wave number $k_{in}$ where strong dissipation sets on. From the maximum at $
k\sim 10^{-3}$ one concludes that in this wavelength range the spectral energy is injected by instability of fast and Alfv\'enic magnetosonic waves which cascade nonlinearly forward  towards shorter wavelengths. {\it Right}: The probability distributions of the parallel and perpendicular magnetic fluctuations. The dashed distributions are log-normal of same maximum. These distributions exhibit extended tails and thus indicate non-stationary and probably intermittent not fully developed magnetic turbulence \citep[][]{Narita2006}.}\label{chap5-fig-narita-turb}
\end{figure}

More mysterious are the higher frequency-short wavelength waves. From comparison with the linear dispersion relation they seem to fit on the right-handed whistler branch (high frequency magnetosonic whistler R$^+$). However, inspection of the dispersion relation indicated that all kinds of polarisation are scattered along the dispersion curve. The curve itself is very irregular even though the propagation angle of the waves seems to be about constant at weakly oblique angles. This wave composition suggests that we are not  dealing here with one single wave mode but rather with the short wavelength part of a turbulent spectrum which has generated all kinds of short wavelength fluctuations with different almost randomly distributed polarisation in a forward cascading process from long to short wavelengths. This idea has been elaborated in more detail in \cite{Narita2006} who found that the shorter wavelength spectrum is indeed about featureless and power-law, close to a $-\frac{5}{3}$-Kolmogorov-spectrum\index{foreshock!Kolmogorov spectrum}\index{Kolmogorov, A. N.} of stationary turbulence with most of the power in the perpendicular (non-compressive) magnetic component providing another argument for Alfv\'enic and magnetosonic turbulence. This is shown on the left in Figure\,\ref{chap5-fig-narita-turb}. The figure also indicates the respective gyro- and inertial wave numbers, $k_{ci}= v_i/\omega_{ci}$ and $k_{in}=c/\omega_{pi}$. There is indication that the spectra change slope at around these number due to onset of ion viscosity and inertia. The large maximum on the perpendicular power curve near $k\sim 10^{-3}\,{\rm km}^{-1}$ corresponds to a wavelength of $\lambda\sim 6000$\,km reported earlier \citep[e.g.,][]{Eastwood2003} and is at the right position for energy injection by instability.\index{foreshock!turbulence}

However, the conclusion that the deep foreshock is subject to fully developed fast magnetosonic/Alfv\'enic turbulence is not fully justified. This becomes obvious from considering the probability distribution\index{foreshock!PDF of fluctuations} of the field fluctuations as given on the right in Figure\,\ref{chap5-fig-narita-turb}. Would the fluctuations be normally distributed then their distribution functions would have the shape of the dashed curves. Instead the distributions are skewed into tails. In the case of the parallel component the distribution lacks symmetry. Such tails may indicate that the turbulence is non-stationary, intermittent or inhomogeneous. All three cases might hold in the foreshock, in particular as the foreshock is a rather limited spatial region which is bounded from two sides and is subject to plasma injections and plasma losses. Nevertheless,  the spectra determined contain signs of strongly nonlinear and turbulent interactions, and the waves, particularly the low-frequency waves, are of large amplitude and interact with the plasma ion component as also with other waves. These waves affect the upstream backstreaming ion component scattering, heating, and accelerating it. On the other hand the upstream ion component is responsible for the existence of the waves as it is the ultimate energy source of the waves. And, to close the cycle, the waves cannot escape upstream very far from the foreshock since in a supercritical shock the Mach number of the flow is higher than any Mach number based on the wave speed and is thus high enough for the flow to advect the wave spectrum towards the shock. As long as the waves do not completely dissipate their energy during this convection, the electromagnetic wave energy accumulates at the location where they arrive at the shock. The consequence of this accumulation is that the waves affect the shock, cause instability of the shock surface, and reorganise the shock front. This makes quasi-parallel shocks non-stationary and subject to some kind of irregular reformation. One may thus expect that a supercritical quasi-parallel shock does not represent a solid shock surface over a large area. It consists of a more or less dense patchwork of areas which together form a shock but which also appear and disappear in an irregular manner, come and go, and organise the shock in a certain volume where the entropy increases but where a multitude of very difficult to handle processes takes place that can be investigated only experimentally or with the help of properly designed numerical simulations. These are, however, more difficult to design than in the quasi-perpendicular case, because of the greater variability of the conditions at a quasi-parallel shock. Since waves and particles are tied to each other, simulations cannot be discussed separately for waves and for particles. We therefore delay the discussion of the simulation results to the later section on quasi-parallel shock reformation. There it will become obvious what the  upstream waves are good for and what their role is in the quasi-parallel shock process.

An important question that has not been addressed anywhere in the investigation of the ion-foreshock dynamics concerns the role of the foreshock electron component which, in the large ion-foreshock domain, consists of two populations, the $\sim 100$\,eV upstream electron population which belongs to the quasi-neutral upstream flow, and the hot $\sim$\,several kev-electron population which is the product of the shock reflected electrons we are going to discuss in the next paragraphs. This component is isotropic but irregular and might contribute to ion-foreshock instabilities thereby affecting the foreshock turbulence and shock formation. Since these questions are difficult to be treated theoretically one has to wait until numerical simulations will become capable of including them into full particle codes in a similar way as has been done \citep[cf.][]{Matsukiyo2003,Matsukiyo2006} for the electron-ion populations in the feet of quasi-perpendicular shocks.   
\begin{figure}[t!]
\hspace{0.0cm}\centerline{\includegraphics[width=0.7\textwidth,clip=]{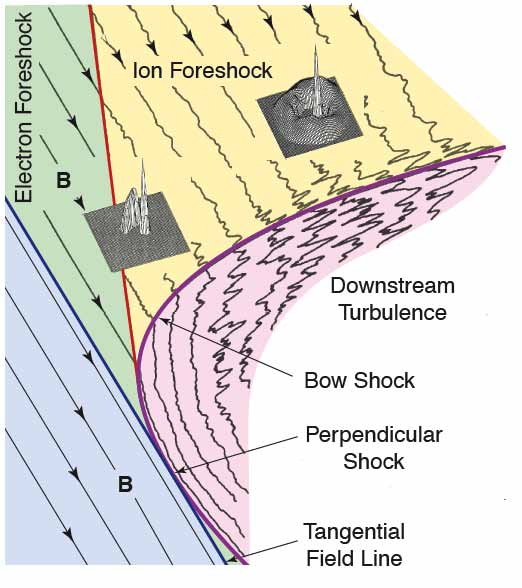}}
\caption[Electron Foreshock]
{\footnotesize Synoptical view of the two foreshocks: the electron and ion foreshocks, respectively.}\label{chap5-fig-fsbc}
\end{figure}
\subsection{Electron foreshock}\noindent\index{foreshock!electron, ion}\index{foreshock!electron waves}Even though the ion-foreshock occupies the larger part of the foreshock and, which is more important, plays the decisive role in the dynamics of the quasi-parallel supercritical shock, it would be an ignorant attitude not to mention the part that aside of the main flow is populated solely by the reflected higher energy electron component, i.e. the electron-foreshock region. Schematically its extent and relation to the ion-foreshock is shown in Figures\,\ref{chap5-fig-foreschem}  and \ref{chap5-fig-fsbc}. 

The electron foreshock has been identified already by \cite{Scarf1971} in the {\footnotesize OGO 5} satellite observations. \index{spacecraft!OGO 5} \cite{Scarf1971} noted the occurrence of enhanced electron fluxes near the bow shock, which were related to observations of high-frequency electric-field spikes at 30\,kHz. They concluded that these spikes resulted from Langmuir waves that had been excited by electron beams arriving from the bow shock along shock-connected magnetic field lines. (A detailed overview of the early observations can be read in \cite{Klimas1985}.) It could be confirmed that these electron fluxes were magnetic-field aligned and of higher energy than the bulk electrons in the main flow. This was interpreted as electron beams emitted into upstream direction from the shock, even though no mechanism was known that could provide the required shock-electron acceleration -- in fact, a de Hoffman-Teller-frame shock-reflection mechanism had already been proposed by \cite{Sonnerup1969} and was reinvented and elaborated on much later by \cite{Wu1984}.

\subsubsection*{Electron beams}\noindent
In analogy to the ion-foreshock boundary the electron-foreshock boundary (which is the ultimate upstream boundary of the foreshock) carries narrow bursts of electron beams, which escape into upstream direction from the shock along the magnetic field and which are slightly displaced by the convective flow into downstream direction. 
\begin{figure}[t!]
\hspace{0.0cm}\centerline{\includegraphics[width=0.95\textwidth,clip=]{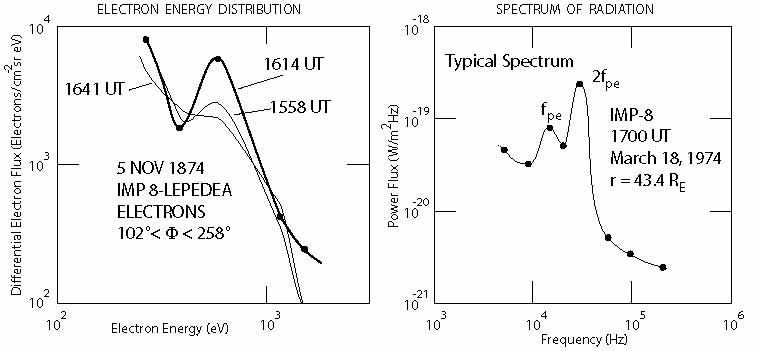}}
\caption[Electron Foreshock]
{\footnotesize {\it Left}: The electron energy distribution as seen by IMP-8 at three times when crossing the foreshock boundary. {\it Right}: The high-frequency wave spectrum of the electromagnetic radiation as measured some distance away from the shock near the foreshock boundary. Two types of emissions are visible, the so-calle `harmonic' emission $f=2f_{pe}$ at twice the electron plasma frequency $f_{pe}=\omega_{pe}/2\pi$, and the weaker so-called `fundamental' emission at $f_{pe}$. \citep[after][]{Gurnett1975}.}\label{chap5-fig-gurnettrad}
\end{figure}

An example of a narrow energy spectrum of such a beam measured by the {\footnotesize IMP-8} satellite is shown on the left in Figure\,\ref{chap5-fig-gurnettrad}.  The spacecraft was in the electron foreshock boundary for a relatively short time only. Being near apogee, the spacecraft was about standing. Thus the short contact time with the beam most probably indicates that the electron-foreshock boundary is fairly narrow. The electron beam also occupies only a small volume in velocity space corresponding to a narrow bump on the distribution \citep{Anderson1981}. Only one energetic electron-gyroradius deeper in the electron foreshock the electron beam becomes depleted and just forms an energetic tail of hot halo electrons on the electron distribution function of the flow. 

\begin{figure}[t!]
\hspace{0.0cm}\centerline{\includegraphics[width=0.95\textwidth,clip=]{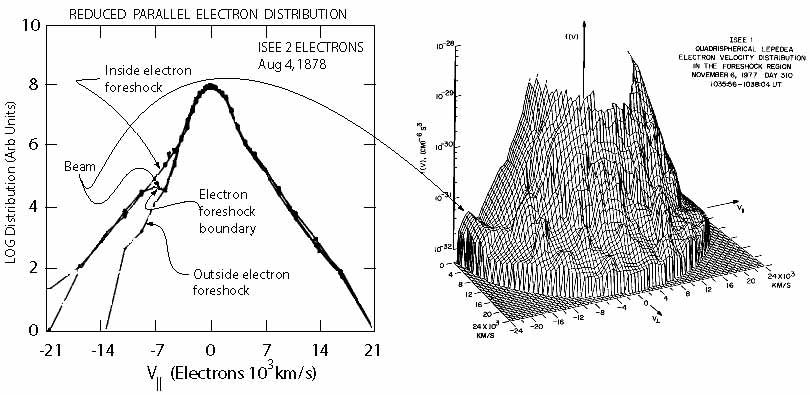}}
\caption[Electron Foreshock]{\footnotesize  {\it Left}:  Three successive reconstructions of the reduced parallel electron distribution function during crossing of the electron foreshock boundary. The distributions are only seconds apart, at 1650:37 UT before touching the foreshock boundary, at 1650:43 UT just crossing it, and at 1650:52 UT being behind it in the electron foreshock. The beam is visible only during the short crossing time. Afterwards the electron distribution shown a heated non-symmetric tail, indicating that the spacecraft sees the depleted beam in the hot electron-foreshock plasma  \citep[after][]{Fitzenreiter1984}. {\it Right}: The full two-dimensional electron distribution function during another crossing of the electron-foreshock boundary at the moment when the beam is visible as a narrow enhancement of the distribution in the negative parallel velocities (upstream along the magnetic field). An instant later, when the spacecraft entered into the electron foreshock the beam had disappeared  \citep[after][]{Anderson1981}.}\label{chap5-fig-fitzen}
\end{figure}

Figure\,\ref{chap5-fig-fitzen} shows electron phase-space observations from the electron instruments on {\footnotesize ISEE 1 \& 2} during crossings of the electron-foreshock boundary.\index{foreshock!electron boundary} The left part of the figure \citep[taken from][]{Fitzenreiter1984} is a much higher time and velocity (respectively energy) resolution plot than that in the former figure. It is nicely seen how the electron distribution evolved from a field-aligned (nearly) Maxwellian distribution prior to contact with the foreshock boundary, through a field-aligned-beam-like distribution at contact, into a distribution with an energetic tail along the magnetic field, where the electron beam has been completely washed out. This transition takes place within a time interval of 15\,s, just allowing to obtain the three measurements. The pseudo-threedimensional plot of the electron distribution function on the right \citep[taken from][]{Anderson1981} was obtained on another day when the spacecraft passed the electron-foreshock boundary and a series of distributions were recorded from upstream, across the foreshock boundary, and into the electron foreshock. Here only the foreshock-boundary distribution is shown when the narrow electron beam has evolved flowing into upstream anti-shockward direction along the magnetic field. Note the narrow angular extension of the beam, its weakness, and its comparably large velocity which allows to identify it just outside the bulk of the hot $\sim\,100$\,eV background electrons which are much hotter than the beam (indicated by the small beam width in $v_\|$).

The conclusion that can be drawn from these observations, which in the follow-up time had also been confirmed by measurements of other spacecraft like {\footnotesize AMPTE IRM and AMPTE UKS},\index{spacecraft!AMPTE UKS} is that similar to the ion foreshock, the electron foreshock consists of a very narrow region along about the shock-tangential upstream-magnetic field line (or flux tube) which is populated by a cold, weak, and fast electron beam, and a broad electron region where the electron distribution exhibits tails and in addition contains a hot, fairly isotropic electron component. The electron beam (like the ion beam at the ion-foreshock boundary) has its origin near the shock-tangential magnetic field line which is connected to the quasi-perpendicular part of the shock. The mechanism of its generation has not yet been unambiguously clarified. Presumably if consists of a combination of the de Hoffman-Teller frame mechanism of \cite{Sonnerup1969} and the bending of the shock surface plus some kind of stochastic acceleration in the electric fields which evolve in the shock transition, ramp and overshoot. Such fields have indeed been reported recently on scales below the electron skin depth $\lambda_e=c/\omega_{pe}$ to be very large, of the order of $\lesssim 100\,{\rm mV\,m}^{-1}$ parallel to the magnetic field and $\lesssim 100\,{\rm mV\,m}^{-1}$ perpendicular to the magnetic field \citep{Bale2007}. The presence of the hot electron component in the foreshock, on the other hand, requires another mechanism that heats  and accelerates electrons at a quasi-parallel shock over a large area of its surface. 

\subsubsection*{Langmuir waves}\noindent\index{waves!Langmuir}So-called `gentle electron beams', which are just those beams one observes in the electron-foreshock boundary -- fast, but not too fast ($V_b$\,few times $V_1$), parallel ($V_b\equiv V_{b\|}, V_{b\perp}=0$), weak ($N_b\ll N_e$), cool ($v_{eb}<v_e\ll V_b$),  are know to be the drivers of Langmuir waves with dispersion relation 
\begin{equation}
\omega^2({\bf k})=\omega_{pe}^2+3v_e^2k^2, \qquad\quad v_{\rm res}>\sqrt{3}\,v_e
\end{equation}
via the resonant kinetic gentle-electron beam instability. The Langmuir resonance conditions is $\omega-{\bf k\cdot v}_{\rm res}=0$, where ${\bf v}_{\rm res}\sim {\bf V}_b$ is the resonant electron velocity. Combining it with the dispersion relation, it is easy to see that  in order for the wave number $k$ to be real the resonant velocity must satisfy the condition on the right in the above equation. Below we will discuss the properties of these waves in more detail in connection with observations.

\paragraph{Stability of the electron beam.} However, there are a number of severe limitations to this instability which theoretically would inhibit its application to the electron-foreshock boundary beam. The most stringent of these limitations is that it is believed that quasi-linear saturation of the gentle-beam instability should quickly, i.e. within a time as short as about $\Delta t\,\omega_{pe}\sim 10$, deplete the beam and transforming it into a plateau on the distribution function. For an upstream plasma density $N_e\simeq 5\times 10^6$\,m$^{-3}$ the plasma frequency is $\omega_{pe}/2\pi\simeq\,20$\,kHz. Hence, the beam should become depleted within $\Delta t\simeq 0.08$\,s. A beam of velocity $V_b\simeq 10^4$\,km\,s$^{-1}$ (which satisfies the above condition on the resonant velocity in an inflow of $T_e\sim 100$\,eV) would thus propagate just 800\,km upstream of the shock along the magnetic field before being depleted, with the implication being that the beam could not escape from its shock-source region. This distance is much less than the distances from the shock at which both the electron-foreshock boundary beams and the beam-excited plasma wave spectra have been observed. Either quasilinear saturation of the Langmuir instability does not work, or the observed plasma waves are no Langmuir waves. The latter is probably not true. So the question arises of how the beam can avoid becoming depleted due to quasilinear effects. 

There are several possibilities of inhibiting beam depletion. It has for long time been believed that collapse of Langmuir waves resulting from modulation instability (also known as oscillating two-stream instability) could shift the beam-excited Langmuir waves out of resonance and remove them from reaction onto the beam. This is a very elegant mechanism which is based on the ponderomotive force the Langmuir waves exert on the plasma. The ponderomotive force is the gradient of the wave pressure which is strongest right in the place where the Langmuir waves are most intense, i.e. in the beam region. For the inert ions of the plasma background the intensity gradient $\nabla|{\bf e}|^2$ of the fast large-amplitude oscillating Langmuir electric field presents a pressure force which pushes them out of place thus decreasing the plasma density locally. Since the electrons react immediately and follow the expelled ions in order to maintain quasi-neutrality, the wave-ponderomotive-pressure force drills holes into the plasma which are filled with Langmuir waves. The density variation produced 
\begin{equation}
\delta N\approx -\frac{\epsilon_0}{4m_ic_{ia}^2}|{\bf e}|^2
\end{equation}
is nothing else but an ion-acoustic wave. The Langmuir waves, when becoming intense enough, can drive ion-acoustic waves of speed $c_{ia}$, which exist only due to the presence of the Langmuir wave pressure gradient. The broad Langmuir waves in this way organise into a large number of density cavities, and the waves are removed from resonance with the beam electrons. 
Enhancement of the outside pressure then forces the holes to shrink. This shrinkage of the hole size shortens the Langmuir wavelength, increases the wave number, reduces the phase velocity and, in this way, shifts the waves into the main electron distribution where they can undergo Landau damping and dissipation. Unfortunately, this kind of modulation instability and collapse of Langmuir waves could never be experimentally approved at shocks. It is therefore quite unlikely (though not impossible) that it occurs and stabilises the beam. \index{process!beam stabilisation}

Another possibility of stabilising the beam by pushing the Langmuir waves out of resonance is scattering them on thermal ions. This process reads $\ell+i\to\ell'+i'$. For purely elastic scattering only the momenta $\hbar{\bf k}_\ell$ of the Langmuir wave $\ell$ and ${\bf p}_i$ of the ion $i$ are changed according to the momentum conservation equation
\begin{equation}
\hbar{\bf k}_\ell +{\bf p}_i \to \hbar {\bf k}'_\ell +{\bf p}_i'
\end{equation}
where the primed quantities are after collision of wave and ion. Loss of momentum decreases the wave number $k_\ell$ and shifts the wave phase velocity to higher values out of resonance with the beam; gain of momentum increases $k_\ell$ and brings the wave phase velocity down into the bulk of the electron distribution where it is again dissipated. The efficiency of this process is, however, low and presumably insufficient for complete stabilisation of the beam. In addition it strongly depends on the available particle number; it is thus most probable to work in the immediate vicinity of the shock only. It does, however, produce a new Langmuir wave $\ell'$ that plays a role in the generation of radio-radiation from a shock. Because of this capacity it nevertheless remains to be of interest in the physics of collisionless shocks.

\cite{Muschietti1991} investigated this nonlinear wave scattering off ions\index{process!scattering off ions} including the polarisation cloud which accompanies each ion. They found that of the ion distribution the ions with velocity $v\simeq 2v_i$ are the most effective scatterers of Langmuir waves. The inclusion of the polarisation cloud increases the scattering efficiency by a factor of three which does not change the above conclusion. However, the wave number of the scattered wave, its direction of propagation and other wave properties change substantially. Nevertheless, the effects are not strong enough to alone explain the long survival of the electron beam over distances very far from the shock.

A review of all the available theories has been given by \cite{Muschietti1990} who performed simulations of a spatially bounded beam. It seems that many effects act together in order to allow the beam to survive. Weak scattering off ions removes some waves from the beam. The remaining waves are only partially reabsorbed because the fast beam front runs them out such that they stay in the trail of the beam only. In this way only the trail part of the beam is depleted and retarded, which affects only the low velocity component of the beam distribution, assuming that the beam has a small but sufficiently large natural beam spread $\Delta\,v_{eb}\ll v_e$, as is suggested by the observations shown, for instance, in Figure\,\ref{chap5-fig-fitzen}. If injection is continuous, this implies that the beam starts pulsating, which may be a reason for the observed strong Langmuir  wave variations along the electron foreshock-boundary field line. Finally, the Langmuir waves, riding on the beam, are generally slower than the beam; since $v_{\,\rm res}^2>3v_e^2$, their group velocity, which carries the energy that can be absorbed, is at most $v_{\ell g}<v_e\sqrt{3}$. So they are more vulnerable to the downstream convection than the beam electrons. The convective ${\bf V\!}_{\!\perp}$\!-motion of the upstream flow thus sweeps them readily out of the beam region at the electron foreshock-boundary field line. All these arguments taken together explain why the beam can survive over long distances and, in addition, why the region across the electron foreshock boundary where the beam is detected is as narrow as found in the observations. 

\paragraph{Electron foreshock-boundary waves.} 
\begin{figure}[t!]
\hspace{0.0cm}\centerline{\includegraphics[width=0.9\textwidth,clip=]{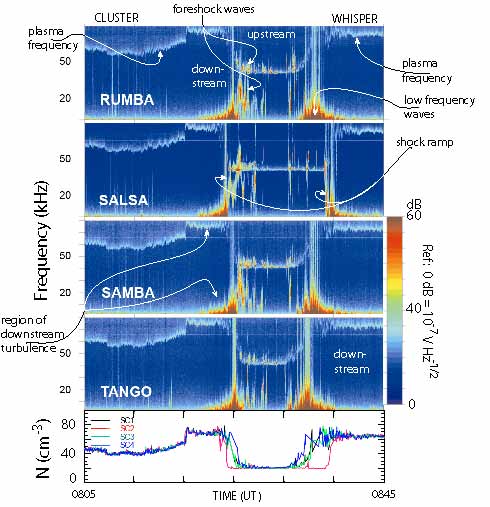}}
\caption[cluster shock emissions]
{\footnotesize A {\CL} spacecraft passage across the Earth's bow shock region on December 22, 2000. The spacecraft are coming from the magnetosphere, pass across the downstream magnetosheath region with its enhanced density, cross the shock into the upstream region and back again and remain in the downstream magnetosheath. The hole region is seen in the light of the broadband plasma wave electric field spectrum from about 0 kHz up to 80 kHz. The figure shows the four panels of all four {\CL} spacecraft (with their somewhat childish names Rumba, Salsa, Samba, Tango). The uppermost light blue emissions in each panel belong to the electron plasma frequency $f_{pe}=\omega_{pe}/2\pi$ which on most of the path is thermal noise mapping the local plasma density. The passage of the shock is signalled by the appearance of intense broadband waves starting at low frequencies and being correlated with a fairly steep drop in the plasma frequency respectively the plasma density. Outside the shock in the low density upstream region the strong spots of intensification of the plasma frequency indicate contact with the electron foreshock boundary beam and excitation of Langmuir waves. the intensification is highly structured, sometimes stretching above, sometime hanging down below $f_{pe}$. Also seen are intermediate frequency emissions near the shock. The lowest panel shown the estimated from $f_{pe}$ plasma density for all four spacecraft \citep[after][]{Decreau2001}.}\label{chap5-fig-clusterwhisper}
\end{figure}

\begin{figure}[t!]
\hspace{0.0cm}\centerline{\includegraphics[width=0.95\textwidth,clip=]{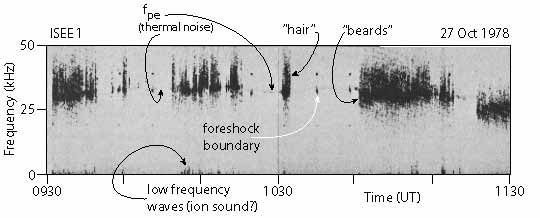}}
\caption[cluster shock emissions]
{\footnotesize An example of  {\footnotesize ISEE 1} high frequency wave observations during contact and passage of the electron foreshock boundary. The thin faint spotty line around 30-40 kHz is the local plasma frequency $f_{pe}=\omega_{pe}/2\pi\sim 30$\,kHz, seen here only as thermal noise with lesser instrumental sensitivity than in the {\CL} observations. The occasional intense dark spots as the one marked by the white arrow are brief contacts with the foreshock boundary field line when the beam occurs. deeper in the foreshock the spectrum broadens non-symmetrically, evolving `hair' and `beards'.  Some moving structures can be identified in the hair. The emissions deeper inside the foreshock are correlated with the occurrence of low frequency emissions below a few kHz which are probably ion-sound waves \citep[after][]{Etcheto1984}.}\label{chap5-fig-etcheto}
\end{figure}

Figure\,\ref{chap5-fig-clusterwhisper}, taken from \cite{Decreau2001}, shows a beautiful double passage of the {\CL}-spacecraft quartet through the Earth's bow shock on December 22, 2000 in the light of the {\CL} plasma wave spectrum recorded by the {\footnotesize WHISPER} plasma wave instrument aboard the four {\CL} spacecraft. Figure\,\ref{chap5-fig-etcheto} shows an example of these high-resolution measurements 

{\CL} came from the magnetosphere, entered the downstream region of the shock, passed the shock to upstream, re-entered the shock, and escaped downstream again. The relevant signatures of each region are seen in all four spacecraft. The uppermost light blue trace is the electron plasma frequency $f_{pe}=\omega_{pe}/2\pi$ which maps the local plasma density. The shock appears as the intense broadband emission with maximum intensity in the low-frequency waves. In the density it is mapped as a steep drop in $f_{pe}$ to the low upstream density values. 

Close to the shock a number of intensifications in $f_{pe}$ can be recognised. These occur at the time of contact with the electron foreshock boundary when intense plasma waves are excited in the Langmuir mode. It is interesting to note that the intensification occurs in spots and is not necessarily narrow-band. It may exceed $f_{pe}$, or it may also drop below it: the plasma frequency may have `hair' or `beards'. Moreover, sometimes intense lower-frequency emissions are detected half way between the plasma frequency and the low frequency emissions in the foreshock. The most intense  of these emissions belong to the region closer to the shock where the density gradient has not yet settled to the upstream values, but weaker emissions of the same kind occur farther away in the upstream low density domain. 

A detailed analysis of all these waves detected by {\CL} has not yet been undertaken and thus is not yet available. It is, however, highly suggestive that the observed intense spots in the plasma frequency, with the `hair' parts exceeding and the `beard' parts hanging down from the local $f_{pe}$, are related to magnetic field-aligned electron beams emanating from the shock and propagating upstream along the electron foreshock-boundary magnetic field as these are the same signatures as those observed with the {\footnotesize ISEE} and {\footnotesize AMPTE} spacecraft, though with much better frequency resolution and higher sensitivity here. For instance, the latter two spacecraft were unable to resolve the plasma frequency which in the {\CL} data is nicely distinguishable even though it represents just thermal noise. It is most interesting that these contacts with the electron foreshock boundary occur quite irregularly throughout the hole period when {\CL} was upstream of the shock and that they do not seem to exhibit a one-to-one correlation between the four spacecraft. Taking into account that the upstream plasma frequency away from the shock is fairly constant this implies that the shock-normal angle $\thetabn$ changes on a irregular and fast time scale, i.e. that the direction of the upstream magnetic field is highly variable even under conditions of apparent quiescence in the upstream medium. The conclusion drawn from this would not be that the upstream magnetic field is subject to violent variations as in the electron foreshock there is no reason for the magnetic field to be strongly affected. Rather this variability points on the temporal variation of the shock area near the tangential field line. Even a small change in the shock normal there will cause a large variation in the location of the tangential magnetic flux tube at a distance upstream of the shock. Reasons for such a variability have been given in the section on quasi-perpendicular supercritical shocks, but to repeat here, the main reason is that the super-critical shock is neither in thermal nor in thermodynamic equilibrium and is thus by its very nature subject to changes in all its physical parameters.    

In addition to the plasma frequency emissions, which hint on contact of the spacecraft with the shock-tangential field line that makes up the electron foreshock boundary, there are a number of most interesting features seen in the wave spectra of Figure\,\ref{chap5-fig-clusterwhisper}. One of them we merely note at this place because we will return it farther below. It is the broadband signals that are occasionally seen in many places, in the plasma frequency as signals covering a frequency band of up to $\sim 20$\,kHz, in the lower frequencies near the shock, an in particular in the shock transition itself where the frequency range of the emission $\Delta f> 80$\,kHz in all cases exceeds the whole range of the instrument. There are no known plasma waves of a comparable bandwidth reaching from $\sim 0$\,kHz deep into the range of free space radiation frequencies. Such spectra are, however, known to be produced from narrow spatially localised electric field structures. The broader the spectrum the narrower will the structures be. We therefore conclude that in all the places where such broadband emissions are encountered, and in particular in the shock transition, we are dealing with very narrow localised and intense electric fields which have been generated by nonlinear processes. Such structures are presumably solitons and/or phase space holes either of electronic or ionic nature.        

The so far most elaborate investigation of the width of the electron-foreshock boundary beam and its relation to the excitation of plasma waves has been performed by \cite{Etcheto1984} and subsequently refined and supported by theory by \cite{Lacombe1985}. An example of their {\footnotesize ISEE 1}observations is shown in Figure\,\ref{chap5-fig-etcheto}. Since the instrument was much less sensitive than {\footnotesize WHISPER} the thermal noise at the plasma frequency around $f_{pe}\sim 30$\,kHz appears only as a very faint spotty line. Occasional strong narrowband intensification of this line has been identified as contacts with the electron foreshock-boundary beam by inspecting the simultaneous electron measurements $>300$\,eV. These emissions were indeed found to be very narrow band, just 1-2\,kHz wide at $f_{pe}$. Deeper inside the foreshock the bandwidth of the emissions broadens substantially in a non-symmetrical way. First, it becomes very noisy, consisting of short emissions clumped together in groups. Second, long `hair' emissions evolve of roughly up to 10\,kHz bandwidth, while `beard' emissions are also found being generally weaker, but sometimes they extend to low frequencies. Low frequency waves below a few kHz are also detected in connection with these deeper-foreshock emission and were found to smoothly merge from below into the beards. \index{foreshock!boundary waves}

\cite{Etcheto1984} and subsequently \cite{Lacombe1985} could show that the spectra changed from very narrowband in the foreshock boundary flux tube to broadband in the foreshock, just a few electron gyroradii away from the foreshock boundary. They also showed that the electron beam distribution flattened over this distance. The waveform of the waves at the plasma frequency in the broadband region away from the foreshock-boundary exhibited modulations, which group the waves into groups of length of a few 10\,ms, a frequency roughly corresponding to the frequency of the low frequency waves seen in the dynamic spectrum of Figure\,\ref{chap5-fig-etcheto}. 

\paragraph{The nature of electron-foreshock waves.} The narrowband electron foreshock-boundary waves detected at the foreshock-boundary flux tube are clearly electron-beam driven Langmuir waves of the surprisingly large amplitude of a few mV\,m$^{-1}$ at the edge of the foreshock and a few \% bandwidth in frequency. 

It is considerably more difficult to infer about the nature of the waves deeper in inside  the electron foreshock. These waves reach large amplitudes of a few 10\,mV\,m$^{-1}$ to a few 100\,mV\,m$^{-1}$ and bandwidths of $\gtrsim 30$\%. (Note that the stationary convection electric field in a $\sim500$\,km\,s$^{-1}$ flow in a $\sim5$\,nT magnetic field is just $E\sim 2.5$\,mV\,m$^{-1}$!) The wave electric field is practically parallel to the upstream magnetic field, and the wavelength is long below and short above $f_{pe}$. Moreover, these waves are modulated and seem spectrally to connect to the intense low frequency waves. For plasma frequencies of $\sim 30$\,kHz the ion plasma frequency is 0.7\.kHz. Thus, accounting for the Doppler-broadening of the low-frequency wave spectrum in the fast flow, the low frequency waves can be tentatively identified with ion-acoustic waves, which accompany the high frequency waves at the plasma frequency. Obviously these wave modulate the latter, it is however not clear whether or not they are created by the high-frequency waves via the modulation instability or whether they are generated in a different interaction between the depleted beam and the upstream ion flow via an ion-acoustic instability. We have already argued that the modulation instability is unlikely under the conditions of the electron-foreshock boundary beam. Moreover, the weak modulation of the wave form of the high -frequency wave noted above does not argue in favour of the modulation instability and caviton formation as the waves are not really bundled in localised groups and the wavelength is not changed appreciably. On the other hand, the spiky broadband nature provides a weak argument for some localisation which can, however, be objected owing to the pronounced asymmetry of the spectrum with respect to $f_{pe}$. 

\cite{Lacombe1985}, ignoring the correlation with the low-frequency waves, interpret the high-frequency waves around $f_{pe}$ on the basis of the kinetic Langmuir wave instability including the beam plasma, i.e. referring to a non-gentle beam of finite temperature and appreciable density. Their simplified dispersion relation then becomes
\begin{equation}
1=\frac{\omega_{pe}^2}{\omega^2}(1+3k^2\lambda_{De}^2)-\frac{N_b}{N}\left[1+i\sqrt{\frac{\pi}{2}}\frac{\omega-kV_b}{kv_{be}}\right], \quad \left|\frac{\omega}{kv_e}\right|\gg1, \left|\frac{\omega}{kV_b}-1\right|<\frac{\sqrt{2}v_{be}}{V_b}
\end{equation}
with $N_b, V_b, v_{be}$ the respective beam parameters, density, velocity, and thermal speed, and $\lambda_{De}=v_e/\omega_{pe}$ is the Debye length of the upstream plasma. The first two terms in this expression are the ordinary Langmuir wave. The second term on the right is the beam contribution. for small gentle-beam densities the contribution to the real frequency can be neglected. However, for larger beam densities we see that $N_b/N$ adds to the unity on the left. Hence the frequency of the wave should decrease for small $k$. Thus long wavelength waves will have frequencies $\omega<\omega_{pe}$. This agrees nicely with the observation. Short wavelength waves, where the term $3k^2\lambda_{De}^2$ comes into play, will have frequencies $\omega>\omega_{pe}$. This is also in accord with observation. Thus, the inclusion of the dense beam into the dispersion relation does reproduce the basic observed spectral properties of the beam excited waves at least qualitatively in a simple linear way without reference to nonlinear effects like the modulation instability and collapse. The occurrence of the low-frequency ion-acoustic waves is then probably due to the electron-ion velocity difference of the hot electron component and the cold upstream ion flow via the electron-ion acoustic instability.

In a series of one-dimensional particle simulation papers, \cite{Dum1990} has investigated the dynamics of beams at the foreshock boundary in an attempt to quantify the above qualitative theoretical conclusions. He considered a pure electron beam on an otherwise neutralising ion background. As expected, a gentle weak beam evolves basically according to quasilinear theory into a plateau which for long times remain unchanged. Even in extremely long simulation times there is practically no evolution in wave energy, and the distribution function remains stable once the quasilinear plateau is formed. This happens approximately after a few hundred plasma periods. For a plasma frequency $f_{pe}\simeq 30$\,kHz this amounts to a relaxation time of the order of just $\sim$1\,ms. Hence, gentle beams at the foreshock should be stabilised within this time, and any nonlinear effects will evolve only at much later times if at all. Excluding the ion dynamics, the simulations have been run until $\sim\,3000\,\omega_{pe}^{-1}$ without any susceptible change in the distribution and wave level, in physical times until $\sim10$\,ms. Moreover, the saturation level of the waves is remarkably low, much lower indeed than the theoretical estimates suggested. One does not expect that weak or strong turbulence effects will evolve which could change or prevent plateau formation. For this to happen, of course, ion dynamics should be included into the runs and should produce small $k$ waves, while in the plateau formation the wave number increases because the plateau widens and the beam front proceeds to smaller and smaller velocities until the plateau is completed and Landau damping stops the further evolution. Therefore, based on these simulations, the dynamics of a clod beam cannot explain the richness of the wave observations in the foreshock boundary.
\begin{figure}[t!]
\hspace{0.0cm}\centerline{\includegraphics[width=0.95\textwidth,clip=]{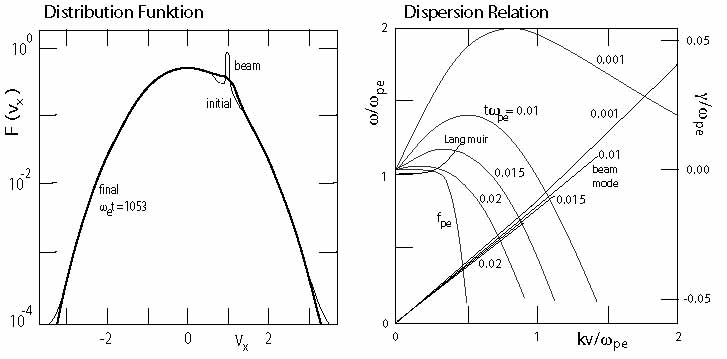}}
\caption[cluster shock emissions]
{\footnotesize Results of a one-dimensional electron-electron beam particle simulation \citep[after][]{Dum1990} intended to explain the high-frequency plasma waves in the electron foreshock. {\it Left}: The electron distribution used. Initially (thin line) it consists of a broad (warm) Maxwellian background distribution plus a narrow (cold) electron beam sitting on the main distribution. All quantities are normalised, velocity is normalised to nominal beam velocity. At end time $t\omega_{pe}=1053$ (solid line) the beam has become depleted and heated but has not disappeared.  {\it Right}: Real dispersion relations $\omega(k)$ (solid lines) and growth rates $\gamma(k)$ (thin lines) at different simulation times $t\omega_{pe}$ as indicated by the numbers. Only the unstable part of the real dispersion relations (corresponding to positive $\gamma>0$) is plotted. The Langmuir wave is shown separately with its growth rate labelled $f_{pe}$ which is weekly positive only at very long wavelengths. The main unstable waves are shown to propagate in the beam mode (about straight lines starting at origin of ${\omega}$ and $k$. Maximum unstable frequencies are well below the plasma frequency $\omega_{pe}$.With progressing time the wavelength of the unstable modes increases (decreasing $k$). The unstable domain in frequency and wave number shrinks. The unstable frequency decreases as a consequence of the decreasing beam velocity which leads to a decrease in the slop of the beam mode.}\label{chap5-fig-dum}
\end{figure}

For a very cold beam the evolution is a bit more complicated as the cold beam initially is not subject to the Langmuir instability but rather to the reactive beam instability which generates a broad spectrum below $f_{pe}$ as inferred above \citep{Lacombe1985}. The maximum growth is close to but below $f_{pe}$, and the spectrum is sharply cut off just above $f_{pe}$. Afterwards the beam evolves readily into the kinetic Langmuir stage described above, forming a plateau and stabilising. This transition proceeds due to intermediate electron trapping in the wave, which heats the beam to a temperature when the kinetic instability can take over. One might thus conclude, that the waves observed below $f_{pe}$ indicated the passage of the cold narrow beam front. This, however, is in contradiction to the observation that these waves are not observed in contact with the electron-foreshock boundary but deeper in the foreshock where no cold beams are present. Hence, another interpretation for the waves is needed. \index{energy!saturation}

One solution is to take into account the bulk streaming distribution \citep{Dum1990}. This is shown in Figure\,\ref{chap5-fig-dum}. Then the interaction becomes an electron-electron beam mode interaction with the possibility to destabilise the electron acoustic wave mode. This mode has frequency sufficiently far below the plasma frequency, $\omega< \omega_{pe}$. A condition is that the beam velocity spread is small. Initially the beam also excites frequencies substantially above $f_{pe}$ with large growth rate, but these are quickly stabilised. The unstable modes are actually beam modes with resonance condition $\omega\lesssim kV_b$ which are destabilised by Landau damping from bulk electrons, i.e. the beam speed must enter the range of bulk velocities for Landau damping and should thus not be displaced far from the bulk distribution like in the gentle beam case. Plateau formation takes very long time in this case such that the wave can reach quite large intensity of the order of $W/NT_e\lesssim10^{-3}$. For an upstream density of $N\sim 5\times10^6\,{\rm m}^{-3}$ and electron temperature of $T_e\sim 100$\,eV this yields an rms electric field amplitude $|{\bf e}|_{rms}\lesssim 1\,{\rm V\,m}^{-1}$ in very good agreement with observation. For larger beam temperatures the general trend as in Figure\,\ref{chap5-fig-dum} remains valid. It is interesting that the unstable frequency readily drops from higher than $\omega_{pe}$ to $\omega<\omega_{pe}$ while the wave number shrinks, i.e. the wavelength of the maximum unstable waves increases. One should thus observe falling tones in the frequency of the emission. It is also interesting that during the evolution of the instability the slope of the beam mode decreases which simply reflects the retardation of the beam during depletion. The final state is low frequency-long wavelength. \index{process!quasilinear saturation}

\begin{figure}[t!]
\hspace{0.0cm}\centerline{\includegraphics[width=0.75\textwidth,clip=]{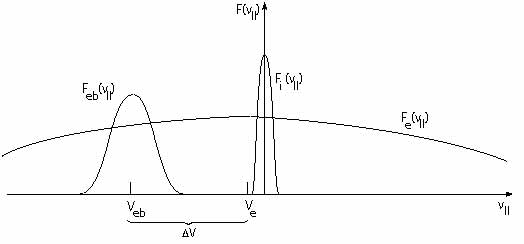}}
\caption[cluster shock emissions]
{\footnotesize Parallel distribution functions in the deep electron foreshock consist of the cold bulk ion distribution and the hot bulk electron distribution on top of which the parallel part of the hot diffuse, i.e. warm upstream electron distribution. This distribution in the upstream frame moves away from the shock. The upstream electron distribution adjusts to the current-free condition by being retarded by the small amount $|V_e-V_i|=V_bN_b/N$. Its maximum is shifted in the direction of the upstream diffuse beam distribution. Such a configuration should be unstable with respect to electron-acoustic waves (dense hot electrons, less dense cool beam electrons). }\label{chap5-fig-fsepdf}
\end{figure}

It seems that this calculation explains the observation of electron foreshock wave emissions. However, the problem about this model is that it should work only right at the electron-foreshock boundary where only emission at the plasma frequency is seen like in the classical case of  a gentle warm low density beam. The broadband waves below $f_{pe}$ are observed inside the electron foreshock, where no electron beams have ever been detected. Moreover, it does neither explain the high intensities and the broadband nature of the waves above $f_{pe}$, nor does it explain the connection of the waves below $f_{pe}$ to the low frequency ion-acoustic waves. Therefore we conclude, that probably a two-temperature counter streaming electron-component plasma with both components warm will be more appropriate to the interior of the foreshock, and probably the cold ion component of the bulk upstream flow must also be included. On the other hand, taking these simulations for serious, the deep electron foreshock region should be filled with many cold and not overwhelmingly fast electron beams propagating along the magnetic field upstream. Possibly the resolution of the current instrumentation is still unable to resolve them. The observations on the right in Figure\,\ref{chap5-fig-fitzen} might indeed indicate the presence of such small beamlets that are distributed over the gyro-angle. So far, however, they can only been regarded as measurement fluctuations. 

The parallel configuration of the distribution in the foreshock is shown in Figure\,\ref{chap5-fig-fsepdf} for a warm beam that is the field-aligned part of the diffuse electron component in the electron foreshock. Because of the vanishing-current condition the bulk electron component will be slightly retarded at the small amount $|\Delta V_e|=V_1N_b/N$. This configuration consists of a hot dense bulk electron component and the warm dilute beam component. In addition to the beam instability it should be unstable with respect to electron-acoustic waves at frequency $\omega\simeq kv_e(N_b/N)^\frac{1}{2}$. Electron-acoustic waves are long-wavelength waves below the electron plasma frequency. Thus the low frequency waves find several explanations, none of them completely satisfying though. For the high frequency waves which exceed the plasma frequency the only available interpretation is that they result from localised electric fields. In frequency space these localised waves have a broadband signature. The distributions in the perpendicular direction are non-symmetric halo distributions with the backstreaming electrons populating the halo tails. They still await an in-depth treatment for inferring their contribution to the wave spectra, particle scattering and plasma heating.\index{waves!electron beam}
\begin{figure}[t!]
\hspace{0.0cm}\centerline{\includegraphics[width=0.95\textwidth,clip=]{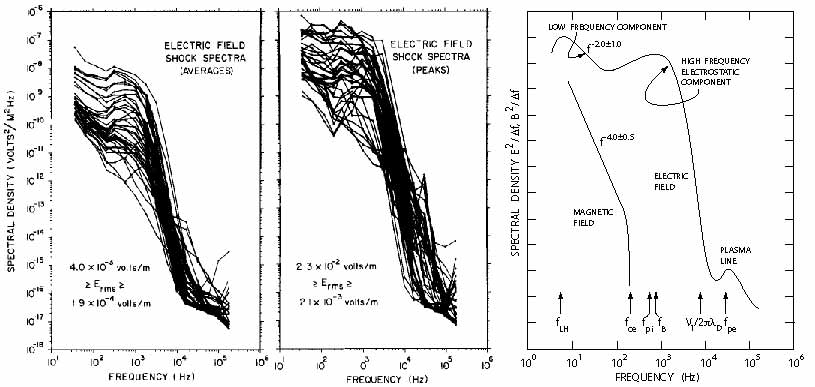}}
\caption[cluster shock emissions]
{\footnotesize A representative distribution of electric shock spectra. {\it Left}:  Average (1 s-averages) spectral values, showing the broad peak in the electric field spectra and the exponential decay of the spectrum toward high frequencies. For intermediate spectral intensities thepeak is well developed, while when the spectral intensity is very high it smoothes out. {\it Centre}: The same spectra but in the peak values (30 ms resolution) and not in the averages The spectra are similar but much more variable and up to two orders of magnitude more intense. This points to the high variability of the electric wave emissions at shocks. In addition indications of the plasma frequency are seen at the high frequency end, suggesting that the emissions in $f_{pe}$  are highly time variable \citep[after][]{Rodriguez1975}.  {\it Right}: A schematic summary of shock spectra showing the magnetic spectra being cut off at the electron cyclotron frequency $f_{ce}$, and the electric spectra containing several maxima at the lower hybrid frequency $f_{LH}$ at its low frequency end, around the ion plasma $f_{pi}$ and Buneman two-stream $f_B$ frequencies, with an absorption dip at the electron cyclotron frequency, a steep exponential decay caused by Doppler-shifted Landau damping of the waves in the Buneman and ion-acoustic modes at frequency $f_{pe}V_1/v_e=V_1/2\pi\lambda_D$, and the little bump at the electron plasma frequency   \citep[after][]{Gurnett1985}.}\label{chap5-fig-rodrig}
\end{figure}

We close this section by presenting in Figure\,\ref{chap5-fig-rodrig} an average synoptic spectral view of the waves detected in the shock transition as was provided by \cite{Rodriguez1975} and \cite{Gurnett1985} from consideration of a large number of shock spectra. There is no new information in this figure except that it summarises at a glance the main features in the higher frequency electric and magnetic field wave spectra. The cut-off of the magnetic spectra at the electron-cyclotron frequency is no surprise. In the electric spectra there is a large temporal variability as can be seen from a comparison of the average and peak values (measured within 1 s measuring time). The variations cover up to two orders in magnitude. Of interest is also that the electric spectra exhibit a maximum at the lower-hybrid frequency, an absorption at the electron cyclotron frequency, and show a broad maximum at the ion plasma/Buneman two-stream frequency before being steeply cut off towards higher frequencies. This cut-off occurs due to Landau damping at the Doppler-shifted frequency of ion-acoustic waves when the waves are shifted into Landau resonance with bulk electrons. In this frequency range all kinds of waves overlap, reaching from ion-acoustic waves and electron-cyclotron harmonics to Buneman waves, localised structures like Bernstein-Green-Kruskal modes and solitons, while the modified two-stream instability provides the large maximum at the lower hybrid frequency. The high variability in the peak values results from the presence of these localised structures which recently have been observed \citep{Bale2007} in situ but could have been concluded also just from the high variability of these spectra in the frequency range of the Buneman-modes respectively Doppler-shifted ion-acoustic waves. These are nothing but the signatures of many microscopic phase-space holes that obviously accumulate in the shock transition region. Support to such an interpretation has been given long ago by the {\footnotesize ISEE} measurements of the electron distribution function during shock transition \citep{Feldman1983} which \cite{Gurnett1985} made responsible already for the high variability of the peak spectral values. In these measurements the electron phase-space distribution function transforms from the upstream streaming Maxwellian to the shock-ramp and downstream flat-topped heated electron distribution, which just in the short time interval of crossing the shock -ramp magnetic-overshoot exhibited a clear signature of an electron beam that was sitting on the upstream edge of the flat top of the distribution (as depicted in Figure\,\ref{chap4-fig-gurnfeld}). The electron beam in that figure is caused by acceleration in the shock potential and has an upstream directed velocity of a few 1000\,km\,s$^{-1}$. Compared to the bulk electron temperature this beam is cool. Thus it may excite electron-beam waves and electron-acoustic waves. However, in combination with the bulk ion flow it is also capable of exciting the Buneman two-stream instability which will readily evolve into electron holes and plays a major role in heating the plasma and making the main electron distribution flat-top. One should also remember that the scale of the overshoot is narrow, of the order of a skin depth, and is most probably due to an intense electron current flowing in the ramp or ramp-transition region which, presumably, is also related to the observed beam. \index{shocks!electron distribution}

At high frequencies the spectra also exhibit the small peak caused by the beam excited Langmuir electron plasma waves, while the lower frequency foreshock waves are buried in the fat bump of the Doppler-shifted ion-acoustic waves. Of course, more cannot be concluded from a picture like this. The more detailed discussion of the spectra requires much higher spectral resolution. To the extent as it was available at the time of writing, we have given it here. But much more work is to be done until the various modes which can be excited in the foreshock and shock transition will be understood better.

\subsubsection*{Radiation}\noindent\index{foreshock!radiation}\index{process!plasma radiation}Shocks are frequently referred to as sources of radiation. Famous examples are supernova shocks, which are visible in almost all wavelengths, from radio through visible light up to x-rays \citep[e.g.,][]{Dickel2004}, and solar type II shocks (cf. Chapter 8) with their main radiation signatures seen in the radio waves. 

Supernova shocks\index{shocks!supernova}\index{shocks!relativistic} are relativistic shocks which are not treated here. Their Mach numbers range from weakly relativistic to highly relativistic, but the energy per particle in them remains to be less than the rest energy of an electron, $\lesssim m_ec^2=0.511$\,MeV, which allows to treat them classically. This does not hold anymore for the central supernova engine which drives the flow and which in some cases results in the generation of ultra-relativistic jets. There the shocks become non-classical, and not only radiation losses but also particle generation must be taken into account in their description.

\paragraph{Observations.} The radiation that is occasionally emitted from nonrelativistic shocks is restricted to the radio wave range. They do not generate x-rays or visible emissions because, first, of their comparably low energy per particle, which is less than the rest energy of an electron $\ll m_ec^2$, second, because of their low energy transmission rate (particles are not retarded from flow speed to rest) and, third, because of the low `emission measure' $EM=\int{\rm d}r^3\Delta N_e^2({\bf r})$, i.e. the\index{radiation!emission measure} number of radiating particles of density $\Delta N_e$ in the volume that contributes to the emission of x-rays is too low for providing any measurable intensity. Moreover, since magnetic fields are weak and the ratio $\omega_{pe}/\omega_{ec}>1$ of electron plasma to electron cyclotron frequency is larger than one, gyro-synchrotron emission is unimportant. Thus, the only means of how free-space radiation of frequency $\omega\gtrsim\omega_{pe}$ can be produced is via plasma wave emission. \index{radiation!harmonic}\index{radiation!x-rays}
\begin{figure}[t!]
\hspace{0.0cm}\centerline{\includegraphics[width=0.98\textwidth,clip=]{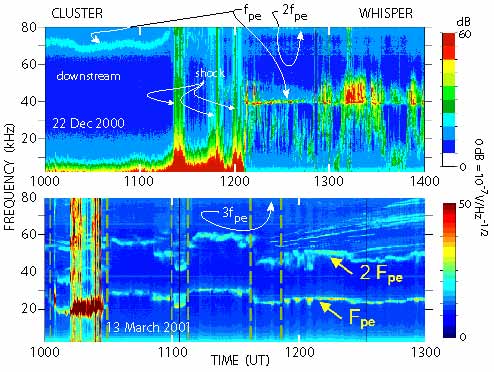}}
\caption[cluster shock emissions]
{\footnotesize CLUSTER observations of electromagnetic radiation from the electron foreshock, following the multiple shock crossings between 1120-1210\,UT that is marked by the intense low frequency noise and its broadband extension through the entire frequency range from 0-80\,kHz.  {\it Upper panel}: Near electron-foreshock boundary typical $f_{pe}$ and lower frequency emissions from 1210-1250\,UT. These are accompanied by a $2f_{pe}$ radiation band near 80 kHz and another weak emission band just above $f_{pe}$ the upper cut-off of which is decreasing in frequency. At later times the spacecraft is deeper in the shock as indicated by the broadband electrostatic emissions at $f_{pe}$. But these are correlated with broadening of the harmonic radio radiation towards lower frequencies. {\it Lower panel}: A sequence of three hours of low upstream density when the spacecraft remained close to the foreshock boundary. With the exception of a few short periods there is no broadening of the plasma frequency, while all the time a harmonic emission of intensity not much less than at $f_{pe}$ accompanies the plasma frequency. During the time of broadening of the plasma frequency one also observes a downward broadening of the harmonic radiation. At high frequencies near 80 kHz a faint third harmonic band $\sim 3f_{pe}$ can be identified. Of interest is the band splitting seen in the harmonic radiation between 1000-1010\,UT. Finally, the many upward drifting equally spaced in frequency narrow-band radiation bands after 1140\,UT are surprising. They do not have any counterpart in the plasma frequency and must thus be from a remote source that might be related to the small but sharp change in density (drop in $f_{pe}$) at 1140\,UT. The dashed vertical lines mark changes in the magnetic field (not shown here), when its direction turned abruptly. Some of these are correlated with small variations also in the magnetic field strength \citep[after][{\rm and Trotignon, private communication}]{Trotignon2001}.}\label{chap5-fig-rad}
\end{figure}

Radio emission from collisionless shocks in the heliosphere is a widely studied field including type II solar radio burst, interplanetary type II burst, CME-driven radio bursts and radio emissions from planetary bow shocks. While the solar bursts, because of their high frequency can be observed from Earth, most of the other emissions have been discovered only from aboard spacecraft. Of the enormous wealth of such observations made by the {\footnotesize ISEE}, {\footnotesize AMPTE}, {\footnotesize P}olar, {\footnotesize WIND}, {\footnotesize GEOTAIL} and other spacecraft, many of them never published, we pick here just a more recent observation from {\CL} \citep{Trotignon2001}.  Figure\,\ref{chap5-fig-rad} shows two cases of such observations on December 22, 2000 and three months later on March 13, 2001, when {\CL} was crossing the bow shock and moved into the electron foreshock. On December 22, 2000 the upstream density was relatively high with plasma frequency near 40 kHz. The intense spots in $f_{pe}$ between 1200-1300 UT indicate touching of the electron foreshock boundary field line as has been discussed above. Broadband wave modes of around $\sim\frac{1}{2}f_{pe}$ frequency indicate beam modes though no frequency drift is detectable. However, the density during this time is low enough for a faint emission to occur at about $\sim80$\,kHz, just two times $f_{pe}$. At this frequency emission can be only in the electromagnetic free space mode. We are thus witnessing local generation of radio emission from the foreshock boundary. At later times when the spacecraft moved deeper into the foreshock -- as visible from the widening of the plasma wave spectrum to both sides, up and down from $f_{pe}$ -- the radio emission becomes more diffuse and more broadband shifting to lower frequencies than the second harmonic of the plasma frequency. This suggests that possibly here the lower frequency modes close to $f_{pe}$ do actively participate in the generation of emission. On the other hand, radiation from the shock ramp might also contribute to these emissions.\index{shocks!harmonic radiation}

The lower panel in Figure\,\ref{chap5-fig-rad}, recorded on March 13, 2001 during a day of much lower upstream density, shows the more typical case of Langmuir waves at $f_{pe}$ and radio emission at almost precisely $2f_{pe}$. Note that near 80\,kHz a faint indication of the presence of an emission at the third harmonic $\sim3f_{pe}$ can be made out. The dashed vertical lines in this panel indicate the times when the magnetic field direction changed abruptly with the changes in magnetic field magnitude and  density remaining comparably small. The emissions in the plasma frequency and harmonic are well correlated during the entire event. Obviously {\CL} was constantly close to the foreshock boundary as only intensification in $f_{pe}$ is seen but no `hair' nor `beards' neither evolve except during a period shortly after 1200 UT in the three yellow spots in $f_{pe}$ when the harmonic emission extends to lower frequencies together with the development of a little beard. 

Two other interesting features can be read from this panel. At early time in the panel, just before the short active phase of the topside sounder at 1020\,UT, splitting of the harmonic emission into two narrow bands is seen, which reminds at the band-splitting in type II bursts. The other interesting feature is the large number of drifting emissions with increasing frequencies, which suddenly evolve right after 1200 UT (following the drop in plasma density at 1140 UT). Both features, the splitting of $2f_{pe}$ and these drifting bursts are not understood yet. The latter might be related to the abrupt changes in density and plasma frequency seen in this panel. These density changes are accompanied also by changes in the magnetic field, which are not shown here. It is worth noting that these banded drifting emissions cannot come from remote simply because the low frequencies arrive first. Radiation from any remote source should be visible first at high frequency.  The drifting emissions must be related to a nearby source, most probably the shock or foreshock. Understanding its generation mechanism should provide valuable information about the radiation source.

\paragraph{Interpretation.}Theory of shock-emitted radiation is based on plasma processes which under the prevailing collisionless conditions in the shock and foreshock plasmas refer to wave-wave coupling as the main generation mechanism. Direct emission from particles is unimportant, because the energy losses a particle experiences when becoming retarded or reflected at a shock, are not transformed into radiation. In all non-relativistic cases radiation losses can completely be neglected compared with all other energy losses. Nevertheless, the observed radiation is of interest because in many cases, where no measurements are possible to be performed {\it in situ}, radiation is the only direct and presumably identifiable signature a shock leaves, when seen from remote. The other signature is the generation of energetic particles, which will be treated in the next chapter, but energetic particles are a more diffuse indicator of a shock, because their propagation is vulnerable to scattering from other particles, obstacles and, in the first place, scattering by magnetic fields. They, thus, do not provide an image of the shock as clear as radiation does. Just because of this reason, investigation and understanding of the mechanisms of emission of radiation from shocks enjoys -- and deserves -- the large amount of attention it receives. 

Expecting that -- presumably -- direct particle involvement into radiation is improbable (maybe with two exemptions,  which we will note later) we are left with a small number of possible mechanisms, which all belong to the class of wave-wave interaction in weak plasma turbulence. The most probable of these are three-wave processes. These can be understood as `collisions' between three `quasi-particles'. Since only these three are involved, the interaction conserves both, energy and momentum, and can symbolically be written as
\begin{equation}
{\textsf L}+{\textsf L'}\to {\textsf T}, \qquad {\textsf L}\equiv \left\{\omega_{\,\textsf L}({\bf k}_{\,\textsf L}), \,{\bf k}_{\,\textsf L}\right\}, \, {\textsf L'}\equiv \left\{\omega'_{\,\textsf L}({\bf k}'_{\,\textsf L}), \,{\bf k}'_{\,\textsf L}\right\}, \, {\textsf T}\equiv \left\{\omega({\bf k}), \,{\bf k}\right\}
\end{equation}
Here ${\textsf L}$ stands for longitudinal, and ${\textsf T}$ for transverse -- as the emitted radiation of frequency $\omega$ and wave number ${\bf k}$ propagates in the free space mode and thus is a transverse electromagnetic wave while its two mother waves are assumed to be longitudinal (i.e. electrostatic) waves of sufficiently high frequency. Moreover, these frequencies $\omega_{\,\textsf L}({\bf k}_{\,\textsf L})$ etc. depend on wave number through the real parts of the electrostatic dispersion relation ${\textsf D}_{\textsf L}(\omega_{\,\textsf L},{\bf k}_{\,\textsf L})=0$ and may become quite complicated expressions.  

In the presence of a plasma there are two free space modes, the ordinary and the extraordinary mode. Naturally, in order to leave the plasma and escape in the form of radiation their frequencies must exceed some lower cut-off frequency. for the ordinary mode 
\begin{equation}
\omega^2=\omega_{pe}^2+k^2c^2
\end{equation}
this is the plasma frequency reached at very long wave lengths $k=0$. However, since the speed of light $c\gg v_e$ is so large, the radiated wave length is much longer than any of the wavelengths of the longitudinal waves involved. This is immediately recognised when comparing the above ordinary wave dispersion relation with the Langmuir wave relation, with ${\textsf L}\equiv\ell$,
\begin{equation}
\omega^2_\ell=\omega_{pe}^2+3k^2_\ell v_e^2
\end{equation}
which is of exactly the same structure. Hence, as long as $v_e\ll c/\sqrt{3}$ we will have $k_\ell \gg k$, and the wave number of the radiated mode is practically zero. Momentum conservation of the three interacting quasi-particles becomes simply ${\bf k}_{\,\textsf L}\approx {\bf k}'_{\,\textsf L}$, implying that the interaction selects counter-streaming electrostatic waves. As for an example, any process that is capable of generating Langmuir waves of comparable wavelengths, propagating in both directions along and opposite to the magnetic field, can in principle contribute to generating escaping radiation. From energy conservation $\hbar\omega_\ell+\hbar\omega'_\ell=\hbar\omega$ of the three `quasi-particles' involved one immediately finds that $\omega\approx 2\omega_{pe}$. This is the origin of and the simplest mechanism for the generation of the $2f_{pe}$-second plasma harmonic radiation and has been proposed more than half a century ago by \cite{Ginzburg1958} as an explanation for the observation of solar type II and type III radio bursts. 

In this simple reasoning we have completely neglected not only the contribution of the Langmuir wave number (which turns out not to be important in magnitude, it just shifts the emitted frequency a tiny amount up in frequency) but also the fact that the electric field of Langmuir waves is polarised along ${\bf k}_\ell=\pm k_\ell {\bf B}/B$ and thus along the ambient  magnetic field ${\bf B}$, while the electric field of the transverse emitted electromagnetic radiation must necessarily be polarised  perpendicular to ${\bf k}$ (because of the absence of space charges at frequencies sufficiently higher than $f_{pe}$). Since the electric field will, after collision and annihilation of the two Langmuir waves involved, remain to oscillate along the magnetic field, the emission is preferably directed perpendicular to the magnetic field. It turns out, then, that it is easier to radiate in the extraordinary than in the ordinary free space mode. The extraordinary mode has a slightly more complicated dispersion relation; in a dense plasma with $\omega_{pe}\gg\omega_{ce}$ (as is encountered in near Earth space where supercritical collisionless shocks evolve) it has a slightly higher cut-off frequency. But the argument about the smallness of $k\ll k_\ell$ holds also in this case. Radiation at the second harmonic $\omega\approx 2\omega_{pe}$ should therefore be polarised perpendicular to the magnetic field in the extraordinary mode. 

Unfortunately, foreshock emission has not been found to show any preference in polarisation \citep{Reiner1996}. Moreover, emission is not only in the second harmonic but has also been detected close to $\gtrsim\omega_{pe}$ and at the third harmonic $\sim3\omega_{pe}$. These emissions require different waves to be involved for which a number of mechanisms have been proposed \citep[cf., e.g.,][]{Cairns1988}. So far none of them could be ultimately verified or even agreed upon, each having its merits and pitfalls. Since radiation is energetically negligible, as we have mentioned above, the whole problem could be put aside. However, since the assumptions made in every radiation mechanism contain important information about the source region, the problem of radiation production in collisionless shocks remains to be tantalisingly urgent and awaits resolution.  

Since there is no agreement yet about the radiation mechanism, we merely note here some of the different proposals. The first is the above mentioned merging of two counter streaming Langmuir waves. The problems about this simple though suggestive mechanism are numerous. First, Langmuir waves are assumed to be generated by the gentle-beam instability. Ignoring the problem of beam survival during its propagation along the shock-tangential field line, which we have discussed already in detail, gentle beams excite only forward Langmuir waves, which requires some mechanism that backscatters a substantial percentage of waves and inverts the direction of their wave numbers. There are three known elegant processes that are capable of doing this: modulation instability respectively collapse, scattering of Langmuir waves off thermal ions, and scattering off ion-sound waves, all three proposed long ago \citep[for an early review of the latter two mechanisms cf., e.g.,][]{Tsytovich1970}.\index{interactions!wave scattering} 

Modulation instability generates ion-sound waves via the ponderomotive pressure force 
${\bf F}_{\rm pmf}=-(e^2/m_e\omega_{pe}^2)\nabla|{\bf e}_\ell|^2$\index{force!ponderomotive}\index{process!plasma wave collapse}\index{waves!Langmuir collapse}
of the high-frequency Langmuir wave, ${\bf e}_\ell({\bf r},t)$. These waves, when becoming locally large amplitude, structure the plasma into a chain of cavities in which the Langmuir waves become trapped. This process generates long wave lengths. It is described by the Zakharov equations for the combined evolution of the Langmuir wave field and the density variation $\delta N$, respectively,
\begin{equation}
\frac{\partial^2 \delta N}{\partial t^2}-c_{ia}^2\nabla^2 \delta N=\frac{\epsilon_0}{m_iN}\nabla^2|{\bf e}|^2, \qquad
\frac{\partial{\bf e}}{\partial t}+\frac{3\omega_{pe}}{2}\lambda_{\bf e}^2\nabla^2{\bf e}=\frac{\omega_{pe}}{2N}{\bf e}\delta N
\end{equation}\index{equation!nonlinear Schr\"odinger}\index{equation!Zakharov}
The first of these equations is a driven wave equation for the density variation which for slow time variations, when the derivative with respect to time is neglected, just gives pressure balance between the ponderomotive pressure on the right and plasma pressure on the left, i.e. proportionality $\delta N\sim -|{\bf e}|^2$. In other words, the density variation anti-correlates with the field pressure, which corresponds to caviton formation. The second equation is a nonlinar Schr\"odinger equation for the evolution of the wave amplitude.\index{waves!cavitons}\index{Zakharov, V. E.}

The Langmuir waves trapped in the cavitons must bounce back and forth, which naturally creates counter-streaming waves of equal intensity with opposite wave numbers. During collapse the cavities shrink in size, the wave numbers and momenta of the waves increase, and the wave energy density increases as well because of the shrinking volume. This yields both, the counter streaming Langmuir waves being localised in the same region and, in addition, a large radio emissivity. Unfortunately, we have already noted it, this process -- as beautiful as it might be -- has not been confirmed experimentally, neither in the observations nor in the full particle simulations. Observed wave intensities are too low in the electron-foreshock boundary and electron foreshock, and the density variations did not indicate the presence of the expected cavities. Simulations, on the other hand support quasilinear evolution and wave scattering off thermal ions. We note, however, that the most recent detection of the very strong electric fields in the shock ramp \citep{Bale2007} might indicate that it is not the electron foreshock where one should expect caviton and collapse to work and cause the most intense radiation, rather it might be the very shock transition where shock radiation is generated by such processes. It is, in this respect, most interesting to remind of the strange radiative behaviour reproduced in Figure\,\ref{chap5-fig-rad} that was detected by {\CL}. We also note that similar observations had been made much earlier with the wave experiment on {\footnotesize AMPTE IRM} in the Earth foreshock (R. A. Treumann \& J. LaBelle, unpublished 1986). The observed band splitting and high intensities might have been caused by Langmuir caviton collapse \citep{Treumann1992}.

Other possibilities to produce counter streaming Langmuir waves are scattering of Langmuir waves off thermal ions \citep[][investigated this process in full detail numerically including the ion polarisation cloud]{Muschietti1991}, a mechanism known since the early sixties. The process reads symbolically ${\textsf L}+{\textsf i}\to {\textsf L}'+{\textsf i}^*$, where the primed quantities are after the collision, and the star on the ion indicates excitation of the ion as it is too heavy for changing momentum during the collision with the Langmuir wave. It is merely excited while the scattered Langmuir wave has changed direction and lost some of its momentum, i.e. attains a longer wave number and lower frequency. The same process does also work with ion-sound waves as ${\textsf L}+{\textsf {IS}}\to {\textsf L}'$. The scattered Langmuir waves then also change direction by absorbing the ion sound. Both processes have been used for radiation generation \citep{Yoon1994}.\index{instability!cyclotron maser}\index{instability!modulation}\index{radiation!electron cyclotron maser}

Radiation at higher frequency, e.g. radiation at the third plasma harmonic can be generated by a four-wave process. This is also favoured by caviton formation and collapse since the waves in this case are all confined to one and the same volume. However, other mechanisms have also been proposed. All these processes are of the kind of wave-wave interactions and thus their efficiencies are proportional to the product of the involved relative wave intensities. Since the latter are usually low, the efficiencies are very small as a  rule. An attempt to increase the growth of Langmuir waves has in the recent past been the idea to consider a statistical theory of growth called `stochastic growth' \citep{Robinson1995,Krasnoselskikh2007}. This attempt takes advantage of the statistically distributed density fluctuations in the foreshock region like in a random medium. Since the Langmuir-wave growth rate is proportional to $\delta N/N$, an average growth rate over the volume of occupation by the Langmuir wave can be calculated. This might be more realistic than using the linear growth rate. Regions of decreased density contribute strongly to the average Langmuir-wave amplitude. In this way an in the average larger emission efficiency is obtained. Moreover, the averaging procedure introduces a statistical element which supports the incoherence of the relation between the detected Langmuir waves and radiation, which is in partial agreement with the observation.   

Generation of radiation at the fundamental $\omega\gtrsim \omega_{pe}$ in a three wave process requires the presence of a low frequency wave. Ion-acoustic waves are one possibility, other possibilities are lower-hybrid waves, Buneman waves, the modified-two stream instability, various kinds of drift waves, and also electron acoustic waves or electron beam waves. In particular the latter are present in the foreshock region and thus can combine with Langmuir waves to generate fundamental radiation slightly above the plasma frequency.  

The two exemptions when particles become involved are the above mentioned scattering of Langmuir waves off ions, and the so-called electron-cyclotron maser instability \citep[for a contemporary review see, e.g.,][]{Treumann2006}. Its advantage is that it operates directly on the free space mode avoiding any intermediate step like three-wave processes or particle scattering. However, it requires a particular form of the electron distribution with a velocity space gradient in the perpendicular direction $\partial F_e(v_\|,v_\perp)/\partial v_\perp>0$, a hot electron distribution and low cold electron density. It is barely known whether such distributions are realised at the shock. However, if they are in some place, then the cyclotron maser instability will outrun all other mechanisms and directly feed the free-space electromagnetic radiation modes. Radiation will then be at a harmonic of the electron cyclotron frequency which is a severe restriction if the magnetic field is low and the density high. Therefore, regions of low density and stronger converging magnetic fields are the best candidates for this radiation source.

\section{Quasi-parallel Shock Reformation}\noindent In quasi-parallel supercritical shocks there is not such a stringent distinction between the region upstream of the shock and the shock itself like in quasi-perpendicular shocks. The foreshock, which we have discussed in some detail in the previous section, and the shock itself cannot be considered separately. This is due to the presence of the reflected and diffuse particle components in the foreshock. These, as we have seen are the source of  a large number of waves. 

The interaction of these waves with the shock is one of the main issues in quasi-parallel shock physics. In the present section this will become clear when we will be dealing with the formation, behaviour and structure of quasi-parallel shocks as it has been inferred less from observation than from numerical simulations.\index{shocks!quasi-parallel reformation} The reason is that the real observations in space do not allow to separate the particles and waves from the shock. They all occur simultaneously and are interrelated and can never be observed in their initial state. The observations to which we will nevertheless occasionally refer will leave the impression of large-amplitude noisy fluctuations. In simulations, on the other hand, it is at least to some degree possible to prepare the system in such a way that a single effect can be studied. For instance, in one-dimensional simulations the direction of wave propagation can be prescribed which allows studying just waves in one direction and their effect on the shock and particles. Moreover, treating the electrons as a neutralising Boltzmannian fluid suppresses their effect on the ion motion and wave genereation. Treating them as an active fluid allows taking account of electron-ion fluid instabilities. Finally, full particle PIC simulations can be performed with low or realistic mass ratios in order to investigate different time and spatial scale dependence and the excitation or coupling to higher frequency waves. Most simulations that have been performed in the past have taken advantage of these possibilities.
\subsection{Low-Mach number quasi-parallel shocks}\noindent
It is usually assumed that low-Mach number shocks are stable, i.e. show no substantial time variation or reformation. It is not completely transparent why this should generally be so. Firstly, the critical Mach number has been shown by \cite{Kennel1985} to become small at narrow shock normal angles $\thetabn\to 0$, in which case even low-Mach number quasi-parallel shocks should become supercritical and reflect ions. Secondly, any fast ions of parallel velocity $v_\| > V_1$ that have been heated in the shock can in principle escape from the quasi-parallel shock upstream along the magnetic field and should appear in the foreshock where they contribute to wave generation. Therefore, it makes sense to investigate the state of quasi-parallel shocks in view of their stability and wave generation even for low Mach numbers. In addition, any waves that are generated in the shock ramp or transition with upstream directed ${\bf k}$ and fast enough parallel phase or group velocities could also escape from the shock in upstream direction. This could, in particular, be possible just for low Mach number quasi-parallel shocks.

To check this possibility \cite{Omidi1990} have performed one-dimensional hybrid simulations finding that initially the quasi-parallel shock consisted of phase-standing dispersive (magnetosonic) whistler waves with the last whistler wave cycle constituting the shock ramp. As expected, the wave vectors of these phase-locked magnetosonic  whistlers are aligned with the shock normal. At later times, backstreaming ions along the upstream magnetic field excite a long-wavelength whistler wave packet upstream of the shock. In the one-dimensional simulation the wave vector is restricted to the shock normal while it is  known from theory that the growth rate is largest along the magnetic field. These oblique whistler waves should thus show up in two-dimensional simulation and may be visible at larger distance for sufficiently large upstream phase velocites. 
\begin{figure}[t!]
\hspace{0.0cm}\centerline{\includegraphics[width=0.8\textwidth,height=0.4\textheight,clip=]{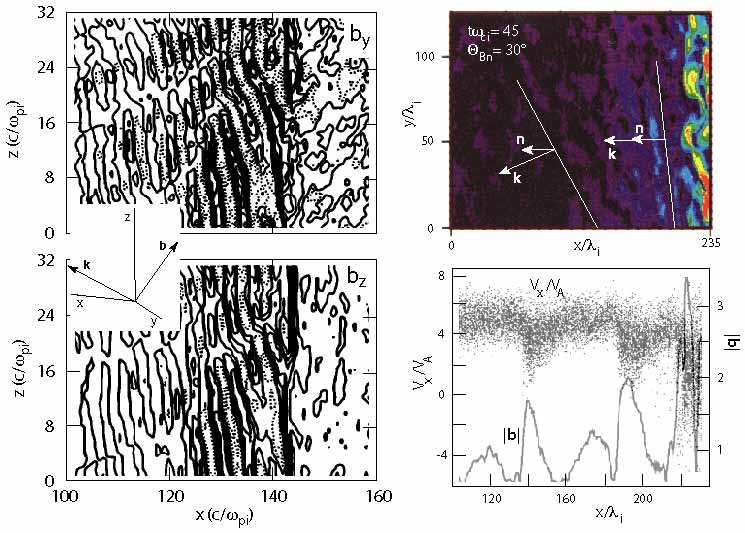} }
\caption[Quasiparallel 2D-upstream waves]
{\footnotesize Two-dimensional hybrid simulations of the evolution of upstream waves. {\it Left column}: Upstream wave in a low-Mach number  ${\cal M}_A=2.2$ quasi-parallel shock \citep[after][]{Scholer1993a}. The nominal shock is in the $(y,z)$-plane. Ions escaping to upstream generate the oblique upstream magnetosonic waves. The contour plot of the two normalised to the upstream magnetic field components of the magnetic fluctuations shown is taken at relatively early times $\omega_{ci}t=68$. It shows the nearly plane magnetic wave fronts inclined against the shock in direction $x$ and having wavelengths of $\sim10\c/\omega_{pi}$ in $z$ while being much shorter in $x$. In the vicinity of the shock the wave fronts turn more parallel to the shock and produce a non-coplanar magnetic component $|b_y|$ which is of same order as the $|b_z|$. Moreover, even though the shock has relatively low Mach number, it is not completely stable but shows structure in $z$ direction which is produced by the presence of the reflected upstream particles and the upstream waves. At places it is impossible to identify one single shock ramp. {\it Right column}: Two-dimensional hybrid simulations of the evolution of giant magnetic pulsations ({\SL}) in front of a quasi-parallel supercritical shock \citep[after][]{Dubouloz1995}. {\it Top}: The simulation plane showing the structure of the (normalised) magnetic fluctuation field $|{\bf b}|$ at time $t\omega_{ci}=45$ and shock normal angle $\thetabn=30^\circ$. The accumulation of the growing wave fronts at the shock transition, their increasing amplitudes, and their turning towards becoming parallel to the shock is clearly visible from the rotation of the two wave fronts and their ${\bf k}$ vectors shown in white. Away from the shock transition the angle between ${\bf k}$ and the shock normal ${\bf n}$ is large. Close to the shock the two vectors are about parallel. The magnetic field is in the wave front, so  $\thetabn$ is close to 90$^\circ$ here. {\it Bottom}: Pulsation amplitude and ion phase space. The fluctuations evolve into large amplitude pulsations when approaching (and making up) the shock. The strong retardation of the upstream flow by the pulsations is visible in the shock-normal velocity component (Mach number ${\cal M}_A$). In the hybrid simulations this slowing done is accompanied by some ion heating.}\label{chap5-fig-schofuji93}
\end{figure}

Such two-dimensional (hybrid) simulations with a non-inertial electron fluid have been performed by \cite{Scholer1993a} for a Mach number of ${\cal M}_A=2.2$ and angles $\thetabn= 20^\circ,30^\circ,45^\circ$ and by \cite{Dubouloz1995} for an angle $\thetabn=30^\circ$ and high Mach number ${\cal M}_A=5$ (Figure\,\ref{chap5-fig-schofuji93}). The lower Mach number simulations\index{simulations!hybrid} show the presence of a substantial number of backstreaming ions which cause an ion-ion instability in the upstream region. However, the excitation and properties of the waves depend strongly on the shock normal angle. Initially, as in the case of \cite{Omidi1990} phase-locked short-wavelength whistlers appear which are replaced at later times by upstream long-wavelength whistlers with phase velocity directed and amplitude growing towards the shock ramp but upstream directed group velocity, i.e. the shock radiates energy away towards upstream, as one would naively expect, because the shock being supercritical must reject the excess inflow of energy which it can do by both, reflecting particles and emitting waves into upstream direction. These waves are excited by the backstreaming ion component in a strongly nonlinear interaction process because of the evolving steep ion-density gradient, which is of the same scale as the whistler wavelength. The ${\bf k}$-vector turns away from the magnetic field having comparable components parallel to ${\bf B}$ and parallel to the shock normal ${\bf n}$. For small $\thetabn$ a remnant of the initial phase-locked whistlers survives but disappears at $\thetabn=30^\circ$. Close to the shock, where the backstreaming ion density is high, the waves have short wavelengths, and ${\bf k}$ is almost parallel to ${\bf n}$. In the high Mach number simulations no shock is produced but instead reflected ions were artificially injected with same Mach number as the incoming flow but with much higher temperature $v_i=14.1\, V_A$, forming a spatially uniform ion beam. The intention was  to investigate the effect of the hot reflected ions. This is shown on the right in the above figure. The result resembles the former one where a shock was generated by reflection at a wall, but the effect in the injected beam case is stronger because of the higher Mach number. Hence it is the hot reflected ion component that is responsible for the wave dynamics and the shock dynamics.

All this can be seen from the two-dimensional intensity contours of these waves in the foreshock, which are plotted at a relatively early time in the shock evolution in the simulations in Figure\,\ref{chap5-fig-schofuji93}. On the right in this figure the geometry is given, with the magnetic field fluctuation vector ${\bf b}$ in the $(y,z)$-plane. The bottom panel on the left shows contours of the $b_z$ fluctuations in the $(x,z)$-plane. The nominal shock ramp is at $x\approx 145\,\lambda_i$ ion inertial lengths $\lambda_i=c/\omega_{pi}$ at this time. The upper panel shows the non-coplanar component $b_y$-contours in the same representation. Behind the shock the fluctuations are irregular and disorganised. However, in front of the shock a clear wave structure is visible with strongly inclined wave fronts and of roughly $\sim10\,\lambda_i$ wavelengths in $z$ parallel to the nominal shock surface. The wavelength in $x$ is about three times as short. \index{foreshock!ion waves}

These waves are seen in both components, $b_z, b_y$, are low amplitude at large distance from the shock but reach very large amplitudes simultaneously in both components during shock approach while, at the same time, bending and assuming structure in $z$-direction that is different from the regular elongated shape at large distance. This deformation of wave front may be due to the residual whistlers near to the shock, but it implies that the shock has structure on the surface in both directions $x$ and $z$ and is not anymore as planar as was initially assumed. The shock becomes locally curved on the scale of the shock-tangential wavelength. The waves deform the shock and, in addition, being themselves of same amplitude as the shock ramp, become increasingly indistinguishable from the shock itself. {\it The shock is}, so to say, {\it the last of the large-amplitude magnetic wave pulsations in downstream direction, and the shock-magnetic field is not anymore coplanar}, because the waves have contributed a substantial component $b_y$ that points out of the coplanarity plane. 
\begin{figure}[t!]
\hspace{0.0cm}\centerline{\includegraphics[width=0.75\textwidth,clip=]{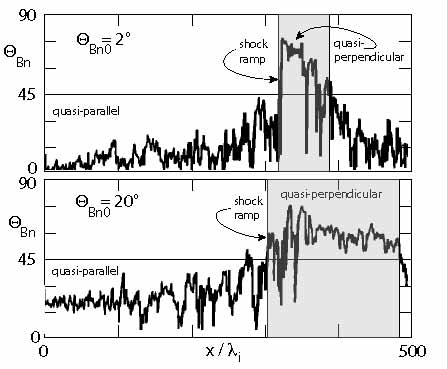} }
\caption[Magnetic Field Evolution in Quasiparallel Shocks: Simulations]
{\footnotesize The evolution of the shock normal angle $\thetabn$ on distance from the shock in two-dimensional hybrid simulations for two initial quasi-parallel shock-normal angles ${\thetabn}_0 = 2^\circ$ and ${\thetabn}_0=20^\circ$, respectively \citep[simulation results taken from][]{Scholer1993a}. The horizontal line at 45$^\circ$ is the division between quasi-perpendicular and quasi-parallel shock normal angles. In both cases $thetabn$ evolves from quasi-parallel direction into quasi-perpendicular direction. The shaded areas identify the quasi-perpendicular domains.}\label{chap5-fig-thetabn}
\end{figure}

We have emphasised this phrase, because it expresses the importance of the low-frequency upstream magnetic waves in quasi-parallel shock physics. Contrary to quasi-perpendicular shocks where the reflected gyrating ions in combination with the reflected-ion excited modified-two stream instability were responsible for the shock dynamics and different kinds and phases of shock reformation, quasi-parallel shock reformation and much of its physics is predominantly due to the presence of large-amplitude and spatially distinct upstream waves. These are the generators of the shock and, due to their presence, the shock changes its character. It is highly variable in time and position along the shock surface and is -- close to the shock transition on a smaller scale -- `less quasi-parallel' (or more perpendicular, i.e. the shock-normal angle $\thetabn$ has increased on the scale of the upstream waves). The latter is due to the out-of coplanarity-plane component of the upstream waves. In spite of concluding this from a hybrid simulation, this conclusion remains basically valid also in full particle simulations. It had been suggested already earlier on the basis of {\footnotesize ISEE 1\& 2} observation of the evolution of the upstream ultra-low frequency wave component \citep{Greenstadt1993}.  \index{shocks!coplanarity}

The gradual evolution of the shock normal angle $\thetabn$ has been demonstrated in other hybrid simulations by \cite{Scholer1993a} and \cite{Dubouloz1995} who investigated the evolution of the shock normal angle in dependence on distance from the shock. This is shown in Figure\,\ref{chap5-fig-thetabn} for two-dimensional hybrid simulations with initial shock-normal angles ${\thetabn}_0 = 2^\circ$ and ${\thetabn}_0=20^\circ$, respectively, which we anticipate here. In both cases $\thetabn$ evolves from quasi-parallel to quasi-perpendicular angles. Qualitatively there is little difference between the two cases. At the shock ramp $\thetabn$ is deep in the domain of quasi-perpendicular shocks. The only difference is that for the nearly parallel case the angle jumps to quasi-perpendicular quite suddenly, just before approaching the shock ramp, while the evolution is more gradual for the larger initial $\thetabn$. In both cases the evolution is not smooth, however, which is due to the presence of large-amplitude foreshock waves. Transition to quasi-perpendicular occurs for the initially nearly parallel case at the nominal shock ramp while for the initially quasi-parallel case it occurs at an upstream distance of about $\lesssim100\lambda_i$ from the shock. 

One notices that this transition is on the ion scale, implying that in the region close to the shock the ions experience the shock occasionally (because of the large fluctuations in $\thetabn$) -- and when ultimately arriving at the shock -- as quasi-perpendicular. It is thus not clear, whether the electrons do also see a quasi-perpendicular shock, here. However, the {\footnotesize ISEE} measurements of the electron distribution function by \cite{Feldman1983} at the shock do not show a difference between quasi-perpendicular and quasi-parallel shocks. This fact suggests, in addition that, close to the shock transition, quasi-parallel shocks behave like  quasi-perpendicular shocks as well on the electron scale, which is just what we have claimed. 

We finally note that the behaviour of the shock normal angle gives a rather clear identification of the location of the shock transition in the quasi-parallel case, as indicated in Figure\,\ref{chap5-fig-thetabn} by shading. Three distinctions can be noticed: 
\begin{enumerate}
\item[$\bullet$] first, at larger initial shock-normal angles the transition to quasi-perpendicular angles occurs earlier, i.e. farther upstream than for nearly parallel shocks. This is due to the stronger effect of the large amplitude upstream waves in this case; 
\item[$\bullet$ ] second, at larger initial shock normal angles the quasi-perpendicular shock transition is considerably broader than for nearly parallel shocks, i.e. it extends farther downstream before the main quasi-parallel direction of the magnetic field in the downstream region takes over again and dominates the direction of the magnetic field:
\item[$\bullet$ ] third,  at an initial shock-normal angle of $20^\circ$, this region is roughly $\sim150\lambda_i$ wide, implying that the magnetic field direction behind a quasi-parallel shock remains to be quasi-perpendicular over quite a long downstream distance measured from the shock ramp. For the nearly parallel shock this volume is only about $\sim50\lambda_i$ wide. 
\end{enumerate}
This observation must have interesting implications for the physics downstream of quasi-parallel shocks. For instance, applied to the Earth's bow shock, where $\lambda_i\sim 10^3$\,km. both distances correspond to regions wider than the order of $>5\,{\rm R_E}$ which is larger than the nominal width of the magnetosheath! Thus, behind the bow shock a substantial part of the magnetosheath plasma should behave as if the bow shock would have been a completely quasi-perpendicular shock.

\begin{figure}[t!]
\hspace{0.0cm}\centerline{\includegraphics[width=0.95\textwidth,clip=]{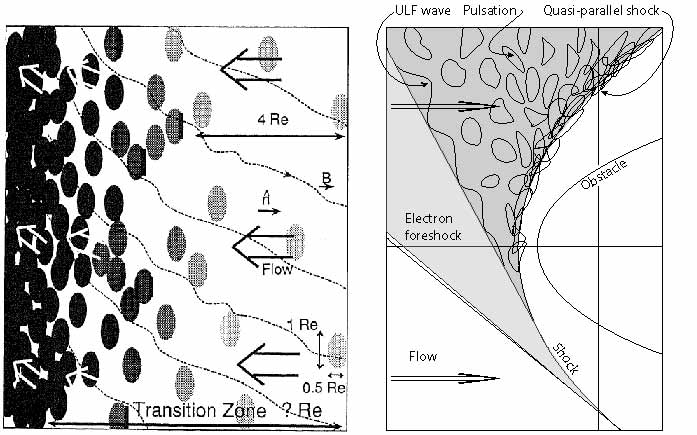} }
\caption[Magnetic Field Evolution in Quasiparallel Shocks: Simulations]
{\footnotesize The patchwork model of \cite{Schwartz1991} of a quasi-parallel supercritical shock mentioned earlier.  {\it Left}: Magnetic pulsations ({\SL}) grow in the ion foreshock and are convected toward the shock where they accumulate, thereby causing formation of an irregular shock structure. Note also the slight turning of the magnetic field into a direction to the shock normal that is more perpendicular, i.e. the magnetic field is more parallel to the shock surface with the shock surface itself becoming very irregular \citep[after][]{Schwartz1991}. {\it Right}: The same model with the pulsations being generated in the relatively broad ULF-wave-unstable region in greater proximity to the ion-foreshock boundary. When the ULF waves evolve to large amplitude and form localised structures these are convected toward the shock, grow, steepen, overlap, accumulate and lead to the build up of the irregular quasi-parallel shock structure which overlaps into the downstream direction.}\label{chap5-fig-patch}
\end{figure}

\subsection{Turbulent reformation}\noindent
When speaking about turbulent reformation we have in mind that a quasi-parallel supercritical shock is basically a transition from one lower entropy plasma state to another higher entropy plasma state that is madiated ba a substantially broad wave spectrum. Such a transition has been proposed by \cite{Schwartz1991} based on the detection of the large amplitude magnetic pulsations ({\SL}) in the foreshock. 

Figure\,\ref{chap5-fig-patch} on the left shows their model assuming that somewhere upstream in the foreshock magnetic pulsations have been excited which become convected downstream toward the shock by the convective flow, grow in amplitude and number and accumulate at the shock transition to give rise to a spatially and temporarily highly variable transition from upstream of the shock to downstream of the shock. An important clue in this argument was the observation that, first, the pulsations grow in amplitude when approaching the shock and that, second, they slow down. this slowing down is effectively an increase in their upstream directed velocity on the plasma frame with growing amplitude such that their speed nearly compensates for the downstream convection of the flow. 

On the right of the figure, which is suggested by the observations of \cite{Kis2007}, a larger volume is seen. Here the pulsations are the result of growing ultra-low-frequency waves which are generated in a volume inside the foreshock but relatively close to the ion-foreshock boundary. These waves grow to large amplitudes until evolving into pulsation which the flow carries toward the shock. Growth, slowing down, and accumulation then lead to the pile up of the pulsations at the shock location and formation of the turbulent shock structure. Of the magnetic field, in this figure on the right, we plotted only the shock-tangential upstream field line. In the left part, several field lines are schematically shown exhibiting the fluctuations imposed by the background level of ultra-low-frequency fluctuations. Moreover, a certain bending of the field lines is included here in approaching the shock transition with the field lines turning more perpendicular the closer they come to the shock. This bending is what we claim to be a parallel shock turning quasi-perpendicular at a scale very close to the shock. In this schematic drawing, however, there would be no reason for the magnetic field to turn this way. What closer observations and simulations show is, however, that the turning of the field is the result of the presence of the large amplitude magnetic pulsations. This will become clearer below.
 \begin{figure}[t!]
\hspace{0.0cm}\centerline{\includegraphics[width=0.95\textwidth,clip=]{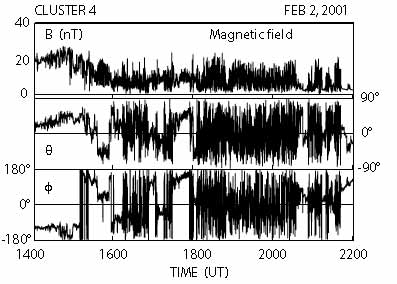} }
\caption[Magnetic Field Evolution in Quasiparallel Shocks: Simulations]
{\footnotesize Eight hours of CLUSTER magnetic field data during a long passage near and across the quasi-parallel supercritical (Alfv\'enic Mach number ${\cal M}_A\sim 12-13$, ion inertial length $\lambda_i\sim 140$\,km) bow shock.  The time resolution was 4\,s. The top panel shows the variation in the magnitude of the magnetic field. The two lower panels are the respective elevation and azimuthal angles $\theta,\phi$ in a GSE coordinate frame \citep[data taken from][]{Lucek2002}. Large variation in the magnetic compression and direction can be seen to be associated with this quasi-parallel shock crossing. Buried in these large variations on this highly time-compressed scale are many magnetic pulsations ({\SL}). The compressive large amplitude fluctuations in the upper panel are typical for a quasi-parallel shock transition. }\label{chap5-fig-qpasCL}
\end{figure}

\subsubsection*{Observations}\noindent
Of course, the model shown in Figure\,\ref{chap5-fig-patch} is a schematic model only which, however, has some merits in explaining the observations. The signature of a quasi-parallel shock in the magnetic field is indeed quite different from that of a quasi-perpendicular shock. We have already seen in the electric recordings reproduced in Figure\,\ref{chap5-fig-rad} that the quasi-parallel shock appears in the electric wave spectrum as a broadband emission of highest spectral intensities at the low frequency end. The magnetic signature of a quasi-parallel shock is quite similar in that it lacks a clear location of the shock front. Rather one detects a broad region of very large amplitude compressive oscillations in magnetic magnitude and in the direction of the magnetic field that subsequently is recognised as a passage across the quasi-parallel shock. 
\begin{figure}[t!]
\hspace{0.0cm}\centerline{\includegraphics[width=0.9\textwidth,clip=]{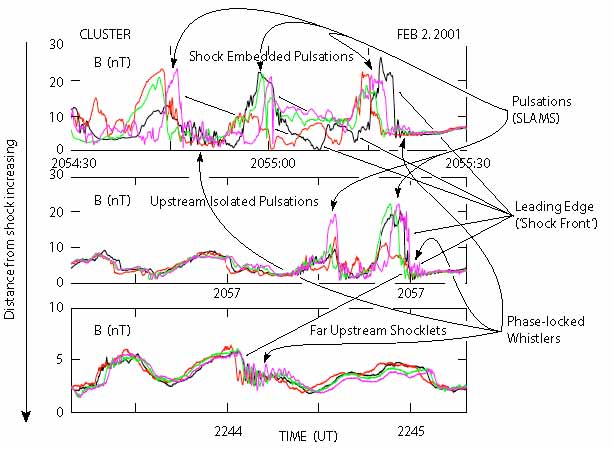} }
\caption[Magnetic Field Evolution in Quasiparallel Shocks: Simulations]
{\footnotesize  CLUSTER magnetic field measurements of magnetic pulsation ({\SL}) near and remote from a quasi-parallel shock on February 2, 2001 \citep[data from][]{Lucek2002}. Only the magnetic field magnitude is shown for all four CLUSTER (colour coded) spacecraft. The spacecraft separation was between a few 100 km and 1000 km. {\it Top}: Clustered pulsations in the shock transition. Three events of large amplitudes are shown. These structures are very irregular with steep fronts. Note that in spite of the small spacecraft separation the shapes of the structures differ strongly from spacecraft to spacecraft. Moreover, determination of the pulsation fronts and normals (not shown) indicates high variability over the spacecraft separation distance. Thus the structures are of relatively small scale and large amplitude. {\it Middle}: Isolated pulsation at greater distance from shock outside compression region. The structures are seen almost simultaneously at the spacecraft and thus must be of larger size. Copared to the embedded pulsations the amplitudes are lower, and the structures are more regular. {\it Bottom}: A shcklet observed outside the pulsation region in the domain of ultra-low-frequency waves. The steep shock-like front is well expressed with the qattached whistler waves it carries with it. Note the much lower amplitude than the pulsations. }\label{chap5-fig-puls}
\end{figure}

An example is shown in Figure\,\ref{chap5-fig-qpasCL} as measured by the {\CL} spacecraft. This figure shows eight hours of observation by {\CL} in the immediate vicinity of the quasi-parallel shock. It is difficult to say where in the figure the shock transition is located as the large fluctuations in the magnetic field magnitude and directions mask the various back and forth passages across the shock that are contained in the data.  Clearly, at the beginning near 1400\,UT the spacecraft was in the downstream region. The fluctuations show that during almost the entire sequence the magnetic field is exhibits compressive fluctuations. These belong to the shock transition. At the same time large fluctuation in the direction of the magnetic field are also observed. In the compressions of the magnetic field buried are also upstream pulsations ({\SL}), and many of the changes in direction belong to the ultra-low-frequency waves present at and near the shock. The changes in direction indicate that the shock does not behave like a stationary flat surface. Instead, it shows structure with highly fluctuating local shock normal directions. \index{waves!SLAMS}\index{waves!pulsations}

\cite{Lucek2002} have checked this expectation by determining the local shock-normal angle $\thetabn$ and comparing it with the prediction for $\thetabn$ estimated from magnetic field measurements by the {\footnotesize ACE} spacecraft \index{spacecraft!ACE} which was located farther out in the upstream flow. The interesting result is that during the checked time-interval of passage of the quasi-parallel shock the prediction for the shock normal was around $20-30^\circ$, as expected for quasi-parallel shocks. However, this value just set a lower bound on the actually measured shock normal angle. The measured $\thetabn$ was highly fluctuating around much larger values and, in addition, showed a tendency to be close to $90^\circ$. This is a very important observation. It strengthens the claim that quasi-parallel shocks are locally, on the small scale, very close to perpendicular shocks, a property that they borrow from the large magnetic waves by which they are surrounded. In fact, we may even claim that locally, on the small scale, quasi-parallel shocks are quasi-perpendicular since the majority of the local shock normal angles was $>45^\circ$. By small-scale a length scale comparable to a few times the ion inertial length or less is meant here.

The data suggest that, indeed, the quasi-parallel shock is the result of a build-up from upstream waves which continuously reorganise and reform the shock. Figure\,\ref{chap5-fig-puls} shows three representative examples of such upstream waves which are far from being continuous wave trains. The upper panel is taken from the large density fluctuation region in the shock transition. This region turns out to consist of many embedded magnetic pulsations ({\SL}) of very large amplitudes. In the present case amplitudes reach $|{\bf b}|\sim 25$\,nT. These pulsations have steep flanks and quite irregular shape, exhibit higher frequency oscillations probably propagating in the whistler mode while sitting on the feet or shoulders of the pulsations. It is most interesting that the different {\CL} spacecraft -- at spacecraft separation $<1000$\,km -- do not observe one coherent picture of a particular pulsation. This implies that the pulsations  in the shock are of shorter scale than spacecraft separation: the different spacecraft observe different structures respectively different pulsations. In addition, the magnetic field directions (not shown in the figure) are very different from spacecraft to spacecraft and thus from pulsation to pulsation,  and even for one pulsation at its front edge and trailing edge different magnetic field directions are observed. The directions of the magnetic normal across a pulsation change on  very short spatial scales. The quasi-parallel shock front has thus a rather irregular shape, which will be bent locally with changing direction of its normal.

The second panel shows an isolated pulsation farther away from the shock transition. This pulsation exhibits a much more coherent way on the four different {\CL} spacecraft. It seems as if the pulsation is still in evolution as  three of the spacecraft see a nearly coherent structure while the fourth which is farther away sees it in a different state. Concluding from this event, isolated pulsations seem to have larger dimensions and lower amplitudes, which would be consistent with the assumed solitary properties of pulsations.

The third panel shows a shocklet, i.e. a structure which presumably has little in common with pulsations. It is embedded into long wavelength ultra-low-frequency wave trains, evolves into steep front and drives whistlers attached to this front across the flow. These waves were already observed by \cite{Russell1971}. Their properties  indeed resemble those of sub-critical little shocks which propagate against the flow, though with slower speed such that they effectively are slowly convected towards the shock. \index{waves!shocklets}
 \begin{figure}[t!]
\hspace{0.0cm}\centerline{\includegraphics[width=0.95\textwidth,clip=]{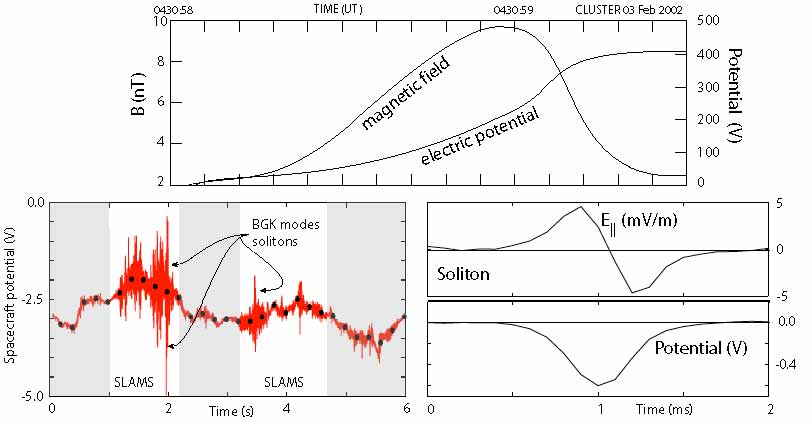} }
\caption[Magnetic Field Evolution in Quasiparallel Shocks: Simulations]
{\footnotesize Observation of electric field structures in large magnetic pulsations ({\SL}) in the quasi-parallel shock transition region \citep[data taken from][]{Behlke2003,Behlke2004}. Structures on three different time scales are shown, corresponding also to three different spatial scales. {\it Top}: CLUSTER passage across on (moderately large amplitude) magnetic pulsation in the shock transition. The (smoothed) magnetic field structure is a slightly steepened magnetic bump. The stationary parallel electric potential field across this structure shows a potential ramp with steep gradient at the leading edge of the pulsation. The potential drop of $\sim400$\,V corresponds to an electric field of $\sim0.47,{\rm V\,m}^{-1}$. Note that the time scale in this panel is 90 s. {\it Bottom left}:  Six seconds of a CLUSTER passage through the shock transition. The black dots show the spacecraft potential variation which maps the local density variation. Overlaid is the high frequency WHISPER trace of the plasma frequeny. In the magnetic pulsation regions (white) the plasma frequency exhibits huge excursions to both sides similar to those seen in the overview Figure\,\ref{chap5-fig-rad} on wave observations. These excursions trace the BGK (nonsymmetric) modes and (symmetric) solitons. {\it Bottom right}: One example of one of the solitons on a 2 ms time scale. It is nicely seen how symmetric the parallel potential trough and the corresponding  bipolar parallel electric field shape look like in the solitary wave structure.}\label{chap5-fig-pulsbgk}
\end{figure}

\cite{Behlke2003} report another interesting property of magnetic pulsations in the shock transition region where they overlap to form the quasi-parallel shock. Measurement of the electric cross-{\SL} potential identify a substantial unipolar drop in the electric potential of several 100 V, corresponding to a potential ramp,  when passing from upstream to downstream across the pulsation. Such a drop signifies the presence of an electric field in one direction across the pulsation. Taking the mean size of a pulsation to be roughly 1000\,km, the mean electric field is $\langle E\rangle\simeq 400\,{\rm mV\,m}^{-1}$. However, this field drops mainly at the leading edge of the pulsation. Such a field presumably corresponds to a steep pressure gradient in the pulsation. It could also be generated by an anomalous collision frequency. This remains to be tested by further observations and data analysis. 

The measurements of \cite{Behlke2003} anticipated the more recent report of strong electric fields in the shock by \cite{Bale2007}. In addition to this observation it was found \citep{Behlke2004} that the single pulsations were subject to a fairly large number of high frequency/Debye scale structures in the electric field seen in the {\footnotesize WHISPER} recordings (bottom panel), which belong to electron holes or solitons which form in the pulsation gradient regions as shown in Figure\,\ref{chap5-fig-pulsbgk}. The bipolar electric field and unipolar potential across one  -- indeed very symmetric -- soliton is seen in this figure. These observations suggest that the pulsations are indeed the main constituents of a quasi-parallel shock with the dynamics on the micro-scale of a quasi-parallel shock going on mainly in the single pulsations of which the shock transition is built. 

The occurrence of these intense nonlinear electrostatic electron plasma waves at the quasi-parallel shock transition is intriguing. It forces one to draw another very important main conclusion from these observations (and other related observations like those of \cite{Bale2007}): that quasi-parallel shocks are sources of electron acceleration into electron beams, which are capable to move upstream along the magnetic field over a certain distance and excite electron plasma waves at intensity high enough to enter into the nonlinear regime, forming solitons and electron holes (BGK modes). \index{waves!BGK modes}\index{waves!solitons}\index{soliton!BGK holes}

Presumably this is possible only when in the supercritical quasi-parallel shock transition region the magnetic field changes from quasi-parallel to quasi-perpendicular on the electron scale $\sim\lambda_e$. Indications of such a change on the ion scale $\lambda_i$ have been noted above at a number of occasions, but the detection of solitary structures in the electron plasma waves in relation to quasi-parallel shock transitions provides a very strong argument for this to be true on a scale which is well below the ion scale.  Only if this is the case, there will be ample reason for electrons to become reflected and accelerated into beams from the transition region in a quasi-parallel shock. As we already noted, we may, therefore, expect that quasi-parallel supercritical shocks on the electron scale are not anymore quasi-parallel but change to become locally quasi-perpendicular, while on the larger ion scale they still maintain properties of quasi-parallelity. \index{shocks!transition to quasi-perpendicular}\index{acceleration!quasi-parallel shock, electrons}

If this conclusion will turn out to be true and will sustain future more sophisticated experimental tests, it will have important consequences for collisionless shock physics. Supercritical collisionless -- nonrelativistic -- shocks will, in fact on the electron scale, always behave quasi-perpendicularly -- and it may be suspected that this conjecture will also hold for relativistic shocks though probably for other reasons (like the generation of transverse magnetic fields by the Weibel instability, which becomes dominant in relativistic shocks \citep[see, e.g.,][]{Jaroschek2004,Jaroschek2005}). This implies also that the true quasi-parallel shock physics cannot be properly elucidated when ignoring electron effects as is, for instance, done in hybrid simulations.\index{shocks!relativistic}

 \subsubsection*{Simulations of quasi-parallel shock reformation}
\noindent Nevertheless, before turning to the very few more realistic full particle PIC simulations with different mass ratios from small to about realistic, we are first going to discuss here hybrid simulations in one and two dimensions of the formation and behaviour of quasi-parallel shocks. Initial particle \citep{Quest1983,Quest1985,Quest1988} and hybrid simulations  \citep{Burgess1989,Thomas1990,Winske1990,Terasawa1990,Scholer1992a} did already illuminate some of the particular interrelations between the dynamics of quasi-parallel shocks: energy dissipation by short wavelength whistlers at the shock transition, the presence of a diffuse hot ion component upstream of the shock, and the importance of upstream waves. Most of these have been the subject of reviews \citep[the interested reader might be directed to the papers by][which more or less systematically discuss selected aspects of these interactions]{Burgess1997,Burgess2005,Lembege2004}. The selection of results we will provide here is guided by the progress that has been achieved in the understanding of the quasi-parallel shock physics and in its relation to the observations.

The first arising question is, whether nearly parallel shocks are at all capable of reflecting ions. This question is neither nonsensical nor academic. It makes sense,  because the investigation of the critical Mach number by \cite{Kennel1985} becomes unreliable at small angles, say $\thetabn\lesssim 30^\circ$, because it is based on linear whistler dispersion and does not take into account the completely modified plasma conditions near the shock ramp. Thus the simple conclusion that the critical Mach number approaches ${\cal M}_{\rm crit}\to1$ for $\thetabn\to 0$ might be an unjustified extrapolation. Theoretically, the reflection of particles at small $\thetabn$ must become entirely due to the electrostatic shock potential drop with the magnetic part of the Lorentz force being obsolete. The potential drop, however, should decrease with ${\cal M}$, and thus the reflection of particles should cease, while being of vital importance for the generation of a collisionless quasi-parallel shock. Hence, the question is also not purely academic. 

Of course, we do already know from the observations that quasi-parallel shocks exist at small $\thetabn$ with their foreshocks being populated by a diffuse ion component that excites upstream waves and mediates the beam-generated upstream ion-foreshock boundary  waves. The impossibility for this diffuse component of being entirely due to shock reflection in the quasi-perpendicular part of the shock, immediately proves that the quasi-parallel (or even the nearly parallel shock) must be able to reflect particles upstream. Hence, either a quasi-parallel shock is capable of generating a large cross-shock potential, or it is capable of stochastically -- or nearly stochastically -- scattering ions in the shock transition region in pitch angle and energy in such a way that part of the incoming ion distribution can escape upstream, or -- on a scale that affects the ion motion -- a quasi-parallel shock close to the shock transition becomes sufficiently quasi-perpendicular that ions are reflected in the same way as if they encountered a quasi-perpendicular shock. 

Observations suggest that the latter is the case, while observations also suggest that large potential drops occur in the large-amplitude magnetic pulsations ({\SL}) where they accumulate in the shock ramp \citep{Behlke2004}. Hence, reflection of ions will be due to the combination of both effects, the electric potential drop and the magnetic deflection. In fact, this can be a quite complicated process for an ion passing across a number of magnetic pulsations, in each of which it is being retarded and at the same time deflected by a small angle until its normal velocity component is decreased sufficiently that a further deflection in pitch angle suffices to let it return into the upstream region. 

\paragraph{Hybrid simulations in 1D.} In agreement with what is known today,  first one-dimensional hybrid simulations in an extended simulation box \citep{Burgess1989} suggested that the reformation of quasi-parallel shocks is about cyclic and is caused by the impact of large-amplitude upstream waves. \cite{Terasawa1990}, using $\thetabn=20^\circ$ and ${\cal M}_A=3.5$ with an upstream ion thermal velocity $v_i=v_A$ (corresponding to $\beta_i=1$) in one-dimensional hybrid simulations (with small numerical resistivity) showed that the reflected ions are not coming from the core of the incident upstream ion distribution but originate in the shell of this distribution, having initial velocities $v\gtrsim 1.7 v_i$. These ions escape from the shock quite far upstream and excite ultra-low frequency waves with upstream directed velocity of $\sim 1.3V_A$ at distances up to $>300\lambda_i$, which are convected downstream to reach the shock. In this one-dimensional hybrid simulation the only mode in which they can propagate is  the compressive fast magnetosonic mode. 

These waves are in fact what in observations has been identified as pulsations ({\SL}) but is not yet recognised as such here. During downstream convection the waves grow and slow down in the interaction with the foreshock ion component. When approaching the shock they generate a large amount of new reflected ions. These slow the incident ion population down and steepen the wave crest, which becomes the new shock front. In the time between the arrival of the compressive waves the shock is about stationary and develops phase-locked upstream whistlers which the arriving next wave crest destroys. From these simulations it could not be concluded what process produced the reflected ions, however, as one-dimensional simulations among suffering from other deficiencies select only one particular direction of wave numbers and are thus not general enough for drawing final conclusions.

The nature, generation and effects of the large-amplitude upstream waves have been further investigated in more detail in one-dimensional \citep[][among others]{Krauss1991,Scholer1992a,Scholer1993}, and in two-dimensional hybrid simulations \citep{Krauss1993,Scholer1993,Dubouloz1995}. Since shocks are three-dimensional, it is clear that two-dimensional numerical simulations at same resolution come closer to reality. However, they suffer from restrictions in size of the simulation box and simulation time. Since reality does not confront us with an initial state, large boxes and long times are needed. However, for the investigation of particular questions, simulations have the great advantage of providing the possibility to prepare them for answering just those questions. Concerning the propagation of upstream waves in one direction with respect to the shock normal, one-dimensional simulations are just good enough. 

\begin{figure}[t!]
\hspace{0.0cm}\centerline{\includegraphics[width=0.9\textwidth,height=0.25\textheight,clip=]{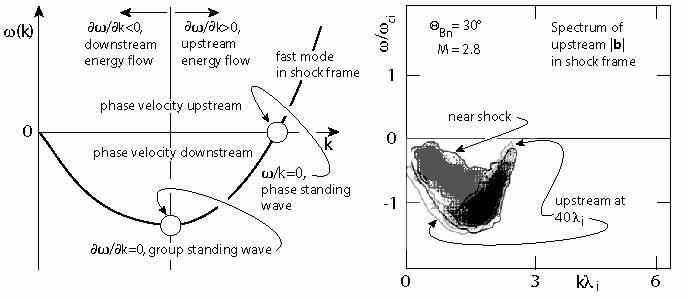} }
\caption[Magnetic Field Evolution in Quasiparallel Shocks: Simulations]
{\footnotesize Fast mode dispersion relation in simulations in the shock frame. {\it Left}:  The Doppler shifted fast mode dispersion relation in a supercritical flow in the shock frame. The Dispersion relation assumes negative frequencies corresponding to the downstream convection by the flow of Mach number ${\cal M}_A=2.8$ and at $\thetabn=30^\circ$. Waves at zero group velocity have energy at rest in the shock frame. Negative group velocities imply downstream transport of energy, positive group velocities imply upstream transport. {\it Right}: Simulated upstream wave dispersion spectra near the shock and upstream of the shock. Near the shock wave energy accumulated around standing and downstream transport. Away from the shock the wave energy still moves upstream \citep[data from][]{Krauss1991}.}\label{chap5-fig-dispfast}
\end{figure}

In order to identify a particular wave mode, the dispersion of the wave must be investigated. This dispersion relation depends on the frame in which it is taken, because the energy/frequency of a wave is not invariant with respect to coordinate transformations; in a medium moving with velocity ${\bf V}$ it is Doppler shifted according to $\omega'= \omega({\bf k})-{\bf k\cdot V}$, where $\omega({\bf k})$ is the dispersion relation in the rest frame of the flow. While the Doppler shift at high frequency is negligible, it completely changes the dispersion of ultra-low frequency waves at large Mach numbers. Figure\,\ref{chap5-fig-dispfast}  on its left shows the deformation of the fast mode dispersion relation in the shock frame at large Mach number ${\cal M}>1$ and for waves propagating upstream in the plasma frame. The deformation causes negative frequencies of the waves which imply downstream directed phase velocities, which is nothing else but the intuitive downstream convection of the waves by the flow. However, the minimum in the dispersion relation implies that waves of a particular frequency and wave number have zero group velocities. In the shock frame the energy of these waves is stationary. Smaller wave numbers have energy moving downstream, larger wave number have energy moving upstream away from the shock.  The right part of the figure shows simulations of upstream waves according to one-dimensional hybrid simulations by \cite{Krauss1991} for $\thetabn=30^\circ$ and a Mach number ${\cal M}_A=2.8$. The entire dispersion of the simulated waves is negative. The waves are all convected toward the shock as their Mach number is less than the streaming Mach number. Near the shock most of the wave energy moves downstream and will cross the shock. Still some shorter wavelength waves (large $k$) move in energy upstream in the shock frame. Farther away from the shock most of the wave energy encountered is seen to move upstream.

\cite{Scholer1993} investigated these waves further in one-dimensional hybrid simulations performing several numerical experiments on them, taking away the shock and instead injecting a diffuse ion component from downstream. The main finding is that the large amplitude upstream magnetic pulsations ({\SL}) evolve out of the ultra-low frequency wave spectrum in the interaction with the diffuse ion component. In accord with observation the pulsations move upstream in the plasma frame. Thereby their upstream leading edge steepens and is right-hand circularly polarised like required for whistlers. However, dispersion is unimportant; the main cause of the evolution of large pulsations is nonlinearity when the wave interacts with the diffuse ion distribution. This distribution has a steeper shock directed density gradient than the pulsation wavelength. Moreover, the flow becomes decelerated at the leading edge of the pulsation (as is seen in Figure\,\ref{chap5-fig-schofuji93}), and here the velocity difference between the flow and the diffuse ion component drastically decreases, which shifts the $k$ vector of the resonant wave to larger values, and the wavelength decreases during convection of the pulsation toward the shock. (Note that no resonant ion beam-whistler interaction exists as the beam is hot and diffuse.) The standing whistlers are in the leading edge are simply generated by the current flowing in the steep edge. It is thus concluded that it is the gradient in the hot diffuse ion component over a length of the same order as the length of the wavelength which produces the pulsations. Ultimately these pulsation cause a quasi-periodic reformation of the shock, as we have described earlier. This is thus proved by one-dimensional hybrid simulations. The same result is obtained when the simulation starts right away without a shock but with an injected hot beam (which is no surprise as the generation of the shock, before it was removed in the former simulations in order to keep with the wave field, was due to the plasma flow-reflected ion beam interaction).

\paragraph{Hybrid simulations in 2D.} The two-dimensional evolution of the pulsation ({\SL})  was studied later by \cite{Scholer1993a} and \cite{Dubouloz1995}  (see Figure\,\ref{chap5-fig-schofuji93}). It basically confirmed the conclusions drawn from one-dimensional simulations with the three important modifications, first, the wave fronts of the pulsations ({\SL}) rotate into a direction that is more parallel to the shock thereby increasing the shock-normal angle locally to become quasi-perpendicular; second, the pulsations have short wavelength in shock normal direction, but are of substantially longer but finite lengths in the direction tangential to the shock, which provides structure to the shock in tangential direction; and third, shock reformation is a result due to the steepening and accumulation of the pulsations and is a quasi-periodic process but the downstream structure of the shock over some distance is caused by the downstream convection of the old shock front, i.e. the bulk of the pulsations that had accumulated at the location of the former shock transition. It is interesting to note that from reformation cycle to the next rather large fluctuations in the magnetic field and density exist in the transition from upstream to downstream which may be capable of trapping particles. The two-dimensional simulations do also confirm the conclusions that the diffuse upstream ion component is responsible for the growth of the pulsations ({\SL}). The simulations by \cite{Dubouloz1995}, in particular, followed the same scheme as the one-dimensional simulations,  injecting hot diffuse ions into upstream in order to control the interaction between the diffuse ion component and large amplitude pulsations ({\SL}).
\begin{figure}[t!]
\hspace{0.0cm}\centerline{\includegraphics[width=0.8\textwidth,clip=]{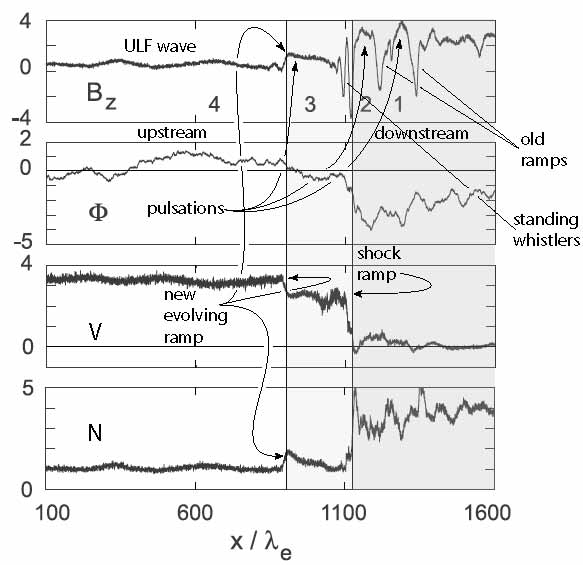} }
\caption[Magnetic Field Evolution in Quasiparallel Shocks: Simulations]
{\footnotesize Full particle PIC simulations of the evolution of a quasi-parallel shock in one dimension only \citep[after][]{Scholer2003}. From top: main magnetic field component $B_z$, electric potential $\Phi$, bulk plasma flow velocity $V$, density $N$, all in simulation units. The numbers indicate three pulsations ({\SL}). Pulsation 1 was the old shock. Pulsation 2 is the actual shock coinciding with the drop in $V$ to zero and the steep increase in density and potential. Pulsation 3 is just evolving. It already has a steep leading upstream edge and decelerates the plasma flow. It will become the next shock ramp. Number 4 indicates a bump in the upstream waves that will become a pulsation. The actual shock ramp has some phase-locked whistlers attached to it which are not well resolved on the scale shown (see the next figure).}\label{chap5-fig-qpasim}
\end{figure}

\paragraph{Full particle PIC simulations.} So far we dealt just with hybrid simulations where the ions are macro-particles while the electrons represent a charge-neutralising background of zero mass. Clearly, such simulations are unrealistic if whistlers become involved. This is, however, the case, as we have discussed above, whenever large amplitude pulsations evolve at the leading edge of which phase-locked whistlers are attached. The question what role the electrons play in the evolution of the pulsations can only be answered by full particle simulations. These require large simulation boxes and at the same time high temporal and spatial resolutions. So far they could therefore only be performed in one dimension \citep{Pantellini1992,Scholer2003,Tsubouchi2004}. \cite{Pantellini1992} used a mass ratio of $m_i/m_e=100$. They had to small a box (just $\sim 30\lambda_i$ in the upstream direction) for following the evolution of upstream waves but stressed the importance of whistlers in shock reformation. One decade later it became possible to substantially enlarge the box and at the same time to switch to a larger mass ratio while staying with one dimension only.  \cite{Scholer2003}, using the same mass ratio at $\thetabn=30^\circ$ and ${\cal M}_A\simeq4.7$, had an upstream entension of $\sim200\lambda_i$ and could follow the shock evolution for a time $t\omega_{ci}\sim100$. Similar simulations with mass ratio $m_i/m_e=50$ have been performed by \cite{Tsubouchi2004}. Here we discuss in greater detail the calculations of \cite{Scholer2003}. An overview of their results is shown in Figure\,\ref{chap5-fig-qpasim} in the fixed lab-frame for the main magnetic field component $B_z$, electric shock potential $\Phi$, bulk stream velocity along the shock normal $V$, and density $N$, all as functions of distance $x$ (measured in electron inertial lengths $\lambda_e$). 
\begin{figure}[t!]
\hspace{0.0cm}\centerline{\includegraphics[width=0.8\textwidth,height=0.4\textheight,clip=]{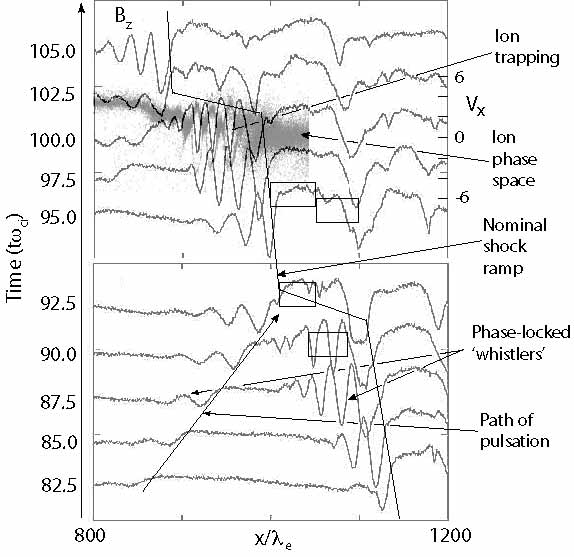} }
\caption[Magnetic Field Evolution in Quasiparallel Shocks: Simulations]
{\footnotesize Full particle PIC simulations of the evolution of a quasi-parallel shock in one dimension \citep[data taken from][]{Scholer2003}. The profile of the main magnetic component $B_z$ is shown for subsequent simulation times shifted by $\Delta\, t\omega_{ci}= 2.5$ upward. The representation is in the simulation frame, i.e. the shock moves to the left into the upstream direction. The simulation shows the reformation of the quasi-parallel shock resulting from the exchange of the shock with an incoming upstream wave which has steepened to become a pulsation ({\SL}). The magnetic field trace at time $t\omega_{ci}= 100$ has been overlaid on the ion phase space at this time. The heavy steps show the location of the nominal shock ramp (where the flow is stopped). It moves slowly upstream until a new pulsation arrives and when it suddenly jumps forward by roughly $100\lambda_e$. Also shown is the fast approach of an upstream pulsation starting at $t\omega_{ci}= 82.5$ and arriving at the shock at $t\omega_{ci}= 92.5$ to take over the role of the shock. Note that in the minima of the `whistler' field fluctuations (at $t\omega_{ci}= 100$) ions are trapped, oscillating back and forth and forming hole vortices in phase space centred around local minima of the electric potential $\Phi$ (not shown). The little boxes indicate where particle (ion or electron) phase space distributions have been determined.}\label{chap5-fig-schoku02}
\end{figure}

Far upstream from the quasi-parallel shock the magnetic field exhibits long wavelength ultra-low frequency waves (4) which, when approaching the shock, start steepening at their leading edges (note that in the plasma frame these waves are moving upstream, as also does the shock, i.e. to the left in the figure). The amplitude of the wave increases (3) , and the wave becomes a pulsation ({\SL}). Its amplitude is large enough to already substantially brake the upstream bulk flow, which causes a drop in $V$ and in $\Phi$, and an increase in density. The pulsation has slowed down the flow to a velocity $\sim2.8V_A$ already here. In fact the leading edge of the pulsation behaves like a `baby shock', which later will become the real `adult shock'. The shock itself is formed further downstream at the local position of the leading steep edge of the previous pulsation (2). In front of this edge (i.e. the genuine shock at this instant and location) between it and the trailing edge of pulsation (3) a standing phase-locked large amplitude whistler has evolved. This whistler is spatially damped by the approaching pulsation (3). An indication of such a whistler has already been seen in front of the leading edge of pulsation (3) as well. Pulsation (2) (the instantaneous shock) has a substantial downstream extension. Further downstream of it the `old shock' is seen, which was formed at an earlier time by pulsation (1); and even farther downstream a remainder of earlier shock ramps (pulsations) is recognised in the trace of $B_z$. The instantaneous shock ramp (pulsation 2) is high enough to completely brake the upstream flow, the velocity of which drops to zero while the density steeply increases and forms a dense wall. 

Figure\,\ref{chap5-fig-schoku02} gives and impression of the shock evolution in higher temporal and spatial resolution. It shows instantaneous magnetic profiles in a box of lengths $400\l_e$ including the shock ramp with time axis running upward. In this simulation frame the shock moves upstream to the left. The heavy step-like line indicates the approximate location of the nominal shock ramp. Also shown is an upstream ultra-low frequency wave that at time $t\omega_{ci}= 82.5$ has begun to evolve into a pulsation while convectively approaching the shock, growing in amplitude and developing a steep leading (upstream) ramp in front of which phase-locked magnetosonic whistlers start growing. When approaching the shock at $t\omega_{ci}= 92.5$, the pulsation kills the phase-locked whistlers that were waiting in front of the ramp by damping them out. At this time the pulsation takes over the role of the shock, and the nominal shock position jumps ahead to upstream by roughly a distance of the width of the pulsation $\sim 100\lambda_e$.  This process repeats itself at $t\omega_{ci}= 102.5$ showing that the quasi-parallel on this time scale shock is not stationary but undergoes nearly periodic reformation which is mediated by the arrival of large amplitude pulsations.
\begin{figure}[t!]
\hspace{0.0cm}\centerline{\includegraphics[width=0.95\textwidth,height=0.3\textheight,clip=]{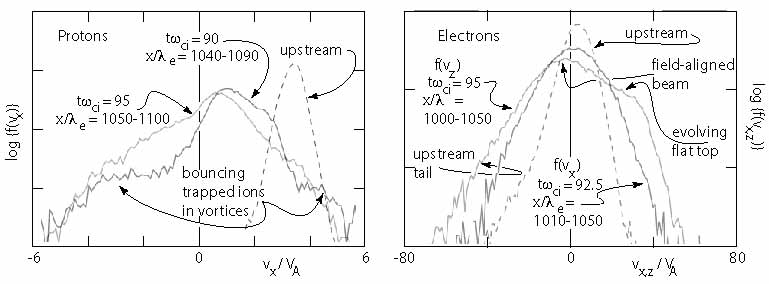} }
\caption[Magnetic Field Evolution in Quasiparallel Shocks: Simulations]
{\footnotesize Velocity distribution functions $f_i(v_x)$ for the full particle PIC simulations of Figure\,\ref{chap5-fig-schoku02} during different phases of shock evolution as indicated by the boxes in that figure \citep[data taken from][]{Scholer2003}. {\it Left}:  Proton distribution functions at times $t\omega_{ci}= 90$ and $t\omega_{ci}= 95$. The first period is in  the arriving pulsation before it takes over to become the shock. Shown is the part of the pulsation where it encounters the whistlers. The upstream distribution has changed here into a heated distribution exhibiting a substantial upstream directed tail which is due to the trapped ion component. The whole distribution assumed a large velocity spread and is strongly heated. At time $t\omega_{ci}= 95$ the trapped ions have damped the whistlers. The distribution has now become very hot but has a two-temperature structure with high-energy tail in the leading part of the pulsation. At this stage the incident plasma is completely slowed down. {\it Right}: Two electron velocity distributions $f_e(v_x)$ at time $t\omega_{ci}= 92.5$ and $f_e(v_z)$ at time $t\omega_{ci}= 95$, respectively. The former is right in the new shock built up of a fresh pulsation showing a hot but otherwise structureless perpendicular electron distribution, the latter is in the well-developed shock but along the main magnetic field. This parallel distribution is strongly nn-symmetric, heated, exhibits an upstream tail, indication of the formation of a top flat distribution and a remaining upstream-beam-like part similar to those distributions as had been measured by \cite{Feldman1983}.}\label{chap5-fig-schoku03}
\end{figure}

In order to illuminate what happens in the shock ramp, the ion phase space data have been overlaid at time $t\omega_{ci}= 100$ just before a new reformation. The shock ramp coincides with the location where the upstream flow is completely braked and the hot downstream ion distribution begins. This location corresponds to the leading edge/ramp of the first broad downstream pulsation. On the other hand, the phase-locked upstream whistlers that are attached to its ramp are seen to trap ions, which form large phase space vortices (holes) with ions bouncing back and forth in the associated potential wells. These are not shown in the figure but resemble the magnetic oscillations, being minimum in the centres of the holes. However, what is really interesting to note is that there is little indication of any back streaming reflected ion beam in this simulation. Instead, a very dilute but broad completely diffuse ion component is seen all over the simulation box, which is denser downstream than upstream but of similar velocity dispersion. Somehow this component has been generated by the shock, and it is which is responsible for the growth of the pulsations.

The evolution of the proton and electron distribution functions is shown in Figure\,\ref{chap5-fig-schoku03}. The two proton (ion) distributions are taken in the phases when fully developed whistlers with trapped ions and phase-space vortex formation exist, and when these whistlers have been completely damped by the trapped component. In the former case at time $t\omega_{ci}= 90$ the (perpendicular) ion shows a substantial heating with respect to the upstream ion distribution, has been slowed down by a factor of roughly $\sim 8$ in velocity and, in addition, exhibits a  nonthermal top-flat part in the upstream direction which is due to the presence of the trapped bouncing ions in the phase space vortices. At time $t\omega_{ci}= 95$ when the whistlers are completely damped and the pulsation has taken over to become the shock, the trapped vortices have disappeared, as has been shown above, and the distribution evolved into a broad and hot distribution at about zero flow velocity. Thus the vortices are entirely due to the interaction between the trailing edge of the shock-arriving pulsation and the whistlers with the latter being compressed, amplified in amplitude until they trap ions, while the pulsation itself is free of those trapped ions. This was already clear from Figure\,\ref{chap5-fig-schoku02} but its effect on the distribution is nicely seen here.

The right part of Figure\,\ref{chap5-fig-schoku03} shows two cases of electron distributions. The first in the perpendicular direction, the second in the parallel direction taken in the arriving pulsation when it becomes the shock, i.e. after the reformation. The boxes are given in Figure\,\ref{chap5-fig-schoku02}. The perpendicular distribution $f_e(v_x)$ simply shows the heated, about symmetric electron distribution in the shock. However, the parallel distribution $f_e(v_z)$ exhibits a number of interesting features. It is also hot, in fact substantially hotter than the perpendicular distribution, exhibiting nonthermal tails in both directions along the magnetic field with a stronger upstream tail. In addition, on the downstream side it shows the evolution of a broad flat top on the distribution with a remainder of an non-flattened upstream directed beam on the upstream side. This is ineresting as it is almost exact the type of distributions that had been reported by \cite{Feldman1983} from crossings of the super-critical quasi-parallel bow shock \citep[see also][]{Feldman1985,Gurnett1985}. 

\paragraph{Conclusions.} The full particle simulations provide much deeper insight into the shock physics than did the hybrid simulations. Still, they have been only one-dimensional and thus are restrict to waves propagating solely into shock normal direction. Moreover, the mass ratio is still not realistic as it competes with the length of the box, and it is obvious that the length of the box is crucial in quasi-linear shock physics for properly investigating the evolution of the waves which build the shock, and for their interaction with the particle component. It is, in particular, of considerable interest that, even though the magnetic field should allow the streaming of particle upstream in a quasi-prarallel shock, no upstreaming beams have been seen in the simulations. Electrons exhibit a signature of an upstreaming beam, but ions do not. The diffuse component which is homogeneously distributed over the entire region depending only on the distance from the shock is obviously participating in  a diffusive process and is not injected into the upstream region. This sheds light on the acceleration and confinement process acting in quasi-parallel shocks.

These simulations have demonstrated that quasi-parallel supercritical shocks are nonstationary on the time scale of the simulations. They are subject to reformation. However, this reformation is quite different from the reformation of a quasi-perpendicular shock as it is not due to the presence of gyro-bunched reflected ions in the foot of the shock. Rather it is caused by the accumulation of large amplitude magnetic pulsations which are not separate entities but grow out of the upstream ultra-low frequency waves in the range where the wavelength of the wave is of the same order as the density-gradient scale of the diffuse ion population in the foreshock. \cite{Scholer1993a}, \cite{Dubouloz1995} and \cite{Scholer2003} have shown that extracting the diffuse ion component results in a lack of upstream pulsations, while injecting additional diffuse ions speeds the evolution of large amplitude pulsations up. Thus, pulsations are the result of the fine-tuning of the resonance of ultra-low frequency upstream waves with the diffuse ion component.  These pulsations play an important role in quasi-parallel shock. Still, their dynamics has not been completely clarified and, in addition, is subject to some controversy. Closer investigation of pulsation dynamics is therefore of vital interest. 

To a certain extent this has been the subject of the one-dimensional full particle PIC simulations performed by \cite{Tsubouchi2004} who use a modest mass ratio $m_i/m_e=50$ \citep[see also][]{Lembege2004}  which, however, allows them to extend the simulation time scale to times as long as $t\omega_{ci}\sim 38$ and to enlarge the simulation box to  $\sim 700\lambda_i$. Their simulation is based on the magnetic piston method instead of the reflecting wall that has been used by \cite{Scholer2003}. In general these simulations confirm the previous results, adding some facts about the structure, size and life times of isolated pulsations whose sizes are  determined to lie in the interval between $(10-20)\lambda_i$  with the tendency of a gradual shrinkage during the reformation process. This shrinkage can simply be attributed to the dynamical compression of the pulsations at the shock transition where many embedded pulsations compete for the available shock volume. 

Also confirmed is the rotation of the magnetic field from quasi-parallel to quasi-perpendicular in the reformation process. {\it Large amplitude pulsations} ({\SL}) {\it behave like localised supercritical quasi-perpendicular shocks of which the quasi-parallel shock is constructed.} We stress the importance of this conclusion here again. Moreover, their evolution is accompanied by the generation of phase-locked whistler precursors at the leading edge of a pulsation, as has been shown above to be the case \citep{Scholer2003}. \cite{Tsubouchi2004} report in addition the observation of large-amplitude spiky electric fields in the leading edge of a pulsation. In view of the observations by \cite{Behlke2004}, their investigation, however, will require even much better time resolution on the electron plasma scale. Structures of this kind require the inclusion of the full electron dynamics and imply accounting for realistic mass ratios in the simulations.

Inclusion of particle dynamics is one of the main problems in quasi-parallel shock physics. We have seen that the evolution of the building blocks of a quasi-parallel shock, i.e. the large amplitude pulsations ({\SL}), is completely determined by the interaction of the ultra-low upstream wave spectrum with the diffuse ion component. On the other hand, the pulsations build up the quasi-parallel shock transition region which is responsible for the generation of the diffuse ion component in a way which has not yet been satisfactorily clarified. The current knowledge will be discussed in the next paper in this series. 

One idea is that the quasi-perpendicular subshocks of which the quasi-parallel shock is constructed -- and which are nothing else but the various pulsations of which the shock consists on the small scale of a few $\lambda_i$ -- reflect sufficiently many ions back upstream. However, because of the complicated magnetic structure of the shock transition region, these ions cannot simply escape along the magnetic field to upstream but remain trapped for long enough time in the shock transition, being scattered in pitch angle and energy until ultimately picking up enough energy in order to leave the shock either upstream or downstream in a diffusion process that works in the shock and partially also in the foreshock regions. The observation of a nearly isotropic diffuse ion distribution both upstream and downstream of the shock in the simulations and the lack of observation of ion beams, both in the measurements and in the simulations, provides a strong argument for a diffusive mechanism to act in the shock transition. The observations by \cite{Kis2007}  also support this conclusion as they show that the diffuse ion component exhibits an exponential density gradient towards the shock with the shock being their source. Such a gradient is typical for a diffusive process. 

Whether electrons undergo a similar process, remains an unresolved question that will be attacked with the increasing capacities of computers when the full electron dynamics can be included into the simulations. There are strong hints on the importance of electron dynamics in the presence of the observed spiky large amplitude electric fields in the pulsation and shock transition. These electric field structures point on the generation of electron beams. Since no such beams have been observed in the deep upstream foreshock, the structure of the shock on the small scale presumably inhibits the escape of electrons to far upstream in the form of beams. This can be due to either a chaotic magnetic field configuration on the electron scale which makes the escape of electrons along the magnetic field impossible. It can also be due to the trapping of electrons in the phase space holes which are related to the smallest-scale electric field structures. These, on the other hand, require strong electron currents to flow in the shock transition and, if present, will stochastically heat the electrons. Moreover, they might be responsible for the electromagnetic radiation generated in the shock ramp, a process that has been observed in solar and interplanetary type II radio bursts but never found a convincing explanation by any known mechanism. The finding in the PIC simulations of the formation of top-flat electron distributions and a rudimentary upstream directed field aligned beam in the shock transition \citep{Scholer2003} might indicate that the electrons are accelerated in the embedded pulsations by electric fields and can form a field aligned upstream beam which apparently cannot leave far upstream but possibly has sufficient free energy in the electrons for driving either a Buneman or modified two-stream instability which causes electron plasma waves, electron holes, and radiation -- and at the same time scatters, heats, and confines the electrons.

\section{Hot Flow Anomalies}
\noindent Would the collisionless upstream flow be stationary at constant Mach number ${\cal M > M}_{\rm crit}$, we could safely terminate the discussion of the supercritical shock transition regions at this place and focus our attention on the downstream part of the shock transition. Unfortunately, this is not the case. The upstream flow is usually in a highly disturbed fluctuating state with the smallest disturbances in it readily evolving into structures, which can appear in various form, either as turbulence on almost all scales, from the electron scales up to scales of a substantial fraction of the macroscopic dimension of the flow, as localised large amplitude waves, as shocklets, or as various kinds of boundaries in the flow, which appear as discontinuities and current layers. 

Then the question arises, what happens when any such forms impact on the shock and interact with it. As long as we are dealing with the interaction of a single wave with the shock one needs to investigate the conditions for wave reflection, refraction and transmission across the shock. This is most easily done for wavelength much longer than the shock transition scale and leads to modifications of Snell's law when taking into account the Rankine-Hugoniot jump conditions at the shock, for instance, when the MHD approximation holds. For shorter wavelengths the continuous change in the wave propagation conditions as given in the dispersion relation can be parameterically treated. The shock may then act as a spectral filter depending on which waves it allows to pass. 

\subsection{Observations}
\noindent Of quite a different kind of problem is the interaction of a current sheet or discontinuity boundary with the shock.  Current sheets and discontinuities are large amplitude distortions of the upstream flow and cannot be treated like waves interacting with a shock. Observations near the Earth's bow shock have shown that they lead to severe distortions of the shock, which have been termed `Hot Flow Anomalies' (HFAs, the term now commonly used even though it does not appropriately refer to the physics involved), `Active Current Sheets', `Diamagnetic Cavities'  or `3D-Plasma Structures with Anomalous Flows'. They have been discovered in spacecraft transitions of the Earth's bow shock from analysing {\footnotesize AMPTE} and {\footnotesize ISEE} measurements \citep{Schwartz1985,Schwartz1988,Thomsen1986,Woolliscroft1986,Woolliscroft1987,Fuselier1987,Paschmann1988}. 
\begin{figure}[t!]
\hspace{0.0cm}\centerline{\includegraphics[width=0.98\textwidth,height=0.5\textheight,clip=]{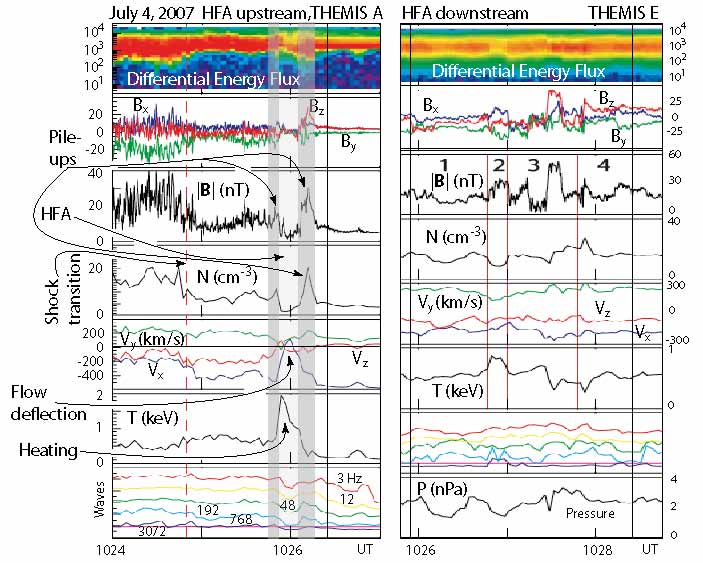} }
\caption[Magnetic Field Evolution in Quasiparallel Shocks: Simulations]
{\footnotesize THEMIS spacecraft measurements of a Hot Flow Anomaly upstream and downstream of the quasi-parallel Earth's bow shock wave \citep[data taken from][]{Eastwood2008}. {\it Left}:  The THEMIS A observations in the foreshock close to the shock transition. Three regions of the HFA have been shaded: darker shading refers to the pile-up regions at the HFA edges, The magnetic depression region in the centre of the HFA is shaded light. The quasi-parallel shock transition is about 1 min before the HFA encounter. The deflection (retardation in $V_x$) of the upstream flow is clearly seen. It coincides with a drop in density and magnetic field and strong plasma heating. Bipolar shape of $B_z, B_y$ indicates the presence of a current sheet. Note also the asymmetry of the HFA pile-up edges. The upstream edge which is exposed to the flow is much stronger piled up. These regions are themselves shocks. The lowest panel is the power spectral density of magnetic waves at different frequencies. {\it Right}: The simultaneous THEMIS E observations downstream of the shock behind the shock transition. The signature of the HFA is very irregular here. The later onset confirms that the HFA is an upstream phenomenon.}\label{chap5-fig-hfa1}
\end{figure}

The first observation of the HFA phenomenon was made by \cite{Schwartz1985} in the plasma and field  observations of {\footnotesize AMPTE UKS} who detected an unusual very hot upstream plasma event resembling a broad layer of plasma in which the upstream magnetic field performed a large rotation as it is known from current sheets. The most surprising observation was, however, that the upstream flow inside this current sheet was sharply deflected by a substantial angle close to $\sim90^\circ$ from its original upstream direction, a deflection far in excess of the deflection the bow shock would provide and in addition not expected to occur in front of the shock in its upstream region. The upstream flow was about stopped from its original flow direction. This deflection was also different from the one predicted by the magnetic stresses across the boundary of the hot flow anomaly, and it was clear that it required a violent momentum exchange between the upstream flow, the shock, and some (unknown) agent. It was also found that the structure was not in its final state but was evolving. Hence,  most probably it did not have come all the way along from the source of the upstream flow, the Sun, and it was speculated that it was either caused by an upstream-flow current sheet interacting with the supercritical bow shock or also could have had its origin in a violent change of the configuration of the obstacle, i.e. the magnetosphere in this case, an idea that was advocated later on \citep{Paschmann1988}. It took about  one decade of measurement and theoretical investigation until a consensus was reached that the HFA phenomenon was a shock phenomenon and was not caused by the obstacle. Clearly, violent changes in the obstacle configuration do also affect the shock \citep{Paschmann1988,Sibeck1999} but most probably to a lesser extent than it happens to be observed in hot flow anomalies. 

The signatures of a Hot Flow Anomaly in the upstream foreshock plasma are: (1) a strong deflection of the bulk flow velocity, (2) reduced magnetic field strength, (3) reduced density, and (4) considerable heating of the plasma. The region where all these changes in the plasma parameters are observed is usually flanked by walls of enhanced plasma density and magnetic field strength when both, the upstream plasma and upstream magnetic fields pile up.

All these properties are nicely seen in the most recent observation of a Hot Flow Anomaly by {\THE} plotted in Figure\,\ref{chap5-fig-hfa1}. \index{spacecraft!THEMIS} Similar observations by the {\CL} spacecraft have been reported by \cite{Lucek2004}, but while all four {\CL} spacecraft were upstream of the shock, the {\THE} spacecraft were both upstream and downstream such that they could follow the evolution of the HFA from upstream to downstream. 

The most spectacular effect in the observation of a Hot Flow Anomaly is seen in the fifth panel on the left of this figure. The velocity component $V_x$ turns positive in the HFA, which corresponds to a complete stopping or even reversal of the flow in the downstream direction. The other two flow velocity components are enhanced simultaneously, indicating a deflection of the flow in the $(y,z)$-direction. 

Inspecting the magnetic field, one finds that its magnitude drops to a small value. However the traces of its transverse components become dipolar, which is the sign of a current that is flowing inside the HFA. To both sides of the HFA current both the magnetic field and density pile up. It is interesting that this pile up is not symmetric but is stronger on the upstream side. This should be so since the HFA must stop the upstream flow, which causes a stronger compression on the upstream than on the downstream sides. In the figure the two pile up regions are shown in shading. They flank the current region where the plasma is dilute and hot, as is seen from the strong increase in temperature and from the ion energy flux in the uppermost panel which exhibits a broadening of the distribution that is comparable to the shock transition at the beginning of the panel. However, though the temperature is high, the mean energy of the ions is low, reflecting the strong deceleration of the flow. 

The right part of the figure shows simultaneous measurements downstream of the shock. These indicate that the HFA downstream is very complex, causing a strong disturbance in the downstream flow. It, moreover, sets on later than in the upstream region which is a confirmation for the HFA being an upstream phenomenon and is not caused by the obstacle or downstream of the shock. This rules out number of models that had been proposed in the past. 

We will not go into the details of this downstream observation other than point on the much more diffuse shape of the HFA after its ``passage" across the shock to downstream. It has become dissolved into a number of different regions which still belong together while each differs from the others. The larger of these regions are numbered from 1 to 4 in the figure. Some of them showing very clear signatures (correlated magnetic plus density compressions, regions 1 and 4) of fast modes, others of slow expansion modes (anticorrelated magnetic and plasma variations plus plasma heating, region 2). In any case, it becomes clear that the shock has split the HFA into different filaments. The HFA has lost the compactness, consistence, and coherence it possessed in the foreshock and during the interaction with the shock. It would contradict the imagination (and the second law) if such a diffuse downstream structure would combine to make up a coherent HFA of the kind we see in front of the shock. In particular, its later downstream than upstream onset, which causes a time delay between the upstream and downstream disturbances, provides a very strong causal argument against a downstream origin of the HFA, at least in this particular case and for the class of HFAs that exhibit similar properties. 

We note that \cite{Thomsen1993} investigated a large number of Hot Flow Anomalies and determined the directions of their normal vectors, applying the minimum variance method. They found that the normal vectors are in good agreement with the assumption that upstream tangential discontinuities are involved into HFAs. This assumption since dominates the interpretation and theory of Hot Flow Anomalies. 
\begin{figure}[t!]
\hspace{0.0cm}\centerline{\includegraphics[width=0.75\textwidth,clip=]{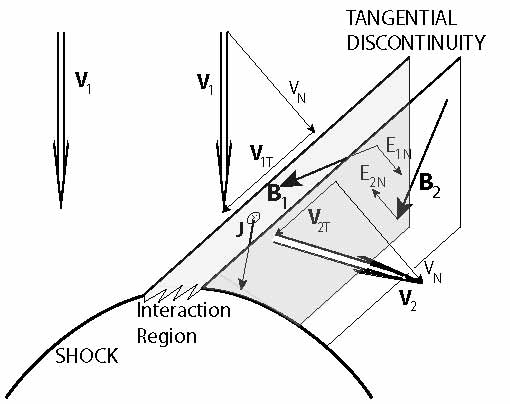} }
\caption[HFA simulation]
{\footnotesize Schematic of the interaction of an upstream tangential discontinuity with the shock. The tangential discontinuity is convected along with the normal component $V_n$ of the upstream flow which, in the shock frame, is thus the same on both sides. However, the tangential velocities and magnetic fields are different on both sides of the discontinuity. The latter is due to the inclined current flow ${\bf J}$ in the discontinuity. Here the rotation of the magnetic field is shown in such a way that the normal components of the electric field $E_n$ point inward on both sides of the discontinuity thus confining any shock-reflected ions to the interior of the discontinuity which here is shown as an extended region. Due to the presence of the discontinuity and interaction with the shock the upstream flow is deflected from its original direction, because the tangential velocity has changed across the discontinuity. For other current direction the tangential magnetic field would rotate differently. $E_n$ can then have other directions, pointing away from the discontinuity on both or also only on one side.}\label{chap5-fig-tdshfa}
\end{figure}

\subsection{Models and simulations}
\noindent The mechanism of hot flow anomalies is believed to be related to the interaction of the shock with an upstream current sheet \citep{Burgess1988,Thomas1991}, most probably a tangential discontinuity, in the near-shock quasi-parallel foreshock. It is this model that is most strongly supported by the recent {\THE} observations shown in Figure\,\ref{chap5-fig-hfa1} and also by the {\CL} observations reported by \cite{Lucek2004}. The current sheet (tangential discontinuity) deflects the upstream reflected ion component, channels the ions back and focusses them along the current sheet. This produces a highly localised hot ion population which starts expanding, blowing off the upstream flow to both sides, deflecting the flow from its original direction, and piling the upstream plasma up to the sides of the current sheet where it causes new shock waves.  

Tangential discontinuities have the property that they are convected with the upstream flow. The model is shown schematically in Figure\,\ref{chap5-fig-tdshfa}. There is no flow across the discontinuity. Thus, in the upstream frame of reference the velocity component $V_n$ perpendicular to the discontinuity is strictly zero; in the shock or laboratory frames it is the same on both sides of the discontinuity. The normal component of the upstream magnetic field $B_n\equiv 0$ vanishes in all frames. The flow velocity ${\bf V}_{\rm T}$ and the upstream magnetic field ${\bf B}_{\rm T}$ are both tangential to the discontinuity to both of its sides but can be rotated by an arbitrary angle and can have different magnitudes on both sides, while the total pressure $P=NT+B^2/2\mu_0$ is constant across the discontinuity in the co-moving upstream frame. Thus, the density, tangential magnetic field, and tangential velocity can change arbitrarily across the discontinuity. The temperature change is then fixed by the continuity of the pressure (we do not consider here an anisotropic pressure, however). 

Any change in the magnetic field, which is a rotation of the field in the discontinuity plane plus a stretching of the tangential component, corresponds to an electric current flowing in the discontinuity. (Note that different magnitudes of the tangential magnetic fields to both sides of the discontinuity simply mean that the current flowing in the discontinuity is not distributed homogeneously over the width of the discontinuity, while a rotation of the field across the discontinuity implies an inclined current.) 
\begin{figure}[t!]
\hspace{0.0cm}\centerline{\includegraphics[width=0.98\textwidth,clip=]{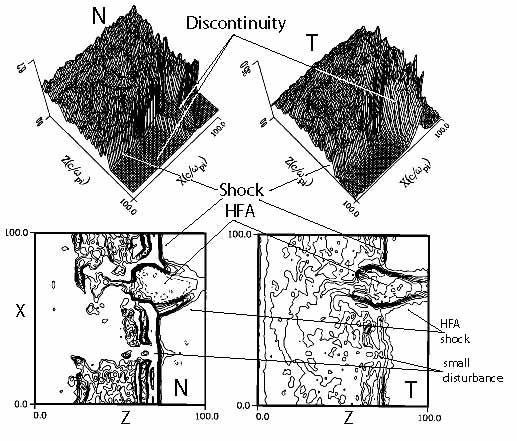} }
\caption[HFA simulation]
{\footnotesize Two-dimensional hybrid simulation of the formation of a Hot Flow Anomaly (HFA) in the interaction of a tangential discontinuity with a quasi-perpendicular shock \citep[after][]{Thomas1991}. The idea is that the tangential discontinuity first touches the shock at the quasi-perpendicular site because the upstream magnetic field is contained in the discontinuity plane, initially interacting with the shock-reflected ions. Two discontinuities are used. One with normal electric field $E_n=E_n^+$ pointing toward the discontinuity, the other with $E_n=E_n^-$ pointing away from the discontinuity. Only the former generates an HFA by capturing the shock reflected ions. The other causes only a minor disturbance of the shock. The figure shows the density $N$ and plasma temperature $T$ in the simulation plane $(x,z)$. {\it Top}: Stacked profiles of density and temperature. The $E_n^+$-discontinuity produces a major disturbance in $N$ and $T$ at the shock surface, consisting of two dense and hot walls surrounding a diluted region. {\it Bottom}: Iso-contours of $N$ and $T$ showing the extension of the HFA disturbance on the shock profile.} \label{chap5-fig-hfa2}
\end{figure}

In addition, tangential discontinuities have one property that turns out to be of vital importance for the formation of a Hot Flow Anomaly. Tangential discontinuities possess a non-vanishing normal electric field component ${\bf E}_n=-{\bf V}_{\rm T}\times{\bf B}_{\rm T}$. This field vanishes only when ${\bf V}_{\rm T}$ and ${\bf B}_{\rm T}$ are either parallel or anti-parallel. However, its direction can be along the external normal to the tangential discontinuity (pointing away from the discontinuity) or anti-parallel to the external normal (pointing toward the discontinuity). In the first case, ions will be removed from the discontinuity by this field, while in the second case ions will be returned to the disontinuity. One expects that only in the second case a Hot Flow Anomaly is created in the interaction with the shock, while the first case will cause only a minor distortion of the shock. Hybrid simulations show that this is indeed the case.

\paragraph{Two-dimensional hybrid simulation.} An example is given in Figure\,\ref{chap5-fig-hfa2} which shows the results of a two-dimensional hybrid simulation of the interaction of two discontinuities with a shock \citep{Thomas1991}. The only difference between the two discontinuities is that the normal electric fields ${\bf E}_n=-{\bf V}_{\rm T}\times{\bf B}_{\rm T}$ to both sides of the discontinuity at $x=25$ points away from the discontinuity while  to both sides on the discontinuity at $x=75$ it  points into the discontinuity. The effect on the interaction between the discontinuities and the shock is dramatic. The $x=25$ discontinuity causes only a minor distortion on the shock while the discontinuity at $x=75$ digs a big hole into the shock and at the same time produces a violent effect in the upstream region which shows all the signs of an HFA. Note that because of technical reasons the simulation was performed assuming that the discontinuity impinges on a perpendicular shock, which is a reasonable assumption for the initial time of the interaction, because the discontinuity, being a tangential discontinuity and therefore containing in its plane the full upstream magnetic field, first touches the shock in the area of the perpendicular shock where the upstream field is tangential to the shock. Only later it is swept over the shock surface by the convective flow sitting for the longest time in the larger quasi-parallel region where HFAs are usually observed. 

The explanation of the simulation result is quite simple even though the complete physics of the interaction is still poorly understood. The quasi-perpendicular supercritical shock reflects ions. When $E_n=E_n^-$ points away from the discontinuity all reflected ions in the interaction region between shock and discontinuity are channelled away from the shock. This has the effect, that shock reformation is inhibited by the discontinuity, and the shock becomes nearly stationary with the exception of minor distortion around the intersection point, as is seen in the lower part of the figure around $x=25$. There is a slight increase in density $N$ at the interaction point in front and a dilution behind the shock surface along the direction of the discontinuity where all the fluctuations are damped out by the discontinuity, and the shock front becomes narrow, weekend, and stable. A similar effect is seen in the temperature $T$.

The more interesting violent distortion in the case $E_n=E_n^+$ around $x=75$ is caused by the channelling of the shock-reflected ions into the discontinuity. Inside of the tangential discontinuity the ions become approximately unmagnetized due to the presence of the current (which generates the discontinuity) and the correspondingly low internal magnetic field. The isotropic ions accumulate and cause a hot dilute ion cloud that compresses the shock in the downstream direction, excavate the upstream flow from the region, deflect it into tangential direction, and cause the density and magnetic field to  pile up at the boundaries of the discontinuity. The shock is hereby violently deformed both in the upstream and downstream directions. In addition, upstream shocks are formed at the discontinuity boundaries (in the non-symmetric case only one shock boundary would form); downstream a HFA bubble structure is caused with sharp, dense and hot walls, as seen from the figure.
\begin{figure}[t!]
\hspace{0.0cm}\centerline{\includegraphics[width=1.0\textwidth,height=0.4\textheight,clip=]{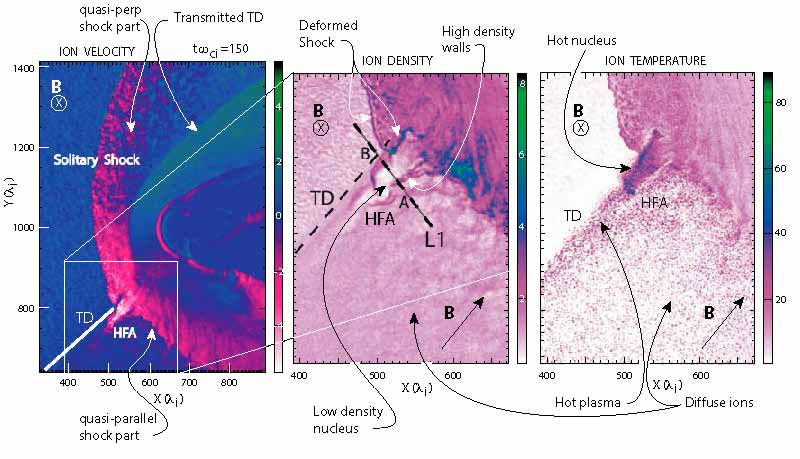} }
\caption[HFA simulation]
{\footnotesize ``Global" hybrid simulations of the interaction of a tangential discontinuity in the upstream medium (solar wind) with a supercritical shock (bow shock) \citep[data taken from][]{Omidi2007}. The obstacle is the magnetosphere. {\it Left}: The bulk velocity in colour coding. The simulation plane is the noon meridian plane (North-South).  The magnetosphere is on the right, upstream is on the left. Originally the magnetic field had a shock normal angle $\thetabn=45^\circ$, i.e. the upper half of the shock is quasi-perpendicular, the lower half quasi-parallel. The tangential discontinuity (TD) is parallel to the field (white fat line in the lower left corner). Approaching the shock from left it moves down in the box from upper left corner to its position shown here at time $t\omega_{ci}=150$. Behind the TD the magnetic field points into the simulation plane. Hence $E_n$ points away from the TD behind and into the TD in front of the discontinuity. The case is non-symmetric. {\it Centre}: Enlargement of the white box in the panel on the left. Shown is the ion density. {\it Right}: Ion temperature in the enlarged box. }\label{chap5-fig-hfa3}
\end{figure}
These simulations are hybrid and thus not completely reliable as the effect of the electrons is not taken into account. Here the electron fluid is simply heated. Moreover, they use a perpendicular shock, i.e. the simulation is relevant only for the initial state of HFA formation. 

\paragraph{`Global' hybrid simulations.} The question arises whether the HFA survives or is modulated when moving into the quasi-parallel shock region. This question has been investigated with the help of a {\it global} two-dimensional hybrid simulation \citep{Omidi2007} which takes account of the entire obstacle, in this particular case the magnetosphere in its interaction with the supersonic solar wind. In such a simulation, because of the finite transverse size of the obstacle (the blunt magnetosphere), the bow shock assumes its natural curvature around and distance from the blunt obstacle. A tangential discontinuity that is convected toward the obstacle thus starts interacting with the self-consistent shock first in the quasi-perpendicular region before reaching the quasi-parallel domain. On the other hand, global simulations suffer, in addition to being hybrid only, from the largeness of the scale that can be resolved. In these simulations the cell size is $\sim1\lambda_i\times 1\lambda_i$. Moreover, global simulations are not completely collisionless, which introduces some non-realistic element through the resistive scale $\ell_\eta=\lambda_i\sqrt{\nu_c/\omega}$, where $\nu_c$ is the numerical collision frequency, and $\tau=\omega^{-1}$ is the time scale of the variation of the field quantities. 

Figure\,\ref{chap5-fig-hfa3} summarises the results of the global simulation. On the left of this figure a large part of the simulation box is shown at time $t\omega_{ci}=150$ when the tangential discontinuity (TD) has arrived in the quasi-parallel shock region. Shown is the bulk flow velocity in colour coding. Upstream directed velocities are from white to red, downstream directed from green to blue. The two panels in the centre and on the right are enlargements of the white box on the left as indicated. They shown in higher spatial resolution the ion density and ion temperature, respectively. The original upstream magnetic field (in front, i.e. downstream of the tangential discontinuity TD) was directed at a shock normal angle $\thetabn=45^\circ$ (as shown in the lower right corners of the central and right panels). Thus the original shock was quasi-perpendicular in the upper half of the left panel, while in the lower half it was quasi-parallel. Behind the TD, i.e. upstream of it the magnetic field points into the simulation plane, thus being rotated by $90^\circ$ in the discontinuity plane. 
\begin{figure}[t!]
\hspace{0.0cm}\centerline{\includegraphics[width=1.0\textwidth,height=0.4\textheight,clip=]{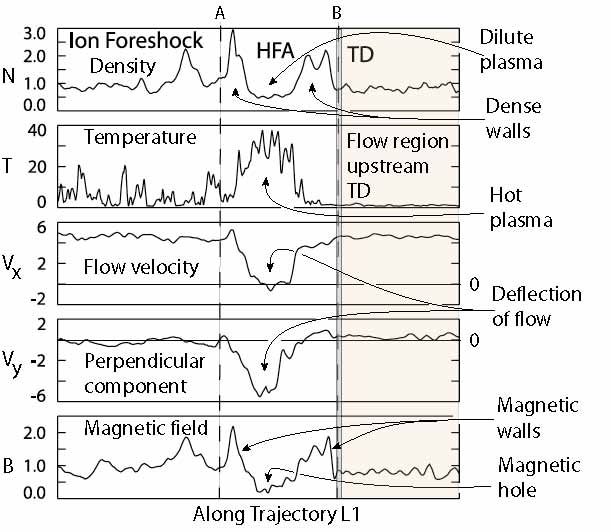} }
\caption[HFA simulation]
{\footnotesize Variation of the plasma parameters and fields along the trajectory \textsf{L1} of a fictive spacecraft passing the HFA in Figure\,\ref{chap5-fig-hfa3}  \citep[data taken from][]{Omidi2007}. The simulated HFA shows all the signs of an observed HFA. }\label{chap5-fig-hfa4}
\end{figure}
The TD being inclined like the original magnetic field at an angle of $45^\circ$ arrived at the shock from above in the left panel touching it first at its quasi-perpendicular part, and then moved down along the shock into the quasi-parallel domain. The first and very interesting observation in this simulation was that during the entire pass along the quasi-perpendicular bow shock no HFA was created. This observation is somewhat disturbing in view of the simulations by \cite{Thomas1991} who used a perpendicular shock. Possibly this is because of the asymmetry of the normal electric field in the ``global" simulation which inhibits accumulation of reflected ions on the backside of the TD.

When the TD arrives at the quasi-parallel shock a Hot Flow Anomaly (HFA in the figure)  is immediately formed. This is shown for the time $t\omega_{ci}=150$ in the figure.  The left panel plots the local ion velocity. White-to-red colouring indicates upstream directed velocities, conversely green-to-blue colouring indicates downstream directed velocities. The magnitude of the velocity has been colour coded (as given by the bar on the right of the left panel in relative simulation units). The Hot Flow Anomaly  appears only on the front side of the TD and is seen as an extended region of fast upstream directed ions. These ions are concentrated and flowing along the TD, which has caused a distortion of the quasi-parallel shock. 

The central and right panels are enlargements of the white box region in the left panel, but showing now the density and temperature variations, respectively. In both cases white is low, green and blue high. Again, in the density and temperature the HFA appears along the front side of the TD as a very dilute hot plasma region that is bounded by dense plasma walls which are approximately parallel to the TD. 

The bow shock is heavily distorted and deformed by the presence of the TD in a way similar to what had been found in the local hybrid simulations of \cite{Thomas1991}, bending into the downstream direction, away from its original shield shape.  On the downstream side of the TD there is no sign of any distortion of the medium other than that the bow shock has become a broad transition region of upstream directed velocities with a sharp, dense upstream edge, which \cite{Omidi2007} identify as a ``solitary shock".  This sharp edge is also seen in the density and temperature plots. 
\begin{figure}[t!]
\hspace{0.0cm}\centerline{\includegraphics[width=1.0\textwidth,clip=]{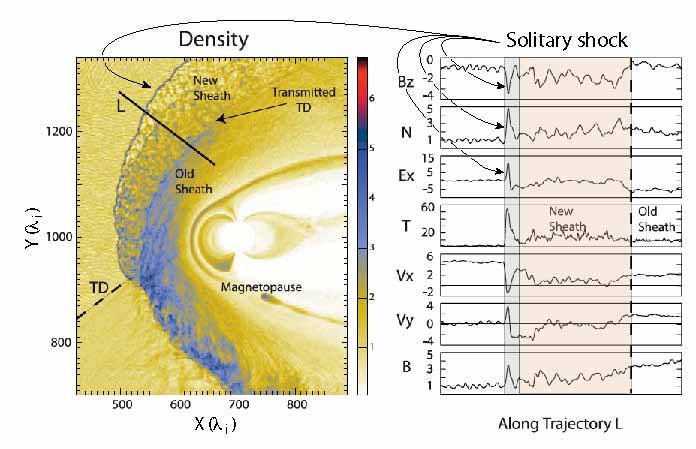} }
\caption[HFA simulation]
{\footnotesize `Global hybrid' simulation of the formation of a `solitary shock' at the bow shock  \citep[data taken from][]{Omidi2007}. {\it Left}: The solitary shock as a replacement of the bos whock in the quasi-perpendicular bow shock region appears as a narrow sharp boundary with a new downstream region behind it (labelled `New sheath'). The `Old sheath' is seen as a dense region. {\it Right}: Plasma data taken along the fictive spacecraft trajectory indicated in the left part of the figure by \textsf{L}. The solitary shock is shaded, while the new sheath region is light coloured.} \label{chap5-fig-solsh1}
\end{figure}

Most interesting is the strong effect of the TD on the ion population. Upstream of the TD, i.e. behind it, the plasma is cold and dilute and of high downstream directed velocity. Downstream, in front of the TD, the plasma is hot and dense, being composed of the hot diffuse ion component that populates the upstream foreshock of the quasi-parallel shock. The TD sweeps this diffuse ion component along when moving with the flow, forming a very sharp boundary between the two regions. It thus seems that the diffuse upstream ion component at the quasi-parallel shock is heavily involved into the formation of the HFA. These ions are absent at the quasi-perpendicular shock. However, the mechanism is not clear in this case, and while it is very interesting to see these differences between the local and global simulations, the global simulations differ so much from the local simulations that it is very difficult to compare them on the same safe grounds.

In order to compare the simulation results with observation \cite{Omidi2007} let a fictive spacecraft fly along the line \textsf{L1} in their simulations. Figure\,\ref{chap5-fig-hfa4} shows the changes in the plasma and field parameters along \textsf{L1}. On the left in this figure is the ion foreshock region (up to point A) with the usual fluctuations in the plasma and field quantities that are typical for the foreshock. The fictive spacecraft encounters the HFA about at point A, detecting a dense wall and strong magnetic field that separate the hot HFA interior from the environment. This interior has the property that the flow velocity $V_x$ has dropped to zero while the transverse flow velocity $V_y$ has taken over the entire moment of the flow. Note that outside the HFA this component was zero. The magnetic field inside the HFA is weak, either corresponding to a magnetic hole or a current sheet. Leaving the HFA close to the tangential discontinuity TD, density and field are again enhanced, and the flow returns to its original direction. All this happens only on the front side of the TD. After passage across the TD, the fictive spacecraft finds itself located in the upstream flow domain well outside the foreshock without any signature of an HFA. The plasma is dilute, cold and quiet and the magnetic field is weak, here. 
\begin{figure}[t!]
\hspace{0.0cm}\centerline{\includegraphics[width=1.0\textwidth,height=0.4\textheight,clip=]{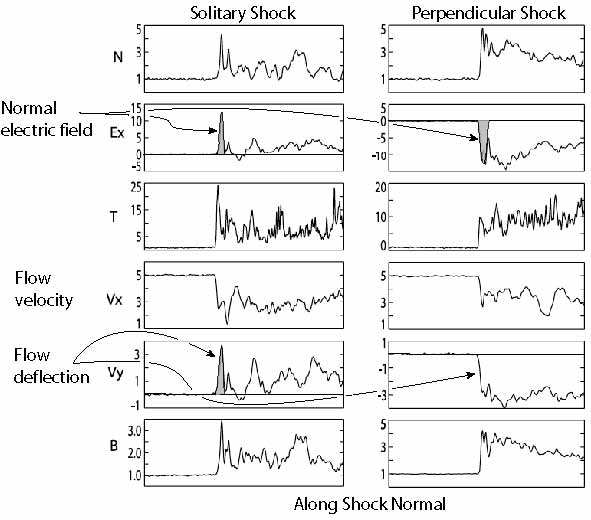} }
\caption[HFA simulation]
{\footnotesize Variation of the plasma parameters and fields along a direction normal to the `solitary shock' and a model `perpendicular shock'  \citep[from][]{Omidi2007}. Indicated are the main differences: the motional electric field and the deflection of the flow which are both in opposite directions. }\label{chap5-fig-solsh2}
\end{figure}

Most of these signatures agree quite well with the observations of Hot Flow Anomalies. As we have noted, the discrepancy is mainly the boundary property of the TD between the foreshock and the quiet upstream flow. Moreover, the HFA appears only on the front side of the HFA lying entirely inside the ion foreshock on the quasi-parallel shock. Whether this is an effect of the asymmetry of the normal electric field at the tangential discontinuity is an open question that cannot be decided at the present time but needs more and specialised simulations. Concerning the tangential discontinuity we note that the `global' simulations seem to indicate that on the perpendicular side of the shock the tangential discontinuity simply passes the shock and survives in a modified form behind the shock in the downstream flow. This seems to be the case in the upper part of the left panel in Figure\,\ref{chap5-fig-hfa3} which advocates such a conclusion. 

\subsection{``Solitary shock"}
\noindent \cite{Omidi2007}, in their `global' hybrid simulations, discovered a new type of a thin shock, which they call a `solitary shock' (because of its narrowness). The signature of this solitary shock is seen in Figure\,\ref{chap5-fig-hfa3} as the sharp boundary between downstream and upstream directed flows behind the tangential discontinuity. 

\cite{Omidi2007} investigated this new type of shock more closely, finding that it is completely unrelated to the presence (or absence) of the tangential discontinuity being a proper phenomenon of formation of a bow shock in front of a blunt obstacle in a magnetised flow (note that the `global' simulations are not completely collisionless, even though in investigating the `solitary shock' \cite{Omidi2007} have tried to determine the dependence of its formation on the resistive scale!). Simulations without a tangential discontinuity also show the formation of a `solitary shock'.

The condition under that it is produced seems to be related to the direction of the motional upstream electric shock  field only. If this has a component that is directed upstream (in the shock frame {\it away} from the shock), then 'solitary shocks' seem to form spontaneously. This motional electric field direction determines the drift and acceleration of the shock-reflected ions. Nonetheless, this observation does not yet lead to a complete understanding of the very mechanism that leads to the formation of this new kind of a shock that under the condition of the upstream directed motional electric field component might evolve and is attached to a bow shock.

Figure\,\ref{chap5-fig-solsh1} gives an example of the `solitary shock' taken from the former simulations (still including the TD) plus a variation of the plasma and field quantities along the cross section \textsf{L} in the figure. The `solitary shock' appears as a narrow line in the density which bounds the new sheath region sharply, separating it from the upstream flow. The plasma and field data on the right show the sharpness of the `solitary shock' (grey shading), the spike in the electric field related to it, and the fluctuating new sheath formed behind it all being separated from the old sheath. 

Simulations performed for a perpendicular shock without the presence of a tangential discontinuity but otherwise under exactly the same conditions with just different directions of the motional upstream electric field are compared in Figure\,\ref{chap5-fig-solsh2}. The `solitary shock' on the left behaves very similar to the perpendicular shock on the right. The two exceptions are the initial assumptions that the direction of the electric field in the `solitary shock' is opposite to the direction of the electric field in the perpendicular shock. Related to this is the opposite direction of the flow deflection in $V_y$. In addition, the solitary shock transition is substantially narrower than the transition region of the perpendicular shock.

We do not intend to discuss these observations/simulations in further detail as neither additional simulations or investigations in depth are currently available, nor is a comparison at hand to observations in real shocks like Earth's bow shock. Moreover, local simulations have demonstrated the importance of electron dynamics for shock formation and reformation. This dynamics is completely ignored in the global hybrid simulations. It is therefore not certain whether this finding will survive the experimental test in space or full particle PIC simulations with large mass ratios. Such `solitary shocks' should, if they exist, be found at the bow shock, because the bow shock is a three-dimensional structure. Hence, at some location on its surface the motional electric field should have the correct direction to generate a `solitary shock' . So far no indication of such a phenomenon has been seen, a lacking result which might be due to having escaped recognition.

\section{The Downstream Region}
\noindent All shocks posses a downstream region which is located between the obstacle and the shock or the shock and the slow stream which has been overturned by a fast stream. This downstream region belongs to the shock transition with its properties being determined by and large by the shock. The downstream region has two boundaries, the shock and the (blunt) obstacle in the wide sense of the meaning, being a solid body, an unmagnetized atmospheric gas, a magnetosphere, a magnetic piston, the driver of a blast wave, the volume of a Coronal Mass Ejection (CME), or simply the slow stream flow. In each case the properties of the downstream region will be different because of the differences in the downstream boundary conditions at the obstacle. Therefore, the properties of the downstream region cannot be considered without including the obstacle. Nevertheless, a few general conclusions  can be noted as far as they concern the effect of the shock boundary on the downstream region. 

It is clear that in the average downstream of the shock, whether quasi-perpendicular or quasi-parallel, the flow is decelerated to Mach number ${\cal M}\lesssim 1$, deflected to flow around the obstacle (independent on whether it can penetrate the obstacle in some cases up to a certain percentage like in the presence of a magnetosphere and diffusive processes or reconnection), the magnetic field direction changes, the plasma and magnetic fields are compressed, and the temperature and pressures are  increased. Moreover, the pressure and temperature anisotropies which were very moderate upstream of the shock change across the shock because of some adiabatic heating in the shock and other effects. The precise processes and the degree to what this heating, change in anisotropy and changes in the parameters are predicted for the different kinds of shocks, shock normal angles, and Mach numbers are not precisely known, but the general trend is well described by the above list of changes. Thus, when asking for the effects of the presence of the shock on the downstream region we are less interested in these average quantities than in the dynamic behaviour of the downstream region as a function of the nature of the shock.

Th dynamic behaviour meant is the turbulent behaviour of the plasma downstream of the shock. What re the turbulent properties? Can turbulence develop downstream of a shock? Is the turbulence, if it exists, a consequence of the presence of the shock, i.e. is it generated by the shock, or does it evolve locally in the downstream region, and the shock provides just the background properties for it such as, for instance, plasma pressure anisotropies? Are there differences in the properties of the turbulence between quasi-parallel and quasi-perpendicular shocks? Of what modes is this turbulence made of? What is the shape of the possible turbulent spectra? These are some of the main questions that arise. 

Few of these questions can currently be answered less, because there would be insufficient observational material but, because turbulence theory is still incomplete and has not been developed for regions of such a limited transverse extent like the transition region between a shock and the obstacle in a high Mach number supercritical plasma flow. As usual, the bet investigated example is the Earth's magnetosheath plasma. But even here no consensus has been reached so far about the state of the turbulence and whether the state of the magnetosheath plasma can at all be called turbulent or not. The evolution of turbulence requires time. In a streaming plasma like the magnetosheath, which is one a few Earth radii in diameter, there is barely sufficient time available for the turbulence to develop up to a stationary state. In particular waves of large wavelength where the energy input is expected to occur at the shock, might not have enough time to cascade down to  form a turbulent spectrum. hence any spectrum will be cut at wave numbers corresponding to the scale of the inverse diameter of the magnetosheath.  

Another reason is that the downstream flow is highly inhomogeneous on all scales larger than a few $\lambda_i$. This implies that the theory of homogeneous turbulence does not apply, and it is not known very well what to make out of it, because at the large scales the effect of the presence of the downstream obstacle boundary cannot be neglected anymore, which inhibits to make general conclusions about the turbulent state of the downstream plasma.

\subsection{Sources of downstream turbulence}
\noindent Let us -- for the moment -- assume that the downstream plasma is indeed in a turbulent state, an assumption we may correct later. What are the sources of this turbulence? 

\subsubsection{Upstream waves}
\noindent The first source that may contribute to the downstream turbulence are the upstream waves. 
If the upstream waves vanquish the shock barrier, they can enter the downstream region, interact with the downstream plasma population until either being damped or amplified; they can decay into other waves which the downstream plasma allows to propagate and, if of sufficiently large amplitude, they can cascade down into a broad spectrum of turbulence. As we have seen, there are several types of upstream waves: periodic ultra-low frequency waves, shocklets, huge pulsations, several types of whistlers from upstream propagating Alfv\'en ion-cyclotron waves to phase-locked whistlers standing in front of the shock and its elements, the pulsations; there are waves which are generated in the ramp, high-frequency electron waves, BGK modes, and the waves in the feet of the quasi-perpendicular shock and, since the pulsations have turned out to behave very similar to quasi-perpendicular shocks, the waves in the feet of pulsations, viz. Buneman and modified-two stream modes, depending on the conditions and plasma composition. Of course, in the quasi-parallel shock upstream region the plasma conditions are modified by the presence of the diffuse ion component, which will change the conditions in the foot region of the pulsations. These questions have not yet been attacked properly, at least not to the extent as they have been investigated for quasi-perpendicular shocks \citep[as was done in the papers by][]{Matsukiyo2003,Matsukiyo2006}.
\begin{figure}[t!]
\hspace{0.0cm}\centerline{\includegraphics[width=1.0\textwidth,clip=]{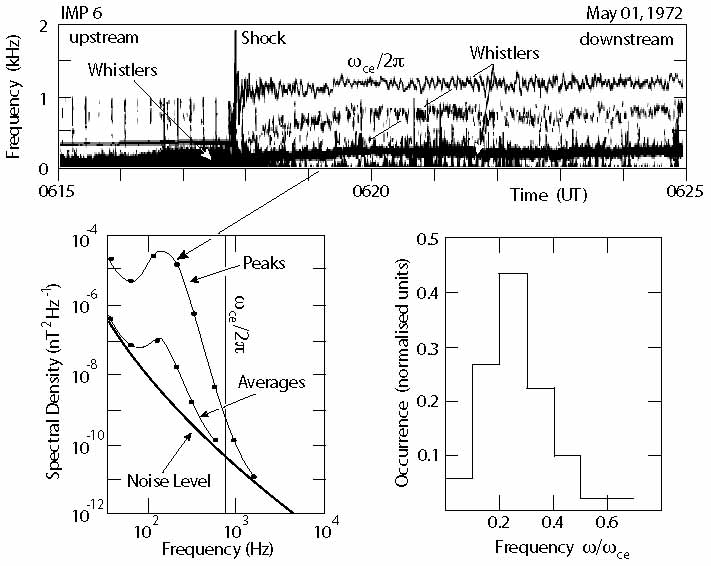} }
\caption[HFA simulation]
{\footnotesize  Connection between upstream (foot) whistlers and downstream (transmitted/excited) long duration whistlers behind the (quasi-perpendicular) shock \citep[data taken from][]{Rodriguez1985}. The whistlers form a well distinct band in the downstream region. {\it Top panel}: An early dynamic spectrum observed by the IMP 6 spacecraft in 1972. The shock is quasi-perpendicular in this case. The vertical lines are time marks. The horizontal line upstream at 300 Hz is the electron gyro-frequency, downstream of the shock it fluctuates strongly around 1 kHz. Waves at the local whistler frequency appear in the shock foot. Some of these waves are whistlers. Behind the shock only the higher frequency part of this whistler band survives. {\it Bottom left}: The downstream magnetic spectrum showing the spectral power density above instrumental noise level. The two curves correspond to few second averages, the upper curve shows the 30 ms peak values within the average measuring time. Both curves show the presence of whistler waves at a fraction of the electron cyclotron frequency. Peak values are up to three orders of magnitude higher indicting the high variability of the downstream waves. {\it Bottom right}: The occurrence frequency of whistler versus frequency. The peak is at about 25\% of the electron cyclotron frequency. } \label{chap5-fig-rodwhist}
\end{figure}
\paragraph{ULF waves.} Quasi-parallel shocks possess a large spectrum of upstream waves that propagates mostly in the upstream direction away from the shock, when seen from the plasma frame but is convected into the shock by the flow. These ultra-low frequency waves propagate in the fast, Alfv\'en-ion cyclotron, and whistler modes. 

In the high-Mach number upstream flow they have no chance of propagating far ahead of the shock. However, arriving at the shock or passing it, the downstream convection becomes sub-magnetosonic and the wave speed has a chance to compensate for the convection. In this case the passing waves will accumulate in the shock transition just behind the ramp. However, the ramp is nonstationary. It reforms quasi-periodically and jumps ahead upstream. In a real system like that of the bow shock where the shock is found at about the same position this reformation implies that the shock ramp oscillates back and forth around its nominal position. It consists of an accumulation of pulsations, and when a new pulsation takes over to become the shock, the old pulsation is expelled downstream with all the upstream waves that have accumulated in it and are attached to it. 

Reformation of a quasi-parallel shock thus implies that quasi-periodically a pulsation and a bunch of upstream waves is added from the shock to the downstream medium where they start their own shock-independent life. They then contribute to the downstream wave population that is in some kind of turbulent state and to wave dynamics, may participate in the resonant or nonresonant interaction with the particle populations or with other waves, or they contribute to the turbulence via the turbulent cascade. 

\paragraph{Whistlers.} The higher frequency branch of the upstream waves propagates in the whistler mode. An example of an observation of (what is believed to be) whistler waves across a (quasi-perpendicular) shock is shown in Figure\,\ref{chap5-fig-rodwhist}. The upstream whistlers in the foot of the shock form a broad intense fluctuation band well below the electron cyclotron frequency (the about stationary horizontal line upstream of the shock ramp). There is a continuous connection between this whistler band across the shock to downstream, now with the low frequency part of it cut off. Only the higher frequencies, which are now found far below the electron cyclotron frequency, survive in the downstream region. It is, however, not known whether these waves have propagated from upstream across the shock (here the bow shock)  to reach the downstream region or whether they are excited about locally in the downstream region (here the magnetosheath). The continuity of the whistler band from upstream to downstream suggests the former: probably most of the downstream spectrum leaks in from upstream. If this is the case, then the lower cut-off of the spectrum is due to the propagation direction of the whistlers. 

The upstream whistler waves are attached to the leading edge of the shock, respectively the pulsation --  in the case of a quasi-parallel shock. We already saw that the arrival of a new pulsation in the quasi-parallel case damps these whistlers out, in the simulations apparently killing them completely. The ejected `old shock front' that has been the `old pulsation' shifts to downstream, but the whistlers which had been attached to it, have gone in the simulation. Hence, any whistlers which occur in the downstream region should have been produced in a different way, either by the downstream particle population, by wave-wave interaction, or in some way by the cascading of the large amplitude pulsation into smaller amplitude waves which build up a turbulent spectrum. Neither of these mechanisms has so far been explored. On the other hand, if the damping of the whistlers is not complete -- which is highly probable in reality --, the pulsation which is ejected to downstream will carry the surviving whistlers along with it and add them to the downstream wave population. In any case the continuity of the whistler band across the shock in Figure\,\ref{chap5-fig-rodwhist} suggests that the downstream whistlers are not independent of the upstream whistlers, at least at the quasi-perpendicular shock. If this is the case, it should not be much different in the quasi-parallel case in particular that everything in the simulations points on the quasi-parallel shock behaving quasi-perpendicular the closer to the shock ramp. 

Local generation of whistlers downstream of the shock requires large electron temperature anisotropies $T_{e\perp}/T_{e\|}>1$ and electron resonant energies ${\cal E}_{e,\,{\rm res}}> \frac{1}{2}m_eV_{Ae}^2$ from linear theory, where $V_{Ae}=B/\sqrt{\mu_0Nm_e}$ is the electron Alfv\'en velocity, here taken in the downstream region. Figure\,\ref{chap5-fig-mshdist1} below indeed indicates an anisotropy in the electron temperature downstream of the shock. It also indicates the presence of counter streaming electron beams in the downstream region. Both, the anisotropy and the beams, might contribute to the excitation of whistlers locally. Whether this is the case has not yet been checked numerically. 
\begin{figure}[t!]
\hspace{0.0cm}\centerline{\includegraphics[width=1.0\textwidth,clip=]{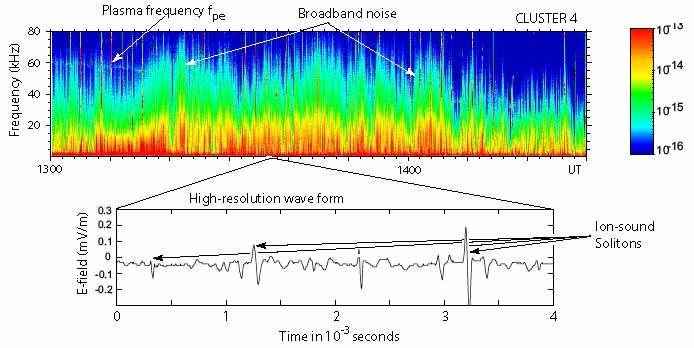} }
\caption[HFA simulation]
{\footnotesize  Downstream magnetosheath high frequency waves observed by the CLUSTER spacecraft \citep[data taken from][]{Pickett2004}. The dynamic spectrum shows indication of the plasma frequency and broadband noise of very spiky nature, similar to that observed earlier by \cite{Rodriguez1975} with the ISEE spacecraft. The high-time resolution wave electric field givn in the bottom panel shows that the latter noise is (as expected) produced from the occurrence of many bipolar electric field structures which indicate the presence of ion acoustic solitons in the downstream region. }\label{chap5-fig-pickett}
\end{figure}
\paragraph{High frequency waves.} Plasma waves like Langmuir waves, Buneman modes and ion-acoustic waves have a rather different behaviour at the shock. The sharp cut-off of Langmuir waves at the local plasma frequency prevents them from passing across the steep density increase at the shock ramp. Only such Langmuir waves can be convected downstream which are generated at the top of the density overshoot in the shock. Langmuir waves from the electron foreshock when being ultimately convected into the shock ramp, assuming that they have not been damped by the diffuse electron population in the foreshock, will stop long before arriving at the shock and will being cut off.

Ion acoustic waves behave differently. As long as they approximately satisfy the linear dispersion relation the increase of the density when approaching the shock and in the shock ramp means that the upstream ion acoustic waves move down on the ion acoustic dispersion curve, i.e. their wave number decreases and they become long wavelength waves which can pass into the downstream region where they ion Landau damping in the warm downstream ion population will absorb their energy and contribute to heating the downstream ions stronger than electrons. It is, however, not clear what happens to the spiky ion acoustic waves and ion BGK-modes or holes which have been observed in the foreshock. Whether these wave survive the passage across the shock cannot be answered yet. 

Figure\,\ref{chap5-fig-pickett} shows an example of high frequency electric waves downstream of the bow shock as has been measured by the {\CL\, \footnotesize 4} spacecraft. The dynamic spectrum shows a weak indication of the plasma frequency in Langmuir waves which must be excited locally either as thermal noise or unstably by weak electron beams. The main signals are, however, the broadband short time emissions which do not show the slightest indication of a dispersive drift and in frequency go even beyond the plasma frequency in several cases. At a plasma frequency of $\sim60$\,kHz taken from the dynamic spectrum, the ion plasma frequency, where the ion acoustic wave branch flattens out, is between 1 and 2 kHz which corresponds to the intense red line at the bottom of the dynamic spectrum. 

Such broadband signals can only be produced by highly localised wave packets in the streaming plasma.  At a streaming velocity of $\sim 100\,{\rm km\,s^{-1}}$ this implies wavelengths of the structures of the order of a few meters. In fact the extremely high time resolution of the wave electric field given in the lower panel confirms the presence of these local wave fields. They form (sometimes non-symmetric) bi-polar signals of amplitudes of some $\sim10^{-4}\,{\rm V\,m}^{-1}$. There are $\sim$ 1 or 2 such structures per ms. At a streaming velocity of $\sim 100\,{\rm km\,s^{-1}}$ this implies that the structures are of transverse size of $<50$\,m, i.e. they are Debye scale structures of the kind of solitons or BGK modes propagating on the ion-acoustic wave branch. These waves and structures may have passed the shock and entered the downstream magnetosheath region. The have similar or even slightly large amplitudes than the similar waves in the foreshock, so they should have been amplified in the interaction with the shock. It is, however, clear that they contribute to the high frequency turbulence in the downstream region.

\subsubsection{Shock generated waves}
\noindent The shock barrier can, on the other hand, permit a limited number of upstream waves to pass downstream, it can refract the waves and it can transform them into other wave modes. In the latter case the shock barrier will itself act as a wave generator becoming a source of downstream turbulence. One type of those waves that is known from the observation and simulation of quasi-parallel shocks are the large amplitude pulsations ({\SL}).  Even though they are not generated in the shock, seen from the downstream region the shock is their source because they reach their largest amplitudes in the shock from where they spread downstream. 

Other proposals are that the shock itself generates low frequency waves due to interface instabilities. Streaming instabilities have been proposed which may arise due to the sheared velocity tangential to the shock. Other proposals concern the generation of interface waves due to the interaction between the downstream hot ion component and the upstream ion flow overlap in the shock ramp region. These interface modes would then be swept downstream by the flow. Typically those waves have long wavelength parallel to the shock surface and short wavelength of the order of the shock width normal to the shock. Along the shock they are included into ripple formation. Quasi-parallel shock reformation by pulsations is one kind of those oscillations of the shock surface. 

The shock transition is of sufficient width for short wavelength electron waves to be excited in the shock by the current flowing in the shock, and by the free energy contained in the shock heated electron distribution. Such waves will be radiated downstream from the shock. However, as we have noted above, Langmuir waves can enter only if their frequency exceeds the high downstream plasma frequency, while ion acoustic waves excited in the shock can be convected downstream.

\subsubsection{Passing particles}
\noindent Part of the downstream turbulence has its origin in the upstream and shock generated waves. In addition, however, the free energy stored in the downstream particle distribution is another source for the excitation of downstream turbulence.  In the first place it is the energetic ions which carry most of the free energy. They are shock-reflected either in the quasi-perpendicular shock or in parts of the quasi-parallel shock. After having been accelerated they can pass the shock ramp and move downstream where they occur as the energetic gyrating ion component which causes a high perpendicular excess in the ion pressure and thus a pressure anisotropy  $P_{i\perp}/P_{i\|}>1$. These ions excite propagating waves on the transverse electromagnetic ion cyclotron branch with frequencies below the ion cyclotron frequency $\omega<\omega_{ci}$ and phase velocities $\omega/k<V_A$. This kind of anisotropy is largest close to the shock where the gyrating ion component has not yet merged into the plasma background. Further down, the plasma flow between shock and obstacle further increases the anisotropy adiabatically. Under certain conditions zero-frequency mirror modes can then be excited which require that  $P_{i\perp}/P_{i\|}-1>\beta_i^{-1}$. Both types of waves have been observed, indeed.

\begin{figure}[t!]
\hspace{0.0cm}\centerline{\includegraphics[width=0.8\textwidth,clip=]{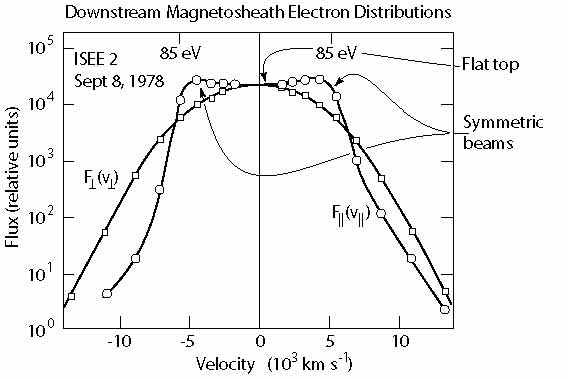} }
\caption[HFA simulation]
{\footnotesize  Downstream magnetosheath electron distributions as measured by ISEE 2 \citep[after][]{Feldman1983}. The perpendicular distribution $F_\perp(v_\perp)$ is a hot symmetric Maxwellian, while the parallel distribution $F_\|(v_\|)$ is cooler, flat top and shows the indication of antiparallel electron beams. }\label{chap5-fig-mshdist1}
\end{figure}

Electrons, on the other hand, do not appear as a transmitted gyrating component in the downstream plasma. They are heated and accelerated in the shock and are seen downstream as a single hot electron component of lesser temperature than the ions. Reasons for this difference in temperature have been given above. Accordingly, electrons do in the average provide a warm background component of little anisotropy. This might, of course, locally be incorrect, when current sheets are formed as the consequence of shock behaviour or by internal processes in the downstream plasma like acceleration in approaching current sheets, current dissipation processes, the internal processes acting in mirror modes, and circumstantially sometimes even reconnection in narrow current sheets. Figure\,\ref{chap5-fig-mshdist1} shows the typical electron distribution in the downstream region as had been measured by the {\footnotesize ISEE} spacecraft \citep{Feldman1983}. Two cuts through this distribution are given, the perpendicular cut is a symmetric Maxwellian $F_{e\perp}(v_\perp)$ of slightly larger temperature than the parallel component. The parallel distribution $F_{e\|}(v_\|)$ has a flat top and shows an indication of the presence of two nearly symmetric but counter streaming electron beams. These measurements are taken close to the shock and are of interest for what concerns the electron wave dynamics of the near shock downstream region. Such weak beams might be involved into the generation of the plasma frequency close behind the shock which is seen in Figure\,\ref{chap5-fig-pickett} as well as for the formation of the BGK modes. They are one of the sources of the high-frequency turbulence in the downstream region. 

The formation of the particular form of the electron distribution shown in the figure has not yet been satisfactorily understood. Electrons are heated in the shock mostly parallel to the field while also being energised  adiabatically in the perpendicular direction in the compressed magnetic fields when crossing the shock. Electron distributions of this kind can well be responsible for the local excitation of whistlers, as we have mentioned above. However, farther downstream of the shock transition the electron distribution looks different, much more isotropic with flat tops in the parallel and perpendicular directions and temperature anisotropies near zero \citep[as seen in the {\CL} measurements analysed by][]{Masood2006}.

In the following we briefly discuss some observations of the downstream turbulence and compare them with simulations of its generation. 

\begin{figure}[t!]
\hspace{0.0cm}\centerline{\includegraphics[width=0.7\textwidth,height=0.6\textwidth,clip=]{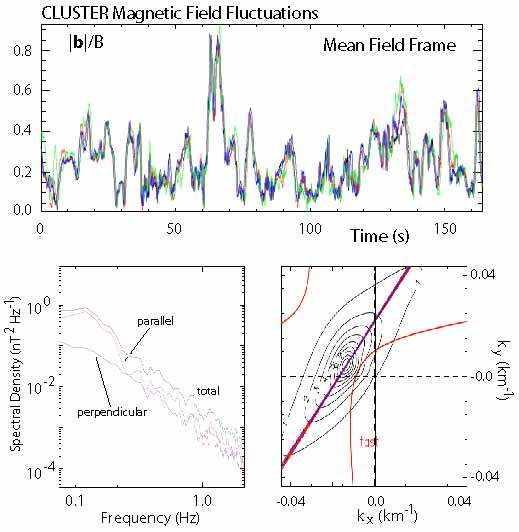} }
\caption[HFA simulation]
{\footnotesize  CLUSTER data of magnetic fluctuations downstream of the bow shock \citep[data taken from][]{Sahraoui2004}. The top panel shows the normalised fluctuation modulus in the mean magnetic field frame for the four spacecraft which are colour coded. Note that on the separation of the spacecraft there is no difference in the fluctuations for any times longer than $\sim 10$\,s. Hence, on these time scales the fluctuations are practically coherent, and there is no turbulence neither on this time scale nor on the corresponding spatial scale. The lower panels show the power spectrum of fluctuations (bottom left) exhibiting some incoherent power law shape that might indicate the presence of turbulence. Based on the spacecraft separation the spectral density in the wave number plane $(k_x,k_y)$ has been determined (bottom right). The power peaks at long wavelengths corresponding to `mirror modes' as indicated by the superimposed projection of the mirror and fast dispersion curves. }\label{chap5-fig-sahraoui}
\end{figure}

\subsection{Downstream turbulence}\noindent
Let us now check whether in the downstream region turbulence can evolve and which are the dominant modes that can be observed. This can be done in two ways, either by referring to observations or by referring to simulations. 

\subsubsection{Observations}
\noindent Figure\,\ref{chap5-fig-sahraoui} shows an example of recent {\CL} observations of magnetic field fluctuations downstream of the Earth's bow shock wave in the deep magnetosheath \citep{Sahraoui2004}.  The top frame is a plot of the normalised fluctuation amplitude for a time period of 160 s and for all four {\CL} spacecraft which were at separation of a few 100 km. The fluctuations measured by the different spacecraft are colour coded, and the main field has been subtracted. For our purposes it is not necessary to know which colour represents which spacecraft. The important observation is, however, that on a time scale longer than $\sim10$\,s there is practically no difference in the signals between the different spacecraft. Thus, on this time scale and the corresponding separation distance which maps to the wave number, the downstream plasma is not in a turbulent state. Any ultra-low frequency waves having entered the magnetosheath or have evolved in the magnetosheath remain as such and have not cascaded down into turbulence while having evolved to nonlinearity, as is signalled by the large relative amplitudes. On the other hand, on the much shorter time scales the signals are incoherent among the spacecraft, and the high frequency fluctuations might have entered some turbulent state. 

This is reflected by the power spectrum in the lower left panel which evolves into a power law tail at frequencies higher than $\sim 0.2$\,Hz. Whether or not this tail signals real turbulence remains to be answered. The time domain power read from this figure is $\sim -2.3\pm0.2$.  If real, it might indicate fractal-dimensional intermittent turbulence, however. 

The panel on the right bottom in the figure is a reconstruction of the $k$-space for the ultra-low frequency waves, the only ones for which it can be done on the basis of the spacecraft distances. The spectral power peaks at very small values of the components $k_x,k_y$ typical for long wavelength, but the structure does not lie on the fast or Alfv\'en modes, they fit better to the mirror mode. Note also that the shape of the wave region in $k$-space is elongated along the mirror dispersion curve. The method of determination of these curves depends strongly on the spacecraft separation and the related resolution and thus remains unaffected from the short wavelength turbulence. \cite{Walker2004} tried to identify the contribution of modes to the higher frequency turbulence. Choosing a frequency of $0.61$\,Hz they found that the spectrum at this frequency was composed of the superposition of an Alfv\'en wave, a weak contribution of a slow mode and some remainder of the mirror mode. It thus seems that the zero frequency mirror mode slowly decays into other wave modes.
\begin{figure}[t!]
\hspace{0.0cm}\centerline{\includegraphics[width=0.95\textwidth,clip=]{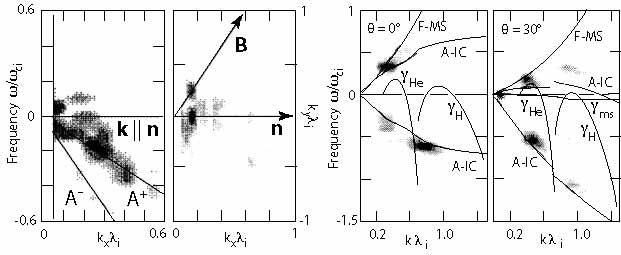} }
\caption[HFA simulation]
{\footnotesize  {\it Left two panels}: Dispersion relation of downstream low frequency waves in two-dimensional hybrid simulations at Mach number ${\cal M}_A=3$ \citep[data taken from][]{Krauss1993}. Shown is the normalised power in the $B_z$ fluctuations plotted versus wave number and frequency. On the left is the dispersion relation in the shock frame indicating the two polarisation branches of the waves, on the right is the wave power in the $(k_x,k_y)$-plane. The waves are mainly in the $A^+$ mode (directed along the shock normal upstream on the plasma frame) splitting into propagation parallel to the magnetic field ${\bf B}$ and the shock normal ${\bf n}$ as seen in the right panel. The waves are thus Alfv\'en-ion cyclotron waves into which the upstream waves had been converted when passing to downstream. {\it Right two panels}:  A two-dimensional confirmation of these results for two wave propagation angles $\theta=0^\circ, 30^\circ$ and two ion populations, protons (H$^+$) and Helium nuclei (He$^{++}$)  \citep[data taken from][]{McKean1996}. }\label{chap5-fig-mshkrauss}
\end{figure}

\begin{figure}[t!]
\hspace{0.0cm}\centerline{\includegraphics[width=0.95\textwidth,clip=]{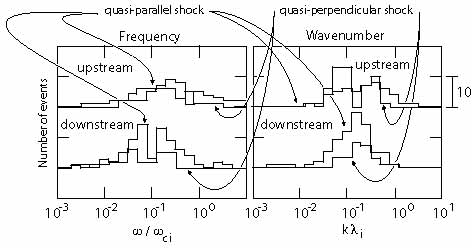} }
\caption[HFA simulation]
{\footnotesize  Statistical behaviour of normalised plasma rest frame frequency $\omega/\omega_{ci}=(\omega_{SC}-{\bf k\cdot V})/\omega_{ci}$ (left panel) and wave number $k\lambda_i$ (right panel) from upstream to downstream in quasi-parallel (solid lines) and quasi-perpendicular (light lines) shocks \citep[data taken from][]{Narita2006}. Note the shift in frequency to lower frequencies which is most prominent in quasi-parallel shocks suggesting that a substantial part of upstream waves is transmitted with about constant frequency. Since $\omega_{ci}$ increases, the ratio will decrease. The wave numbers  increase instead. }\label{chap5-fig-statwave}
\end{figure}

\cite{Narita2005} analysed the plasma rest frame dispersion relation of downstream ultra-low frequency waves in dependence on the downstream distance from the shock (using {\CL} observations across the bow shock when the spacecraft were on short separation of few 100 km). They confirmed the finding that close to the shock the waves seemed to be a mixture of Alfv\'en ion cyclotron and mirror modes. Farther away from the shock closer to the obstacle (the magnetosphere) the mirror mode dominated over the Alfv\'en ion cyclotron mode. \cite{Narita2006} extended this analysis to many observed events for the regions downstream of the quasi-parallel as well as quasi-perpendicular bow shock. They found a downshift in the frequency $\omega/\omega_{ci}$ normalised to the local cyclotron frequency and a decrease in $k\lambda_i$ from upstream to downstream in both cases (see Figure\,\ref{chap5-fig-statwave}). Generally, the downstream region of a quasi-parallel shock is more disturbed than the downstream region of a quasi-parallel shock, i.e. the number of fluctuation events is larger. Moreover, from upstream to downstream the direction of wave propagation changes from along the background (main) magnetic field to perpendicular to the magnetic field, upstream being in the fast-Alfv\'en ion cyclotron mode, downstream showing properties of the slow or mirror modes. From this it might be concluded that the upstream waves (mainly in the case of a quasi-parallel shock) are not transmitted downstream \citep{Narita2006}. The wave vector direction downstream seems to support this conclusion and contradicts the two-dimensional hybrid simulations of quasi-parallel shocks \citep{Krauss1993} which are shown in Figure\,\ref{chap5-fig-mshkrauss}. These simulations suggest that waves can be transmitted and that some of the downstream waves propagate like the upstream waves in the Alfv\'en-ion cyclotron mode in upstream direction and some in the fast mode in downstream direction. 

\subsubsection{Simulations}
\noindent \cite{Krauss1993} in their two-dimensional hybrid simulations consider a quasi-parallel shock at different Mach numbers from ${\cal M}_A<2$ to ${\cal M}_A>3$. As expected at the small Mach numbers upstream standing whistlers are initially attached to the shock. Later the backstreaming ion component adds long wavelength waves to them which are excited by the right-hand resonant ion-ion instability and propagate along the magnetic field. for the low Mach numbers they can escape upstream, but when the Mach number is large they are convected back to the shock and can be transmitted downstream. During the transmission they convert into upstream propagating (directed along the shock normal) Alfv\'en-ion cyclotron modes. Because of the refraction of the magnetic field their wave vector assumes a substantial perpendicular component (Figure\,\ref{chap5-fig-mshkrauss}).  These results have essentially been confirmed by two-dimensional simulations at quasi-perpendicular shocks also shown in Figure\,\ref{chap5-fig-mshkrauss}.  

To study the transmission, \cite{Scholer1997} performed one-dimensional high-Mach number hybrid simulations at a nearly parallel $\thetabn=5^\circ$ shock. These authors wanted to infer about the role of the shock transition region in the transmission process. They found that the shock transition plays a role at medium super-critical Mach numbers ${\cal M}_A<8$ similar to what had been described by \cite{Krauss1993} (who did not see the formation of pulsations because of the small number of diffuse ions in their simulations). For larger Mach numbers right hand polarised waves of both positive and negative helicity were found behind the shock ramp which could have been excited by the resonant ion-ion and the non-resonant right hand ion-ion instabilities in the very shock transition region. Linear theory predicts such a mechanism in the overlap region of the incoming fast ion beam and the dense downstream slow ion component in the very ramp.  Here the relative velocities are large between the two ion components, and the ambient downstream ion density is high, as it should be for the instability to work. Hence the shock transition behaves actively in generating waves in addition to those which are already present. At high Mach numbers these waves have larger wavelengths and, because their phase velocities are large, they are weakly damped and can survive when propagating downstream. Downstream they may undergo an inverse cascade to generate longer wavelength waves in the course of a parametric decay instability. In the very high Mach number regime ${\cal M}_A>10$ no damping of these waves has been observed. This implies that for such high Mach numbers the waves will survive undamped over the entire downstream transition region between shock (for instance the bow shock) and obstacle (for instance the magnetosphere). One would expect to observe monochromatic waves all over the transition region in this case. 
\begin{figure}[t!]
\hspace{0.0cm}\centerline{\includegraphics[width=0.95\textwidth,clip=]{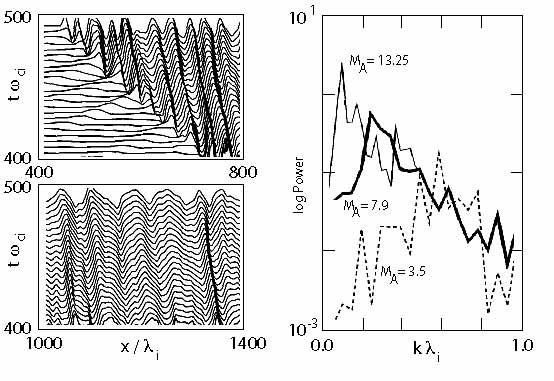} }
\caption[HFA simulation]
{\footnotesize Shock generated downstream waves as inferred from one-dimensional hybrid simulations at high Mach numbers \citep[data taken from][]{Scholer1997}. {\it Left}:  Wave profiles of the negative helicity component of the magnetic field $B_y$ for high Mach number ${\cal M}_A=13.25$  and two different regions $400<x/\lambda_i>800$ and $1000<x/\lambda_i<1400$ in the downstream frame as function of time (time is upward on the ordinate). The first domain  contains the shock location. It is seen that the waves are generated in the shock, are moving in upstream direction but much slower than the shock such that they effectively are moving downstream toward the obstacle. They are large amplitude and practically undamped. The lower panel is taken in the downstream region. The waves have flattened and attain a longer wavelength. Some shorter waves have been produced such that the wave profile is less sinusoidal but still exhibits coherence. {\it Right}: Shock interface generated wave power spectra as function of wave number for three Mach numbers. The wave power increases with Mach number. At the same time the interface waves assume longer wavelengths with the spectral peak shifting to small $k$. Note also that for large ${\cal M}_A\gtrsim 5$ the spectrum has an exponentially decaying tail towards larger $k$. Hence there is no remarkable cascading.}\label{chap5-fig-downwave}
\end{figure}

Figure\,\ref{chap5-fig-downwave} shows the negative helicity waves seen in this simulation at high Mach number. Clearly the waves are generated in the shock ramp as is visible from the upper left panel. The waves assume quite large amplitude and are moving upstream in the downstream frame, however much slower than the shock. They are thus flowing downstream toward the obstacle. With downstream propagation they broaden and attain longer wavelengths. The spectrum on the right shows that the wave power and wavelength increases with Mach number. At large Mach number the spectrum exhibits an exponentially decaying tail which suggests that some shorter wavelengths are generated but that the waves do not participate substantially in a cascade. 
The effects observed in the latter simulations were not present in those of \cite{Krauss1993} because of their much larger shock normal angles and further restrictions on the simulations. It thus seems that quasi-parallel shocks can themselves produce waves in their transition regions. It is, however, difficult to identify these waves in real observations if waves from upstream are mixed in into the transmitted waves. This will necessarily be the case for more oblique shocks that generate large amplitude pulsations upstream which themselves become the shock ramps and during reformation expel the old pulsation-old ramp downstream where it becomes downstream turbulence, respectively it becomes large amplitude coherent downstream waves with broad low frequency spectrum. In addition, whistlers might be attached to these pulsations and escape from them simply because the downstream Mach number is small enough for letting them go. Hence, the downstream turbulence is rather in a mixed state of some large amplitude waves than in a turbulent state, with the mixed state consisting of pulsations, detached whistlers, and interface waves from the shock transition, as well as waves which are excited by  the instability of the diffuse downstream ion component in the downstream plasma background. These downstream diffuse ions should as well interact with the downstream wave component, either damping them, or exciting them, or stimulating their decay in a nonlinear wave-particle interaction process. Turbulence, on the other hand,  will hardly have time to evolve in the narrow transition region between the shock and the obstacle for high Mach numbers, because the transition time for the waves will in general be shorter than the decay time of the waves or the typical time to become involved into a turbulent cascade. Therefore any conclusions about real turbulence must be taken with care.

Another interesting conclusion that can be drawn from these considerations and simulations concerns the possibility of mirror modes near the shock. The simulations have all shown that the release of pulsations from the shock to downstream is accompanied by a sequence of large amplitude magnetic fluctuations and magnetic holes with the latter containing groups of hot nearly isotropic ions which compensate for the pressure balance. These holes are similar to mirrors but are in fact generated not by the mirror instability but by the shock process of accumulation of pulsations and trapping of ions between the pulsations. They are in fact enforced structures that have not evolved through an instability but in the violent impact process of an upstream pulsation onto the shock ramp. Their observation downstream of the shock implies that the pulsations can survive for quite a long distance downstream.  

We are not entering into any discussion of any processes in the downstream region farther away from the shock close to the obstacle. Such processes are strongly affected by the nature of the obstacle and have relatively little in common with the shock which for them provides an input source and a boundary condition. The latter consists for any waves generated near the obstacle in the inhibition of propagation across the shock into the region upstream of the shock. As we will see in the section on particle acceleration such a propagation is possible only for fast particles and, of course, for high frequency radiation. Fast particles are produces in almost all shocks to some degree. However, in non-relativistic shocks no efficient mechanism  is known that would be capable of producing electromagnetic radiation downstream of the shock.

\section{Summary and Conclusions}
\noindent Supercritical quasi-parallel shocks behave completely different from quasi-perpendicular shocks. They are subject to a very high fluctuation level with the shock ramp often very hard to distinguish from the environment in observations. They possess extended foreshocks which are divided into electron and ion foreshocks. Electron foreshocks contain a shock-reflected and accelerated upstream electron population that is flowing in upstream direction. These electrons are believed to have their origin at the quasi-perpendicular part of a bent shock. So far it remains unclear of how these electrons are generated and reflected. The shock potential is attracting and not repulsing for electrons. Therefore, the mechanism of generating reflected electrons is probably kind of a mirror mechanism in converging magnetic fields. Reflected electrons return into the upstream flow along the tangential field line as narrow fast electron beams. These can generate a spectrum of plasma waves via several instabilities but can survive due to the combined action of convection and escape from the self-excited wave population. Probably only part of the electron distribution is subject to quasi-linear depletion. 

The ion foreshock is less extended upstream. Similar to electrons an ion beam is flowing along the ion-tangential field line and is responsible for generation of ultra-low frequency waves. These waves fill the ion foreshock. In addition, the ion foreshock is filled with a diffuse energetic ion population which interacts with the wave spectrum. The upstream waves are instrumental in shaping the quasi-parallel shock. During their interaction with the upstream ion population they evolve into `shocklets' and large amplitude `pulsations', so-called `{\SL}', which have a particular structure consisting of upstream leading fronts and downstream trails. In the plasma frame they move upstream at slow speed. in the shock frame they are convected toward the shock front, interact with the diffuse ion background gradient in a resonant way, steepen, become very large amplitude and, when arriving at the shock ramp, clump together with other pulsations, accumulate, retard to zero speed and form the new shock ramp. They are responsible for quasi-linear shock reformation, which is completely different from quasi-perpendicular shock reformation. Quasi-parallel shocks reform due to accumulation of pulsations, while quasi-perpendicular shocks reform due to bunched gyrating foot ions.

During reformation the old pulsation/ramp is expelled downstream. Whistler that had been standing in front of it become partially damped. These and the old pulsation form the initial source of downstream turbulence. however, other contributions to turbulence downstream are shock-interfac generated waves, leaking whistlers, and locally excited plasma waves due to anisotropy in the particle distribution and due to inhomogeneity. 

Tow important facts concerning quasi-parallel shocks should be noted here. The first is that the generation of pulsations in the foreshock implies that the upstream magnetic field is turned about into tangential direction to the nominal shock surface. This happens on the scale of a few ion inertial length or less. On the large scale the quasi-parallel shock remains to behave like a quasi-parallel shock. however, on the short scale the quasi-parallel shock becomes locally quasi-perpendicular, a fact that should be very important in considering particle dynamics, in particular the reflection and acceleration of electrons. It should be noted in this context that no ion beams have been seen in front of the quasi-parallel shock neither in observations nor in the available simulations. This suggests that quasi-parallel shocks remain to behave quasi-parallel for ions, which implies that the scale on that the shock becomes quasi-perpendicular leaves not enough space for ion reflection -- or is strongly fluctuating such that ions become scattered but not really reflected. No consensus has yet been reached on this fact.

The second fact is that the foreshock is populated not by beams but by energetic diffuse ions the density of which increases exponentially towards the shock. There is a close relation between the upstream wave growth and this density scale. It is most interesting that this diffuse component has maximum density at the shock ramp and is otherwise about homogeneously distributed both towards upstream and downstream of the shock. This implies that it is generated in the shock transition. 

All these observations leave a large number of question unanswered of which the production of diffuse energetic ions is only one. How are these ions generated? How are they injected? Why does the quasi-parallel shock (apparently) not generate ion beams, why not on the short scale? How does a quasi-parallel shock look like in two or three dimensions? It cannot behave stationary because it reforms continuously. It jumps back and forth around its nominal position that is governed by the arrival of new pulsations at the shock. How is the ejection of the old pulsation/ramp to downstream going on? We have seen that in between ion holes are formed and the magnetic field shows minima. What kind of structures are these? Are they related to mirror structures, as is naively believed, or aren't they rather violently produced ion holes instead of growing from an instability? What kind are the different particle distributions just upstream and just downstream of the shock ramp? Which waves can they drive unstable? What is the mechanism that generates the electron holes and spiky electric field structures that have been observed in quasi-parallel shocks? Is the generation of radiation at the shock ramp possible at all? 

The list of these question could be extended. It is clear that their answering requires considerably more observational work, mostly on the small scales. In addition, it requires performing full particle PIC simulations with realistic mass ratio in two or even three dimensions in order to include the variability of the shock along its surface and investigating the reaction of the various particle components to the shock ramp variations.


\end{document}

%% file: calbook_a_defs.tex
\def\3{\ss}

\def\q0{\phantom{1}}

\def\thetabn{\Theta_{Bn}}

\def\parp2{\frac{\partial^{2}}{\partial p^{2} }}

\def\CL{\footnotesize CLUSTER}
\def\SL{\footnotesize SLAMS}
\def\IRM{\footnotesize AMPTE IRM}

\def\m21{$2^{\circ}\times 1^{\circ}$}

\def\ts{\thinspace}

\def\ne8{Ne\ts{$\scriptstyle {\rm VIII}$} }

\def\CL{\footnotesize CLUSTER}
\def\SL{\footnotesize SLAMS}
\def\IRM{\footnotesize AMPTE IRM}

\def\THE{\footnotesize THEMIS}

\newcommand{\asr}{{Adv. \ Space. \ Res.}}

\newcommand{\ag}{{Ann.\ Geo\-phy\-si\-c\ae}}

\newcommand{\grl}{{Geo\-phys.\ Res.\ Lett.}}

\newcommand{\jgr}{{J.\ Geo\-phys.\ Res.}}

\newcommand{\pss}{{Planet.\ Space Sci.}}

\newcommand{\ssr}{{Space Sci.\,Rev.}}

\newcommand{\npg}{{Nonlin.\ Process.\ Geophys.}}

\newcommand{\pop}{{Phys.\ Plasmas}}

\def\ion[#1 #2]{#1\,{\sc #2}}
\def\lamb[#1]{#1\,{\AA}}
\def\serts89{SERTS-89}
\def\fe12{Fe\,{\sc xii}}

\def\mb[#1]{\makebox[0.15cm][l]{#1}}


%% file: SHOCK_5a.bbl
\begin{thebibliography}{} 
{\scriptsize
\bibitem[\protect\citeauthoryear{\it Anderson et al.}{1981}]{Anderson1981}
Anderson R. R. et al.: 1981,
{\it \jgr}  {\bf 86}, 4493-4510.

\bibitem[\protect\citeauthoryear{\it Archer et al.}{2005}]{Archer2005}
Archer M. et al.: 2005,
{\it \jgr}  {\bf 110}, A05208, doi:10.1029/2004JA010791.


\bibitem[\protect\citeauthoryear{\it Asbridge et al.}{1968}]{Asbridge1968}
Asbridge J. R., Bame S. J. \& Strong I. B.: 1968,
{\it \jgr}  {\bf 73}, 5777-5782.

\bibitem[\protect\citeauthoryear{\it Axford et al.}{1977}]{Axford1977}
Axford W. I., Leer E. \& Skadron G.: 1977,
{\it Proc. XVth Int. Conf. Cosmic Rays}  {\bf 11}, 132-134.

\bibitem[\protect\citeauthoryear{\it Bale \& Mozer}{2007}]{Bale2007}
Bale S. D. \& Mozer F. S.: 2007,
{\it \prl}  {\bf 98}, 205001, doi:10.1103/PhysRevLett.98.205001.


\bibitem[\protect\citeauthoryear{\it Barnes}{1970}]{Barnes1970}
Barnes A.: 1970,
{\it Cosmic Electrodyn.}  {\bf 1}, 90-114.

\bibitem[\protect\citeauthoryear{\it Behlke et al.}{2003}]{Behlke2003}
Behlke R. et al.: 2003,
{\it \grl}  {\bf 30}, 1177, doi:10.1029/2002GL015871.

\bibitem[\protect\citeauthoryear{\it Behlke et al.}{2004}]{Behlke2004}
Behlke R. et al.: 2004,
{\it \grl}  {\bf 31}, L16805, doi:10.1029/2004GL019524.

\bibitem[\protect\citeauthoryear{\it Blanco-Cano \& Schwartz}{1997}]{Blanco1997}
Blanco-Cano X. \& Schwartz S. J.: 1997,
{\it \ag}  {\bf 15}, 273-288.

\bibitem[\protect\citeauthoryear{\it Burgess}{1989}]{Burgess1989}
Burgess D.: 1989,
{\it \grl}  {\bf 16}, 345-348.

\bibitem[\protect\citeauthoryear{\it Burgess}{1997}]{Burgess1997}
Burgess D.: 1997,
{\it \asr}  {\bf 20}, 673-682.

\bibitem[\protect\citeauthoryear{\it Burgess \& Schwartz}{1988}]{Burgess1988}
Burgess D. \& Schwartz S. J.: 1988,
{\it \jgr}  {\bf 93}, 11327-11340.

\bibitem[\protect\citeauthoryear{\it Burgess et al.}{2005}]{Burgess2005}
Burgess D. et al.: 2005,
{\it \ssr}  {\bf 118}, 205-222, doi:10.1007/s11214-005-3832-3.

\bibitem[\protect\citeauthoryear{\it Cairns}{1988}]{Cairns1988}
Cairns I. H.: 1988,
{\it \jgr}  {\bf 93}, 858-866, 3958-3968.

\bibitem[\protect\citeauthoryear{\it Czaykowska et al.}{2001}]{Czaykowska1998}
Czaykowska A., Bauer T. M., Treumann R. A. \& Baumjohann W.: 1998,
{\it \jgr}  {\bf 103}, 4747-4753.


\bibitem[\protect\citeauthoryear{\it Czaykowska et al.}{2000}]{Czaykowska2000}
Czaykowska A., Bauer T. M., Treumann R. A. \& Baumjohann W.: 2000,
arXiv:physics/0009046v1 [physics.space-ph].

\bibitem[\protect\citeauthoryear{\it Czaykowska et al.}{2001}]{Czaykowska2001}
Czaykowska A., Bauer T. M., Treumann R. A. \& Baumjohann W.: 2001,
{\it \ag}  {\bf 19}, 275-287.

\bibitem[\protect\citeauthoryear{\it D\'ecr\'eau et al.}{2001}]{Decreau2001}
D\'ecr\'eau P. M. E. et al.: 2001,
{\it \ag}  {\bf 19}, 1241-1258.

\bibitem[\protect\citeauthoryear{\it Dickel \& Wang}{2004}]{Dickel2004}
Dickel J. R. \& Wang S.: 2004,
{\it \asr}  {\bf 33}, 446-449, doi:10.1016/j.asr.2003.08.023.

\bibitem[\protect\citeauthoryear{\it Dubouloz \& Scholer}{1995}]{Dubouloz1995}
Dubouloz N. \& Scholer M.: 1995,
{\it \jgr}  {\bf 100}, 9461-9474.

\bibitem[\protect\citeauthoryear{\it Dum}{1990}]{Dum1990}
Dum C. T.: 1990,
{\it \jgr}  {\bf 95}, 8095-8131.


\bibitem[\protect\citeauthoryear{\it Eastman et al.}{1981}]{Eastman1981}
Eastman T. E., Anderson R. R., Frank L. A. \& Parks G. K.: 1981,
{\it \jgr}  {\bf 86}, 4379-4395.


\bibitem[\protect\citeauthoryear{\it Eastwood et al.}{2003}]{Eastwood2003}
Eastwood J. P., Balogh A. \& Lucek E. A..: 2003,
{\it \ag}  {\bf 21}, 1457-1465.

\bibitem[\protect\citeauthoryear{\it Eastwood et al.}{2002}]{Eastwood2002}
Eastwood J. P. et al.: 2002,
{\it \grl}  {\bf 29}, 2046, doi:10.1029/2002GL015582.

\bibitem[\protect\citeauthoryear{\it Eastwood et al.}{2004}]{Eastwood2004}
Eastwood J. P. et al.: 2004,
{\it \grl}  {\bf 31}, L04804, doi:10.1029/2003GL018897.

\bibitem[\protect\citeauthoryear{\it Eastwood et al.}{2005a}]{Eastwood2005a}
Eastwood J. P. et al.: 2005a,
{\it \jgr}  {\bf 110}, A11220, doi:10.1029/2004JA010618.


\bibitem[\protect\citeauthoryear{\it Eastwood et al.}{2005b}]{Eastwood2005}
Eastwood J. P. et al.: 2005b,
{\it \ssr}  {\bf 118}, 41-94.

\bibitem[\protect\citeauthoryear{\it Eastwood et al.}{2008}]{Eastwood2008}
Eastwood J. P. et al.: 2008,
{\it \grl}  {\bf 35}, L17S03, doi:10.1029/2008GL033475.

\bibitem[\protect\citeauthoryear{\it Edmiston et al.}{1982}]{Edmiston1982}
Edmiston J. P., Kennel C. F. \& Eichler D.: 1982,
{\it \grl}  {\bf 9}, 531-534.

\bibitem[\protect\citeauthoryear{\it Etcheto \& Faucheux}{1984}]{Etcheto1984}
Etcheto J. \& Faucheux M.: 1984,
{\it \jgr}  {\bf 89}, 6631-6653.

\bibitem[\protect\citeauthoryear{\it Fairfield}{1974}]{Fairfield1974}
Fairfield D. H.: 1974,
{\it \jgr}  {\bf 79}, 1368-1378.

\bibitem[\protect\citeauthoryear{\it Feldman et al.}{1982a}]{Feldman1982a}
Feldman W. C. et al.: 1982a,
{\it \prl}  {\bf 49}, 199-201.

\bibitem[\protect\citeauthoryear{\it Feldman et al.}{1982b}]{Feldman1982b}
Feldman W. C. et al.: 1982b,
{\it \jgr}  {\bf 87}, 632-642.

\bibitem[\protect\citeauthoryear{\it Feldman et al.}{1983}]{Feldman1983}
Feldman W. C. et al.: 1983,
{\it \jgr}  {\bf 87}, 96-110.

\bibitem[\protect\citeauthoryear{\it Feldman}{1985}]{Feldman1985}
Feldman W. C.: 1985,
in {\it Collisionless Shocks in the Heliosphere: Reviews of Current Research}  (B. T. Tsurutani \& R. G. Stone, eds., AGU Washington D.C.) pp. 195-205.



\bibitem[\protect\citeauthoryear{\it Fitzenreiter et al.}{1984}]{Fitzenreiter1984}
Fitzenreiter R. J., Klimas A. J. \& Scudder J. D.: 1984,
{\it \grl}  {\bf 11}, 496-499.

\bibitem[\protect\citeauthoryear{\it Fuselier et al.}{1987}]{Fuselier1987}
Fuselier S. A. et al.: 1987,
{\it \jgr}  {\bf 92}, 3187-3194.

\bibitem[\protect\citeauthoryear{\it Gary}{1993}]{Gary1993}
Gary S. P.: 1993,
{\it Theory of Space Plasma Instabilities}  (Cambridge Univ. Press, Cambridge, UK).

\bibitem[\protect\citeauthoryear{\it Ginzburg \& Zheleznyakov}{1958}]{Ginzburg1958}
Ginzburg V. L. \& Zheleznyakov V. V.: 1958,
{\it Sov. Astron.}  {\bf 2}, 653-.

\bibitem[\protect\citeauthoryear{\it Gosling et al.}{1978}]{Gosling1978}
Gosling J. T. et al.: 1978,
{\it \grl}  {\bf 5}, 957-960.

\bibitem[\protect\citeauthoryear{\it Gosling et al.}{1984}]{Gosling1984}
Gosling J. T. et al.: 1984,
{\it \jgr}  {\bf 89}, 5409-5418.
 
\bibitem[\protect\citeauthoryear{\it Greenstadt \& Baum}{1986}]{Greenstadt1986}
Greenstadt E. W. \& Baum L. W.: 1986,
{\it \jgr}  {\bf 91}, 9001-9006.

\bibitem[\protect\citeauthoryear{\it Greenstadt et al.}{1993}]{Greenstadt1993}
Greenstadt E. W. et al.: 1993,
{\it \grl}  {\bf 20}, 1459-1462.


\bibitem[\protect\citeauthoryear{\it Gurnett \& Frank}{1975}]{Gurnett1975}
Gurnett D. A. \& Frank L. A.: 1975,
{\it Solar Phys.}  {\bf 45}, 477-493-9006.

\bibitem[\protect\citeauthoryear{\it Gurnett}{1985}]{Gurnett1985}
Gurnett D. A.: 1985, in
{\it Collisionless Shocks in the Heliosphere: Reviews of Current Research}  (B. T. Tsurutani \& R. G. Stone, eds., AGU Washington D.C.) pp. 207-224.


\bibitem[\protect\citeauthoryear{\it Hasegawa}{1972}]{Hasegawa1972}
Hasegawa A.: 1972,
{\it \jgr}  {\bf 77}, 84-90.

\bibitem[\protect\citeauthoryear{\it Hoppe \& Russell}{1980}]{Hoppe1980}
Hoppe M. M. \& Russell C. T.: 1980,
{\it \asr}  {\bf 1}, 327-332.


\bibitem[\protect\citeauthoryear{\it Hoppe \& Russell}{1981}]{Hoppe1981}
Hoppe M. M. \& Russell C. T.: 1981,
{\it Nature}  {\bf 287}, 417-420.

\bibitem[\protect\citeauthoryear{\it Hoppe \& Russell}{1983}]{Hoppe1983}
Hoppe M. M. \& Russell C. T.: 1983,
{\it \jgr}  {\bf 88}, 2021-2027.

\bibitem[\protect\citeauthoryear{\it Hoppe et al.}{1982}]{Hoppe1982}
Hoppe M. M., Russell C. T., Eastman, T. E. \& Frank L. A.: 1982,
{\it \jgr}  {\bf 87}, 643-650.

\bibitem[\protect\citeauthoryear{\it Hoshino \& Terasawa}{1985}]{Hoshino1985}
Hoshino M. \& Terasawa T.: 1985,
{\it \jgr}  {\bf 90}, 57-64.

\bibitem[\protect\citeauthoryear{\it Jaroschek et al.}{2004}]{Jaroschek2004}
Jaroschek C. H., Lesch H. \& Treumann R. A.: 2004,
{\it \apj}  {\bf 616}, 1065-1071.

\bibitem[\protect\citeauthoryear{\it Jaroschek et al.}{2005}]{Jaroschek2005}
Jaroschek C. H., Lesch H. \& Treumann R. A.: 2005,
{\it \apj}  {\bf 618}, 822-831.



\bibitem[\protect\citeauthoryear{\it Kennel et al.}{1985}]{Kennel1985} 
Kennel C. F., Edmiston J. P. \& Hada T.: 1985, in
{\it Collisionless Shocks in the Heliosphere: A Tutorial Review, R. G. Stone \& B. T. Tsurutani, eds., AGU, Washington D. C.} pp. 1-36. 

\bibitem[\protect\citeauthoryear{\it Kecskem\'ety et al.}{2006}]{Kecskemety2006}
Kecsk\'emety K. et al.: 2006,
{\it \asr}  {\bf 38}, 1587-1594.



\bibitem[\protect\citeauthoryear{\it Kis et al.}{2004}]{Kis2004}
Kis A. et al.: 2004,
{\it \grl}  {\bf 31}, L20801, doi: 10.1029/2004GL020759.

\bibitem[\protect\citeauthoryear{\it Kis et al.}{2007}]{Kis2007}
Kis A. et al.: 2007,
{\it \ag}  {\bf 25}, 785-799.

\bibitem[\protect\citeauthoryear{\it Klimas}{1985}]{Klimas1985}
Klimas A. J.: 1985, in
{\it Collisionless Shocks in the Heliosphere: Reviews of Current Research}  (B. T. Tsurutani \& R. G. Stone, eds., AGU Washington D.C.) pp. 237-252.

\bibitem[\protect\citeauthoryear{\it Koval et al.}{2005}]{Koval2005}
Koval A.,  {\v S}afr{\'a}nkov{\'a} J. \& N{\v e}me{\v c}ek Z.: 2005,
{\it \pss}  {\bf 53}, 41-52.


\bibitem[\protect\citeauthoryear{\it Krasnoselskikh et al.}{2007}]{Krasnoselskikh2007}
Krasnoselskikh et al.: 2007,
{\it \jgr}  {\bf 112}, A10109, doi:10.1029/2006JA012212.

\bibitem[\protect\citeauthoryear{\it Krauss-Varban \& Omidi}{1991}]{Krauss1991}
Krauss-Varban D. \& Omidi N.: 1991,
{\it \jgr}  {\bf 96}, 17715-17731.

\bibitem[\protect\citeauthoryear{\it Krauss-Varban \& Omidi}{1993}]{Krauss1993}
Krauss-Varban D. \& Omidi N.: 1993,
{\it \grl}  {\bf 20}, 1007-1010.

\bibitem[\protect\citeauthoryear{\it Krauss-Varban et al.}{1994}]{Krauss1994}
Krauss-Varban D., Omidi N. \& Quest K. B.: 1994,
{\it \jgr}  {\bf 99}, 5987-6009.


\bibitem[\protect\citeauthoryear{\it Kucharek et al.}{2004}]{Kucharek2004}
Kucharek H. et al.: 2004,
{\it \ag}  {\bf 22}, 2301-2308.

\bibitem[\protect\citeauthoryear{\it Lacombe et al.}{1985}]{Lacombe1985}
Lacombe C., Mangeney A., Harvey, C. C. \& Scudder J. D.: 1985,
{\it \jgr}  {\bf 90}, 73-94.


\bibitem[\protect\citeauthoryear{\it Lee}{1982}]{Lee1982}
Lee M. A.: 1982,
{\it \jgr}  {\bf 87}, 5063-5080.

\bibitem[\protect\citeauthoryear{\it Le \& Russell}{1992}]{Le1992}
Le G. \& Russell C. T.: 1992,
{\it \pss}  {\bf 40}, 1203-1225. 

\bibitem[\protect\citeauthoryear{\it Lemb\`ege et al.}{2004}]{Lembege2004}
Lemb\`ege B. et al.: 2004,
{\it \ssr}  {\bf 110}, 161-226. 

\bibitem[\protect\citeauthoryear{\it Lin et al.}{1974}]{Lin1974}
Lin R. P., Meng C. I. \& Anderson K. A.: 1974,
{\it \jgr}  {\bf 79}, 489-498.

\bibitem[\protect\citeauthoryear{\it Lin}{2002}]{Lin2002}
Lin Y.: 2002,
{\it \pss}  {\bf 50}, 577-591.


\bibitem[\protect\citeauthoryear{\it Lucek et al.}{2002}]{Lucek2002}
Lucek E. A. et al.: 2002,
{\it \ag}  {\bf 20}, 1699-1710.

\bibitem[\protect\citeauthoryear{\it Lucek et al.}{2004}]{Lucek2004}
Lucek E. A. et al.: 2004,
{\it \jgr}  {\bf 109}, A06207, doi:10.1029/2003JA010016.


\bibitem[\protect\citeauthoryear{\it Mandt \& Kan}{1985}]{Mandt1985}
Mandt M. E. \& Kan J. R.: 1985,
{\it \jgr}  {\bf 90}, 115-121.

\bibitem[\protect\citeauthoryear{\it Masood et al.}{2006}]{Masood2006}
Masood W., Schwartz S. J., Maksimovic M. \& Fazakerley A. N.: 2006,
{\it \ag}  {\bf 24}, 1725-1735.


\bibitem[\protect\citeauthoryear{\it Matsukiyo \& Scholer}{2003}]{Matsukiyo2003} 
Matsukiyo S. \& Scholer M.: 2003, 
{\it \jgr} {\bf 108}, 1459. 

\bibitem[\protect\citeauthoryear{\it Matsukiyo \& Scholer}{2006}]{Matsukiyo2006} 
Matsukiyo S. \& Scholer M.: 2006, 
{\it \jgr} {\bf 111}, A06104. 


\bibitem[\protect\citeauthoryear{\it Mazelle et al.}{2000}]{Mazelle2000}
Mazelle C., LeQu\'eau D. \& Meziane K.: 2000,
{\it \npg}  {\bf 7}, 185-190.



\bibitem[\protect\citeauthoryear{\it Mazelle et al.}{2003}]{Mazelle2003}
Mazelle C. et al.: 2003,
{\it \pss}  {\bf 51}, 785-795.

\bibitem[\protect\citeauthoryear{\it McKean et al.}{1996}]{McKean1996}
McKean M. E., Omidi N. \& Krauss-Varban D.: 1996,
{\it \jgr}  {\bf 101}, 20013-20022.


\bibitem[\protect\citeauthoryear{\it Mellott}{1986}]{Mellott1986}
Mellott M. M.: 1986,
{\it \asr}  {\bf 6}, 25-32.

\bibitem[\protect\citeauthoryear{\it Meziane \& d'Uston}{1998}]{Meziane1998}
Meziane K. \& d'Uston C.: 1998,
{\it \ag}  {\bf 16}, 125-133.

\bibitem[\protect\citeauthoryear{\it Meziane et al.}{2004}]{Meziane2004}
Meziane K. et al.: 2004,
{\it \ag}  {\bf 22}, 2325-2335.

\bibitem[\protect\citeauthoryear{\it Muschietti}{1990}]{Muschietti1990}
Muschietti L. : 1990,
{\it Solar Phys.}  {\bf 130}, 201-228.

\bibitem[\protect\citeauthoryear{\it Muschietti \& Dum}{1991}]{Muschietti1991}
Muschietti L. \& Dum C. T.: 1991,
{\it Phys Fluids}  {\bf B3}, 1968, doi:10.1063/1.859665.

\bibitem[\protect\citeauthoryear{\it Narita \& Glassmeier}{2005}]{Narita2005}
Narita Y. \& Glassmeier K.-H.: 2005,
{\it \jgr}  {\bf 110}, A12215, doi: 10.1029/2005JA011256.


\bibitem[\protect\citeauthoryear{\it Narita et al.}{2006}]{Narita2006}
Narita Y. et al.: 2006,
{\it \jgr}  {\bf 111}, A01203, doi: 10.1029/2005JA011231.


\bibitem[\protect\citeauthoryear{\it Narita}{2007}]{Narita2007}
Narita Y.: 2007,
{\it \pss}  {\bf 55}, 243-244.


\bibitem[\protect\citeauthoryear{\it Narita et al.}{2004}]{Narita2004}
Narita Y. et al.: 2004,
{\it \ag}  {\bf 22}, 2315-2323.

\bibitem[\protect\citeauthoryear{\it Narita et al.}{2003}]{Narita2003}
Narita Y. et al.: 2003,
{\it \grl}  {\bf 30}, 1710, doi:10.1029/2003GL017432.

\bibitem[\protect\citeauthoryear{\it Narita et al.}{2006}]{Narita2006}
Narita Y., Glassmeier K.-H. \& Treumann R. A.: 2006,
{\it \prl}  {\bf 97}, 191101.

\bibitem[\protect\citeauthoryear{\it Olson et al.}{1969}]{Olson1969}
Olson J. V., Holzer  R. E. \& Smith E. J.: 1969,
{\it \jgr}  {\bf 74}, 2255-2262.

\bibitem[\protect\citeauthoryear{\it Omidi et al.}{1990}]{Omidi1990}
Omidi N., Quest K. B. \& Winske D.: 1990,
{\it \jgr}  {\bf 95}, 20717-.

\bibitem[\protect\citeauthoryear{\it Omidi \& Sibeck}{2007}]{Omidi2007}
Omidi N. \& Sibeck D. G.: 2007,
{\it \jgr}  {\bf 112}, A01203, doi:10.1029/2006JA011663.


\bibitem[\protect\citeauthoryear{\it Pantellini et al.}{1992}]{Pantellini1992}
Pantellini F. G. E., Heron A., Adam J. C. \& Mangeney A.: 1992,
{\it \jgr}  {\bf 97}, 1303-1311.


\bibitem[\protect\citeauthoryear{\it Parks et al.}{2006}]{Parks2006}
Parks G. K. et al.: 2006,
{\it \pop}  {\bf 13}, 050701, doi:10.1063/1.2201056.



\bibitem[\protect\citeauthoryear{\it Paschmann et al.}{1981}]{Paschmann1981}
Paschmann G. et al.: 1981,
{\it \jgr}  {\bf 86}, 4355-4364.

\bibitem[\protect\citeauthoryear{\it Paschmann et al.}{1988}]{Paschmann1988}
Paschmann G. et al.: 1988,
{\it \jgr}  {\bf 93}, 11279-11294.

\bibitem[\protect\citeauthoryear{\it Peredo et al.}{1995}]{Peredo1995}
Peredo M., Slavin J. A., Mazur E. \& Curtis S. A.: 1995,
{\it \jgr}  {\bf 100}, 7907-7916.

\bibitem[\protect\citeauthoryear{\it Pickett et al.}{2004}]{Pickett2004}
Pickett J. S. et al.: 2004,
{\it Nonlin. Proc. Geophys.}  {\bf 12}, 1-14.

\bibitem[\protect\citeauthoryear{\it Quest}{1985}]{Quest1985}
Quest K. B.: 1985, in
{\it Collisionless Shocks in the Heliosphere: Reviews of Current Research}  (B. T. Tsurutani \& R. G. Stone, eds., AGU Washington D.C.) pp. 185-194.

\bibitem[\protect\citeauthoryear{\it Quest et al.}{1983}]{Quest1983}
Quest K. B., Forslund D. W., Brackbill J. U. \& Lee K.: 1983,
{\it \grl}  {\bf 10}, 471-474.

\bibitem[\protect\citeauthoryear{\it Quest}{1988}]{Quest1988}
Quest K. B.: 1988,
{\it \jgr}  {\bf 93}, 9649-.

\bibitem[\protect\citeauthoryear{\it Reiner et al.}{1996}]{Reiner1996}
Reiner M. J. et al.: 1996,
{\it \grl}  {\bf 23}, 1247-1250.

\bibitem[\protect\citeauthoryear{\it Robinson}{1995}]{Robinson1995}
Robinson P. A.: 1995,
{\it \pop}  {\bf 2}, 1466-1479.

\bibitem[\protect\citeauthoryear{\it Rodriguez \& Gurnett}{1975}]{Rodriguez1975}
Rodriguez P. \& Gurnett D. A.: 1975,
{\it \jgr}  {\bf 80}, 19-31.

\bibitem[\protect\citeauthoryear{\it Rodriguez}{1985}]{Rodriguez1985}
Rodriguez P.: 1985,
{\it \jgr}  {\bf 90}, 241-248.


\bibitem[\protect\citeauthoryear{\it Russell}{1988}]{Russell1988}
Russell C. T.: 1988,
{\it \asr}  {\bf 8}, 147-156.

\bibitem[\protect\citeauthoryear{\it Russell \& Hoppe}{1981}]{Russell1981}
Russell C. T. \& Hoppe M. M.: 1981,
{\it \grl}  {\bf 8}, 615-617.

\bibitem[\protect\citeauthoryear{\it Russell \& Hoppe}{1983}]{Russell1983}
Russell C. T. \& Hoppe M. M.: 1983,
{\it \ssr}  {\bf 34}, 155-172.

\bibitem[\protect\citeauthoryear{\it Russell et al.}{1971}]{Russell1971}
Russell C. T., Childers D. D. \& Coleman P. J., Jr.: 1971,
{\it \jgr}  {\bf 76}, 845-861.

\bibitem[\protect\citeauthoryear{\it Sahraoui et al.}{2004}]{Sahraoui2004}
Sahraoui F. et al.: 2004,
{\it \ag}  {\bf 22}, 2283-2288.


\bibitem[\protect\citeauthoryear{\it Sanderson et al.}{1981}]{Sanderson1981}
Sanderson T. R., Reinhard R. \& Wenzel K.-P.: 1981,
{\it \jgr}  {\bf 86}, 4425-4434.

\bibitem[\protect\citeauthoryear{\it Sanderson et al.}{1983}]{Sanderson1983}
Sanderson T. R. et al.: 1983,
{\it \jgr}  {\bf 88}, 85-95.

\bibitem[\protect\citeauthoryear{\it Scarf et al.}{1971}]{Scarf1971}
Scarf F. L., Fredricks R. W., Frank L. A. \& Neugebauer M.: 1971,
{\it \jgr}  {\bf 76}, 5126-.



\bibitem[\protect\citeauthoryear{\it Scholer}{1985}]{Scholer1985}
Scholer M.: 1985, in
{\it Collisionless Shocks in the Heliosphere: Reviews of Current Research}  (B. T. Tsurutani \& R. G. Stone, eds., AGU Washington D.C.) pp. 287-301.

\bibitem[\protect\citeauthoryear{\it Scholer}{1990}]{Scholer1990}
Scholer M.: 1990,
{\it \grl}  {\bf 17}, 1821-1824.


\bibitem[\protect\citeauthoryear{\it Scholer}{1993}]{Scholer1993}
Scholer M.: 1993,
{\it \jgr}  {\bf 98}, 47-57.

\bibitem[\protect\citeauthoryear{\it Scholer \& Burgess}{1992}]{Scholer1992a}
Scholer M. \& Burgess D.: 1992,
{\it \jgr}  {\bf 97}, 8319-8326.

\bibitem[\protect\citeauthoryear{\it Scholer \& Terasawa}{1990}]{Terasawa1990}
Scholer M. \& Terasawa T.: 1990,
{\it \grl}  {\bf 17}, 119-122.


\bibitem[\protect\citeauthoryear{\it Scholer \& Fujimoto}{1993}]{Scholer1993a}
Scholer M. \& Fujimoto M.: 1993,
{\it \jgr}  {\bf 98}, 15275-15283.

\bibitem[\protect\citeauthoryear{\it Scholer et al.}{1997}]{Scholer1997}
Scholer M., Kucharek H. \& Jayanti V.: 1997,
{\it \jgr}  {\bf 102}, 9821-9833.


\bibitem[\protect\citeauthoryear{\it Scholer et al.}{2003}]{Scholer2003}
Scholer M., Kucharek H. \& Shinohara I.: 2003,
{\it \jgr}  {\bf 108}, 1273, doi:10.1029/2002JA009820.

\bibitem[\protect\citeauthoryear{\it Schwartz et al.}{1985}]{Schwartz1985}
Schwartz S. J. et al.: 1985,
{\it Nature}  {\bf 318}, 269-271.

\bibitem[\protect\citeauthoryear{\it Schwartz et al.}{1988}]{Schwartz1988}
Schwartz S. J. et al.: 1988,
{\it \jgr}  {\bf 93}, 11295-11310.

\bibitem[\protect\citeauthoryear{\it Schwartz et al.}{1992}]{Schwartz1992}
Schwartz S. J. et al.: 1992,
{\it \jgr}  {\bf 97}, 4209-4227.

\bibitem[\protect\citeauthoryear{\it Schwartz \& Burgess}{1991}]{Schwartz1991}
Schwartz S. J. \& Burgess D.: 1991,
{\it \grl}  {\bf 18}, 373-376.


\bibitem[\protect\citeauthoryear{\it Schwartz et al.}{1996}]{Schwartz1996}
Schwartz S. J., Burgess D. \& Moses J. J.: 1996,
{\it \ag}  {\bf 14}, 1134-1150.

\bibitem[\protect\citeauthoryear{\it Sckopke et al.}{1983}]{Sckopke1983}
Sckopke N. et al.: 1983,
{\it \jgr}  {\bf 88}, 6121-6136.

\bibitem[\protect\citeauthoryear{\it Sckopke}{1995}]{Sckopke1995}
Sckopke N.: 1995,
{\it \asr}  {\bf 15}, 261-269.

\bibitem[\protect\citeauthoryear{\it Scudder et al.}{1984}]{Scudder1984}
Scudder J. D., Burlaga L. F. \& Greenstadt E. W.: 1984,
{\it \jgr}  {\bf 89}, 7545-7550.

\bibitem[\protect\citeauthoryear{\it Sentman et al.}{1981a}]{Sentman1981a}
Sentman D. D., Kennel C. F. \& Frank L. A.: 1981a,
{\it \jgr}  {\bf 86}, 4365-4373.

\bibitem[\protect\citeauthoryear{\it Sentman et al.}{1981b}]{Sentman1981b}
Sentman D. D., Edmiston J. P. \& Frank L. A.: 1981b,
{\it \jgr}  {\bf 86}, 7487-7497.

\bibitem[\protect\citeauthoryear{\it Sentman et al.}{1983}]{Sentman1983}
Sentman D. D. et al.: 1983,
{\it \jgr}  {\bf 88}, 2048-2056.

\bibitem[\protect\citeauthoryear{\it Shevyrev et al.}{2006}]{Shevyrev2006}
Shevyrev N. N., Zastenker G. N., Eiges P. E. \& Richardson J. D.: 2006,
{\it \asr}  {\bf 37}, 1516-1521.

\bibitem[\protect\citeauthoryear{\it Shin et al.}{2008}]{Shin2008}
Shin K., Kojima H., Matsumoto H. \& Mukai T.: 2008,
{\it \jgr}  {\bf 113}, A03101, doi:10.1029/2007JA012344.

\bibitem[\protect\citeauthoryear{\it Sibeck et al.}{1999}]{Sibeck1999}
Sibeck D. G. et al.: 1999,
{\it \jgr}  {\bf 104}, 4577-4593.

\bibitem[\protect\citeauthoryear{\it Skadron et al.}{1988}]{Skadron1988}
Skadron G., Holdaway R. D. \& Lee M. A.: 1988,
{\it \jgr}  {\bf 93}, 11354-11362.



\bibitem[\protect\citeauthoryear{\it Sonnerup}{1969}]{Sonnerup1969}
Sonnerup B. U. \"O.: 1969,
{\it \jgr}  {\bf 74}, 1301-1304.


\bibitem[\protect\citeauthoryear{\it Strangeway \& Crawford}{1995}]{Strangeway1995}
Strangeway R. J. \& Crawford G. K.: 1995,
{\it \asr}  {\bf 15}, 29-42.

\bibitem[\protect\citeauthoryear{\it Tanaka et al.}{1983}]{Tanaka1983}
Tanaka M., Goodrich C. C., Winske D. \& Papadopoulos K.: 1983,
{\it \jgr}  {\bf 88}, 3046-3054.

\bibitem[\protect\citeauthoryear{\it Terasawa et al.}{1985}]{Terasawa1985}
Terasawa T. et al.: 1985,
{\it \grl}  {\bf 12}, 373-376.

\bibitem[\protect\citeauthoryear{\it Thomas et al.}{1990}]{Thomas1990}
Thomas V. A., Winske D. \& Omidi N.: 1990,
{\it \jgr}  {\bf 95}, 18809-.

\bibitem[\protect\citeauthoryear{\it Thomas et al.}{1991}]{Thomas1991}
Thomas V. A., Winske D., Thomsen M. F. \& Onssager T. G.: 1991,
{\it \jgr}  {\bf 96}, 111625-11632.

\bibitem[\protect\citeauthoryear{\it Thomsen}{1985}]{Thomsen1985}
Thomsen M. F.: 1985, in
{\it Collisionless Shocks in the Heliosphere: Reviews of Current Research}  (B. T. Tsurutani \& R. G. Stone, eds., AGU Washington D.C.) pp. 253-270.

\bibitem[\protect\citeauthoryear{\it Thomsen et al.}{1985}]{Thomsen1985a}
Thomsen M. F., Gosling J. T., Bame S. J. \& Mellott M. M.: 1985,
{\it \jgr}  {\bf 90}, 137-148.


\bibitem[\protect\citeauthoryear{\it Thomsen et al.}{1986}]{Thomsen1986}
Thomsen M. F. et al.: 1986,
{\it \jgr}  {\bf 91}, 2961-2973.

\bibitem[\protect\citeauthoryear{\it Thomsen et al.}{1988}]{Thomsen1988}
Thomsen M. F. et al.: 1988,
{\it \jgr}  {\bf 93}, 11311-11325.

\bibitem[\protect\citeauthoryear{\it Thomsen et al.}{1993}]{Thomsen1993}
Thomsen M. F. et al.: 1993,
{\it \jgr}  {\bf 98}, 15319-15330.



\bibitem[\protect\citeauthoryear{\it Tidman \& Northrop}{1968}]{Tidman1968}
Tidman D. A. \& Northrop T. G.: 1968,
{\it \jgr}  {\bf 73}, 1543-1553.

\bibitem[\protect\citeauthoryear{\it Trattner \& Scholer}{1994}]{Trattner1994a}
Trattner K. J. \& Scholer M.: 1994,
{\it \jgr}  {\bf 99}, 6637-6650.

\bibitem[\protect\citeauthoryear{\it Trattner et al.}{1994}]{Trattner1994}
Trattner K. J. et al.: 1994,
{\it \jgr}  {\bf 99}, 13389-13400.

\bibitem[\protect\citeauthoryear{\it Treumann}{2006}]{Treumann2006}
Treumann R. A.: 2006,
{\it Astron. Astrophys. Rev.}  {\bf 13}, 229-315.

\bibitem[\protect\citeauthoryear{\it Treumann \& LaBelle}{1992}]{Treumann1992}
Treumann R. A. \& LaBelle J.: 1992,
{\it \apj}  {\bf 399}, L167-L170.

\bibitem[\protect\citeauthoryear{\it Trotignon et al.}{2001}]{Trotignon2001}
Trotignon J. G. et al.: 2001,
{\it \ag}  {\bf 19}, 1711-1720.

\bibitem[\protect\citeauthoryear{\it Tsubouchi \& Lemb\`ege}{2004}]{Tsubouchi2004}
Tsubouchi K.  \& Lemb\`ege B.: 2004,
{\it \jgr}  {\bf 109}, A021114, doi:10.1029/2003JA010014.


\bibitem[\protect\citeauthoryear{\it Tsytovich}{1970}]{Tsytovich1970}
Tsytovich V. N.: 1970,
{\it Nonlinear effects in plasmas}  (Plenum Press, New York; Original in Russian, Moscow 1966).

\bibitem[\protect\citeauthoryear{\it Walker et al.}{2004}]{Walker2004}
Walker S. et al.: 2004,
{\it \ag}  {\bf 22}, 3021-3032.


\bibitem[\protect\citeauthoryear{\it Watanabe \& Terasawa}{1984}]{Watanabe1984}
Watanabe Y. \& Terasawa T.: 1984,
{\it \jgr}  {\bf 89}, 6623-6630.

\bibitem[\protect\citeauthoryear{\it Winske \& Leroy}{1984}]{Winske1984}
Winske D. \& Leroy M. M.: 1984,
{\it \jgr}  {\bf 89}, 2673-2688.

\bibitem[\protect\citeauthoryear{\it Winske et al.}{1990}]{Winske1990}
Winske D., Omidi N., Quest K. B. \& Thomas V. A.: 1990,
{\it \jgr}  {\bf 95}, 18821-.

\bibitem[\protect\citeauthoryear{\it Woolliscroft et al.}{1986}]{Woolliscroft1986}
Woolliscroft L. J. C. et al.: 1986,
{\it \asr}  {\bf 6}, 89-92.

\bibitem[\protect\citeauthoryear{\it Woolliscroft et al.}{1987}]{Woolliscroft1987}
Woolliscroft L. J. C. et al.: 1987, in
{\it Proc. {\rm 21st ESLAB} Symp.},  ESA-SP {\bf 275}, pp. 193-198.

\bibitem[\protect\citeauthoryear{\it Wu}{1972}]{Wu1972}
Wu C. S.: 1972,
{\it \jgr}  {\bf 77}, 575-587.

\bibitem[\protect\citeauthoryear{\it Wu}{1984}]{Wu1984}
Wu C. S.: 1984,
{\it \jgr}  {\bf 89}, 8857-8862.

\bibitem[\protect\citeauthoryear{\it Yoon et al.}{1994}]{Yoon1994}
Yoon P. H.: 1994,
{\it \jgr}  {\bf 99}, 23481-23488.

}


\end{thebibliography}
